\documentclass[sort&compress]{svjour3} % onecolumn (standard format)
\smartqed % flush right qed marks, e.g. at end of proof
\usepackage[utf8]{inputenc}
\usepackage{amsmath}
\usepackage{amsfonts}
\usepackage{amssymb}
\usepackage{listings}
\usepackage{mathrsfs}
\usepackage{newtxtext,newtxmath} % unified Times-compatible text+math font (replaces times+mathpazo)
\usepackage{xcolor}
\usepackage{soul}
\usepackage{eucal}
\usepackage{enumerate}
\usepackage[title]{appendix}
\usepackage{array}
\usepackage{calligra}
\usepackage{multirow}
\usepackage{placeins}
\newcolumntype{M}[1]{>{\centering\arraybackslash}m{#1}}

\usepackage[
pdfstartview=XYZ,
bookmarks=true,
colorlinks=true,
linkcolor=blue,
urlcolor=blue,
citecolor=blue,
pdftex,
linktocpage=true, % make the page number as hyperlink in table of content
hyperindex=true
]{hyperref}

\usepackage{subfig}
\usepackage{natbib}
\setcitestyle{numbers}
\setcitestyle{square}
\setcitestyle{comma}
\usepackage[pdftex]{graphicx}
\usepackage{epstopdf}
\usepackage{booktabs}
\usepackage{tikz}
\usetikzlibrary{shapes.geometric}
\usetikzlibrary{arrows}
\usetikzlibrary{trees}
\usetikzlibrary{positioning, calc, decorations.markings, fit, arrows.meta, backgrounds, patterns, shapes.misc}
\usepackage{indentfirst}
\usepackage{caption}
\usepackage{enumitem}
\usepackage[para]{threeparttable}
\usepackage{upgreek}

\usepackage[letterpaper,margin=3cm]{geometry}

\journalname{Computer Methods in Applied Mechanics and Engineering}

\newcommand{\tensor}[1]{\ensuremath{\boldsymbol{#1}}}

\DeclareMathOperator{\diag}{diag}

\usepackage{algorithm}
\usepackage{algpseudocode}
\newcommand\StateX{\Statex\hspace{\algorithmicindent}}
\algdef{SE}[SUBALG]{Indent}{EndIndent}{}{\algorithmicend}
\algtext*{Indent}
\algtext*{EndIndent}

\newcommand*{\affaddr}[1]{#1}
\newcommand*{\affmark}[1][*]{\textsuperscript{#1}}

\begin{document}

\title{Geometry-informed neural atlas for boundary value problems of complex 3D geometries}

\titlerunning{geometry-informed neural charts}

\author{WaiChing Sun \protect\affmark[a]}

\institute{
\affaddr{\affmark[a] Department of Civil Engineering and Engineering Mechanics, Columbia University, New York, NY 10027, USA. \email{wsun@columbia.edu}} \\
%\affaddr{\affmark[b] Terminal Effects Division, DEVCOM Army Research Laboratory, Aberdeen Proving Ground, MD 21005, USA} \\
%\affaddr{\affmark[c] Department of Chemistry and Materials Science \& Engineering Institute, University of Missouri, Columbia, MO 65211, USA} \\
}

\date{Received: [DATE] / Accepted: [DATE]}

\maketitle

\begin{abstract}
When three-dimensional bodies contain thin features, non-trivial topology, or scan-derived surfaces, volumetric meshing can become the dominant bottleneck in simulation workflows. We replace this step with a learned geometric representation: overlapping volumetric coordinate charts, each equipped with a neural decoder and Jacobian, trained from point-cloud or level-set data to form a differentiable atlas.
Governing equations are pulled back to chart-local reference coordinates via the Piola identity, and local solutions are coupled through multiplicative Schwarz iterations on the overlap graph. Because the atlas is constructed independently of the downstream discretization, one frozen geometric substrate can support fundamentally different solvers (for example, a meshfree physics-informed neural network and a conventional finite-element method) without re-meshing or re-parametrization. Benchmark and verification studies show that the learned atlas preserves expected finite-element convergence behavior and enables both forward and inverse analyses on geometries that would otherwise require solver-specific volumetric meshing.
\end{abstract}


\section{Introduction} \label{sec:introduction}
The conventional computational-mechanics pipeline proceeds from a CAD model or surface description through volumetric meshing to a PDE solver \citep{de_souza_neto_computational_2011,simo_computational_1998,zienkiewicz_finite_2000}. When the spatial domain is geometrically complex---containing thin features, strongly curved interiors, non-trivial topology, or surfaces imported from scanning data---meshing itself becomes a substantial engineering task \citep{quey_large-scale_2011}. 
This paper introduces an alternative approach that uses neural networks to represent complex geometries, including domains that are not homeomorphic to a sphere.
We refer to this representation as a neural atlas. It is designed to define the computational domain, provide the change-of-variables formulas required by the PDE operator, and organize inter-subdomain coupling within one geometric construction.
A neural atlas is an overlapping collection of volumetric coordinate charts parameterized by neural networks. Each chart is equipped with a decoder map $\varphi_i:\widehat{\Omega}_i\to\Omega_i$ and its Jacobian $\mathbf{J}_i$, and is constructed as a geometric object before any PDE is posed. By ``geometry-first'', we mean that an atlas generated by this pipeline serves as the computational substrate for local solvers---neural or classical---without re-meshing or re-parametrization. Governing equations are then pulled back to chart-local reference coordinates via the Piola identity, and chart-local solutions are coupled through multiplicative Schwarz iterations with partition-of-unity blending.

\subsection{Relation to prior work}
The Piola-mapped operators used in this work are standard in isogeometric analysis \citep{hughes2005isogeometric,cottrell2009isogeometric} and mapped finite elements \citep{ciarlet2002finite}. In the multi-patch IGA setting, NURBS patches with explicit parametrizations are coupled through mortar or penalty methods on conforming interfaces \citep{brivadis2015isogeometric}. Nitsche-type coupling \citep{apostolatos2014nitsche} and Schwarz-type overlapping methods \citep{dittmann2019isogeometric} have also been developed for multi-patch IGA on complex geometries.
Accordingly, our contribution is not the Piola mapping itself, but the geometric substrate on which mapped operators are defined. We replace CAD-derived NURBS patches with neural volumetric charts learned directly from point-cloud or level-set data. Chart overlap is an intrinsic part of the representation, so the Schwarz decomposition is induced directly by the atlas construction.

Relative to overlapping Schwarz and domain-decomposition methods \citep{quarteroni1999domain,toselli2005domain}, the present framework borrows the coupling logic but changes the source of the subdomains: the atlas defines both the geometry and the decomposition simultaneously, rather than partitioning an already-discretized domain. Relative to decomposition-based neural solvers such as XPINNs and FBPINNs \citep{jagtap2020xpinns,shukla2021parallel,moseley2021fbpinns}, which partition a pre-defined domain to improve training, here the partition emerges from the charts that form the atlas.

In the neural operator literature, DeepONet \citep{lu2021deeponet} and the Fourier Neural Operator \citep{li2021fno} learn solution maps across parameterized PDE families but require training data from an existing solver and do not supply coordinate chart structure. Physics-informed neural operators (PINO) \citep{li2024pino} reduce the data requirement by incorporating PDE residuals during training but still operate on a fixed spatial grid. The present atlas approach is complementary: it provides the geometric infrastructure on which either data-driven or physics-informed solvers can operate.

In data-driven computational mechanics, \citet{kirchdoerfer2016data} proposed a model-free approach that bypasses constitutive equations entirely, while kernel-based constitutive learning has also been developed by \citet{kanno2021kernel}. Physics-constrained local convexity data-driven modeling for anisotropic nonlinear elastic solids was developed by \citet{he2020physics} where local Euclidean space is constructed to replace classical constitutive updates. Neural-network constitutive models \citep{vlassis2021sobolev,fuhg2024review} learn material response from data while enforcing thermodynamic constraints. \citet{bahmani2022manifold} introduced manifold embedding in a data-driven mechanics setting, showing how low-dimensional geometric structure can regularize constitutive data and improve robustness. In \citet{xiao2022geometric}, geometric priors for multi-resolution yielding manifolds are combined with local closest-point projection to handle nearly non-smooth plasticity. In \citet{jian2024prediction}, geometric learning is integrated with discrete dislocation dynamics data to predict the yield surface of single-crystal copper. This viewpoint is conceptually aligned with our use of geometric structure as a first-class computational object. The inverse identification benchmarks in the present work (Examples~4--5) use parametric constitutive laws rather than learned energy functionals, but the chart-local FEM infrastructure could accommodate neural constitutive models in the same reference-coordinate framework.

Neural implicit representations \citep{park2019deepsdf,mescheder2019occupancy} reconstruct surfaces from data but do not by themselves furnish solver-ready volumetric atlases with chart Jacobians, overlap structure, and transition operators for mapped PDE evaluation. The idea of using neural networks to parametrize coordinate charts is aligned with the deep geometric prior of \citet{williams2019deep}, which learns chart-like neural surface parameterizations for reconstruction. Our contribution is to introduce mapped differential operators on these learned charts and to leverage their overlap-induced decomposition in a multiplicative Schwarz solver. Geometry-aware positional encodings based on Laplace-Beltrami eigenfunctions ($\Delta$-PINNs, \citealt{sahli2024deltapinns}) improve PINN training on complex shapes but still require a background mesh to compute the eigenfunctions and operators, whereas the present atlas provides both the coordinate system and the PDE operator pullback in a single geometric object. 

\subsection{Key Contributions}
The paper makes the following contributions:
\begin{enumerate}[leftmargin=*]
\item \textbf{Neural chart-based geometric decomposition.} Overlapping volumetric charts, equipped with learned decoder maps and quality-assured through geometric gate metrics, define \textbf{both} the domain representation and the Schwarz overlap structure. This same atlas pipeline supports multiple local solver families without re-parametrization. The chart count is a tunable parameter analogous to the number of subdomains in a domain decomposition method. (Examples~1--3).
\item \textbf{Inter-chart consistency through Schwarz coupling.} The multiplicative Schwarz iteration maintains interface displacement jumps at $O(10^{-7})$ throughout history-dependent cyclic loading, and per-chart material parameters converge to machine-precision consensus in the inverse setting, confirming that the coupling enforces global consistency without additional constraints.
\item \textbf{Validations on PDE solvers applied on complex geometries.} On a twelve-chart rabbit Poisson problem, a compact PINN achieves relative $L^2$ error $2.21\times10^{-2}$, while a P1 tetrahedral FEM on an eight-chart atlas built by the same pipeline recovers the expected $O(h^2)$ convergence rate across five refinement levels. This result confirms that the mapped operator, Schwarz coupling, and partition-of-unity blending introduce no order-of-accuracy penalty. Meanwhile, the resultant atlas is a reusable geometric scaffold that can be incorporated in different PDE solvers (Example~2). An eight-chart Schwarz forward BVP solves $J_2$ elastoplasticity with kinematic hardening under cyclic loading on the torus, showing that the atlas infrastructure supports nonlinear equilibrium solves where the displacement field is unknown and must be determined from the balance of momentum (Example~3). Constitutive identification on the same domain recovers Neo-Hookean parameters with machine-precision inter-chart consensus ($\mu$ std $=1.47\times10^{-13}$), and an elastoplastic extension identifies yield stress and hardening modulus from cyclic data (Examples~4--5). These problems show that the proposed method is capable of solving inverse and forward problems in bodies with complex details and not homomorphic to sphere. 
\end{enumerate}

The paper is organized as follows. Section~\ref{sec:methodology} introduces the volumetric atlas and mapped PDE operator. Section~\ref{sec:solver_method} presents the fixed-atlas solvers, including the Schwarz PINN and chart-local FEM realizations. Section~\ref{sec:implementation} summarizes the geometry pipeline and quality gates. Section~\ref{sec:benchmark_formulations} presents numerical evidence, and Section~\ref{sec:discussion} discusses the relation to isogeometric analysis, current limitations, and future directions.

\section{Geometry and Operator Mapping}
\label{sec:methodology}

 We first define the notation used in this section. We use $\boldsymbol{x}$ for a point in the physical body, $\boldsymbol{\zeta}$ for a chart-local reference coordinate on one atlas patch, and $\boldsymbol{\xi}$ for the global reference-ball coordinate used only in the special simply connected examples. This two-level notation is enough for the methodology while avoiding the earlier conflict in which one symbol was asked to represent both a local chart coordinate and a global reference coordinate.

\subsection{Geometric setting: manifold, coordinate charts, and atlas}
\label{sec:atlas_geometry}
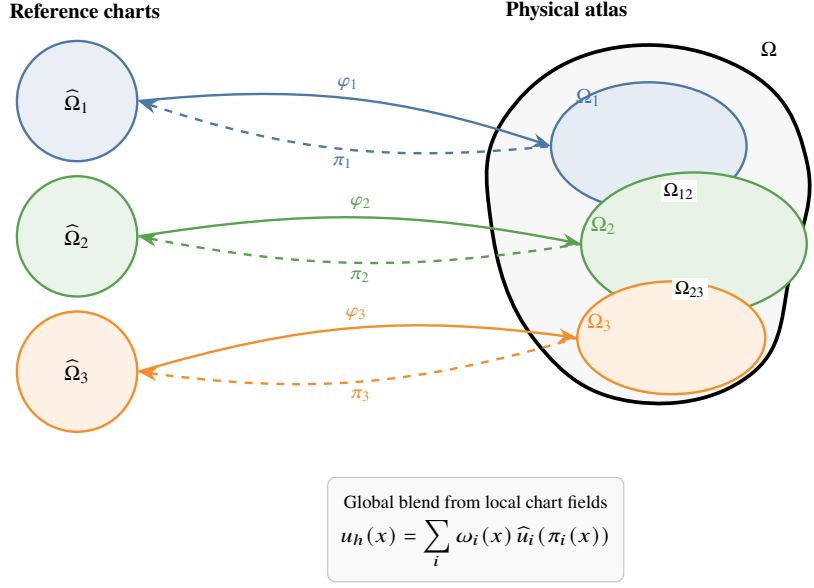
\begin{figure}[htbp]
\centering
\begin{tikzpicture}[
  >=Stealth,
  every node/.style={font=\small},
  refball/.style={circle, draw, line width=0.9pt, minimum size=1.6cm, inner sep=0pt},
  formulabox/.style={draw=gray!50, fill=gray!5, rounded corners=3pt, inner sep=5pt, font=\footnotesize},
]
% --- Color definitions ---
\definecolor{cBlue}{HTML}{4C78A8}
\definecolor{cGreen}{HTML}{59A14F}
\definecolor{cOrange}{HTML}{F28E2B}

% --- Section titles ---
\node[font=\small\bfseries, anchor=west] at (-3.8, 3.4) {Reference charts};
\node[font=\small\bfseries, anchor=west] at (2.8, 3.4) {Physical atlas};

% --- Reference patches (left): circles = unit balls ---
\node[refball, draw=cBlue, fill=cBlue!12] (R1) at (-2.8, 2.2) {$\widehat{\Omega}_1$};
\node[refball, draw=cGreen, fill=cGreen!12] (R2) at (-2.8, 0.4) {$\widehat{\Omega}_2$};
\node[refball, draw=cOrange, fill=cOrange!12] (R3) at (-2.8,-1.4) {$\widehat{\Omega}_3$};

% --- Physical body (right): smooth Bezier outline ---
\begin{scope}[shift={(4.8,0.4)}]
  \draw[line width=1.4pt, fill=gray!6]
    plot[smooth cycle, tension=0.72] coordinates {
      (-1.9, 1.6) (-0.5, 2.5) (1.3, 2.2) (2.1, 1.0)
      (1.9,-0.4) (1.4,-1.8) (-0.2,-2.2) (-1.5,-1.4) (-2.1, 0.2)
    };
  \node[font=\small\bfseries] at (1.6, 2.5) {$\Omega$};

  % --- Chart images inside body: overlapping ellipses ---
  \draw[cBlue, line width=0.9pt, fill=cBlue!14]
    (0.0, 1.2) ellipse (1.3cm and 0.85cm);
  \node[color=cBlue, font=\small\bfseries] at (-0.8, 1.85) {$\Omega_1$};

  \draw[cGreen, line width=0.9pt, fill=cGreen!14]
    (0.6, -0.1) ellipse (1.5cm and 0.95cm);
  \node[color=cGreen, font=\small\bfseries] at (-0.6, 0.15) {$\Omega_2$};

  \draw[cOrange, line width=0.9pt, fill=cOrange!14]
    (0.3, -1.35) ellipse (1.25cm and 0.75cm);
  \node[color=cOrange, font=\small\bfseries] at (-0.65,-1.15) {$\Omega_3$};

  % --- Overlap labels ---
  \node[fill=white, inner sep=1pt, font=\scriptsize] at (0.4, 0.6) {$\Omega_{12}$};
  \node[fill=white, inner sep=1pt, font=\scriptsize] at (0.55,-0.7) {$\Omega_{23}$};
\end{scope}

% --- Bidirectional arrows: phi_i (top, forward) and pi_i (bottom, inverse) ---
% Chart 1
\draw[->, cBlue, line width=0.9pt, bend left=12]
  (R1.east) to node[above, font=\scriptsize, text=cBlue] {$\varphi_1$} ($(4.8,0.4)+(0.0,1.2)+(-1.3,0)$);
\draw[<-, cBlue, line width=0.9pt, bend right=12, dashed]
  (R1.east) to node[below, font=\scriptsize, text=cBlue] {$\pi_1$} ($(4.8,0.4)+(0.0,1.2)+(-1.3,0)$);

% Chart 2
\draw[->, cGreen, line width=0.9pt, bend left=10]
  (R2.east) to node[above, font=\scriptsize, text=cGreen] {$\varphi_2$} ($(4.8,0.4)+(0.6,-0.1)+(-1.5,0)$);
\draw[<-, cGreen, line width=0.9pt, bend right=10, dashed]
  (R2.east) to node[below, font=\scriptsize, text=cGreen] {$\pi_2$} ($(4.8,0.4)+(0.6,-0.1)+(-1.5,0)$);

% Chart 3
\draw[->, cOrange, line width=0.9pt, bend left=12]
  (R3.east) to node[above, font=\scriptsize, text=cOrange] {$\varphi_3$} ($(4.8,0.4)+(0.3,-1.35)+(-1.25,0)$);
\draw[<-, cOrange, line width=0.9pt, bend right=12, dashed]
  (R3.east) to node[below, font=\scriptsize, text=cOrange] {$\pi_3$} ($(4.8,0.4)+(0.3,-1.35)+(-1.25,0)$);

% --- Formula box (bottom center) ---
\node[formulabox, anchor=north] at (2.5, -2.8) {%
  \begin{tabular}{@{}c@{}}
  {\scriptsize Global blend from local chart fields}\\[2pt]
  $\displaystyle u_h(x)=\sum_{i}\omega_i(x)\,\widehat{u}_i(\pi_i(x))$
  \end{tabular}
};
\end{tikzpicture}
\caption{Conceptual view of the volumetric neural atlas. The left side shows chart-local reference patches (unit balls) parameterized by $\boldsymbol{\zeta}$; the right side shows their mapped physical images covering overlapping parts of $\Omega$. Solid arrows represent the chart maps $\varphi_i$; dashed arrows represent the local inverses $\pi_i$. The lower formula indicates the global blend assembled from the chart-local fields. The drawing is a planar slice through a three-dimensional construction.}
\label{fig:atlas_slice}
\end{figure}

Let $\Omega \subset \mathbb{R}^3$ be a bounded body with sufficiently regular boundary $\partial\Omega$. From a computational standpoint, the closure $\overline{\Omega}$ may be viewed as a compact three-dimensional manifold with boundary embedded in the ambient Euclidean space \citep{lee2012introduction}. While a single low-distortion volumetric parametrization is not always sufficient to represent such a body, it may still be described by several compatible local volumetric coordinates, known as coordinate charts. The implementation therefore works with a finite computational atlas of bounded patches in $\mathbb{R}^3$.  This manifold viewpoint provides the minimal structure needed to treat these overlapping patches as a single geometric object (see Fig. \ref{fig:atlas_slice}).

\subsubsection{Coordinate charts and local coordinates}
A coordinate chart is, in general, a pair $(U_i,\varphi_i)$ where $U_i\subset\mathbb{R}^n$ is an open coordinate domain and $\varphi_i:U_i\to\mathcal{M}$ is a smooth map into an $n$-dimensional manifold (or manifold with boundary) \citep{lee2012introduction}. For 3D domains, we set $n=3$, such that each coordinate chart is volumetric. As such, the downstream solver must sample interior collocation points, evaluate volumetric PDE residuals, and communicate information through overlapping subregions. The basic chart family is written as
\begin{equation}
\mathcal{A}=\{(\widehat{\Omega}_i,\varphi_i)\}_{i=1}^{M},
\qquad
\widehat{\Omega}_i \subset \mathbb{R}^3,
\qquad
\varphi_i:\widehat{\Omega}_i \rightarrow \mathbb{R}^3,
\label{eq:atlas_def}
\end{equation}
with physical patches
\begin{equation}
\Omega_i := \varphi_i(\widehat{\Omega}_i),
\qquad
\overline{\Omega}\subset \bigcup_{i=1}^{M}\overline{\Omega}_i.
\label{eq:atlas_cover}
\end{equation}
where $M\in\mathbb{N}$ denotes the total number of charts in the atlas. Assuming that the 3D body is sufficiently smooth, the Jacobian of chart $i$ exists and can be denoted by
\begin{equation}
\mathbf{J}_i(\boldsymbol{\zeta}) := D_{\boldsymbol{\zeta}} \varphi_i(\boldsymbol{\zeta})\in\mathbb{R}^{3\times 3},
\qquad
j_i(\boldsymbol{\zeta}):=\det \mathbf{J}_i(\boldsymbol{\zeta}).
\label{eq:atlas_jac}
\end{equation}
For the mapped PDE derivations below, each chart is \textbf{assumed} to be locally nondegenerate:
\begin{equation}
j_i(\boldsymbol{\zeta})\ge j_{\min}>0
\qquad \text{for all sampled }\boldsymbol{\zeta}\in \widehat{\Omega}_i.
\label{eq:atlas_nondegenerate}
\end{equation}
This condition guarantees a locally invertible change of variables. It does \emph{not} by itself imply global injectivity. Whenever we refer to a chart inverse
\begin{equation}
\pi_i := \varphi_i^{-1}:\Omega_i\rightarrow \widehat{\Omega}_i,
\label{eq:atlas_inverse}
\end{equation}
we mean the inverse on a patch where the chart is assumed or verified to be injective. In the implementation, positive Jacobian determinant is enforced or monitored at sampled points, while global injectivity is treated as a training objective rather than an established geometric property from \eqref{eq:atlas_nondegenerate}.
Each chart is anchored at a seed point $\tensor s_i$ with an orthonormal local frame $\{\mathbf{t}_{1i},\mathbf{t}_{2i},\mathbf{n}_i\}$, and a physical sample $\tensor x$ is first expressed in rigid local coordinates by
\begin{equation}
\boldsymbol{\zeta}^{\mathrm{rig}}_i(\tensor x)
:=
\left(
(\tensor x-\tensor s_i)\cdot \mathbf{t}_{1i},
\,
(\tensor x-\tensor s_i)\cdot \mathbf{t}_{2i},
\,
(\tensor x-\tensor s_i)\cdot \mathbf{n}_i
\right).
\label{eq:atlas_rigid_local_coords}
\end{equation}
This is the coordinate transform used in the atlas-training stage and in the Schwarz solver before the learned decoder adds a bounded residual correction to the rigid frame map. As such, each chart is equipped with a different mapping to a Euclidean coordinate system pre-trained before the solver is applied for the boundary value problem. Unless PINN is used as the local solver, there is no additional training needed to solve the boundary value problem. 

\subsubsection{Atlas, overlaps, and transition maps}
The atlas serves two roles here. Geometrically it represents the body by local volumetric parametrizations.  Algorithmically, it provides the overlap graph and the coordinate transformations needed for local residual evaluation and inter-chart communication. 
An atlas is a collection of such charts whose physical images must cover the body, while overlapping is allowed. The overlapping nature of coordinate chart is leveraged such that it enables neighboring local PDE solves to exchange information \textbf{without} building a global volumetric mesh. 
On overlaps we define
\begin{equation}
\Omega_{ij}:=\Omega_i\cap \Omega_j,
\qquad
\widehat{\Omega}_{ij}:=\pi_i(\Omega_{ij})\subset \widehat{\Omega}_i,
\label{eq:atlas_overlap}
\end{equation}
and the corresponding transition map
\begin{equation}
\psi_{ij}:=\pi_j\circ \varphi_i:\widehat{\Omega}_{ij}\rightarrow \widehat{\Omega}_j,
\qquad
\varphi_i(\boldsymbol{\zeta})=\varphi_j(\psi_{ij}(\boldsymbol{\zeta})).
\label{eq:atlas_transition}
\end{equation}
The transition map $\psi_{ij}$ identifies the same physical point through two neighboring coordinate systems. It is therefore the object that makes overlap consistency meaningful: if $x\in\Omega_{ij}$ and $x=\varphi_i(\boldsymbol{\zeta})$, then $\psi_{ij}(\boldsymbol{\zeta})$ is the coordinate in chart $j$ representing that same $x$. This local compatibility is the geometric counterpart of the chart-to-chart transmission used later by the alternating Schwarz solver.
The overlaps also define the atlas adjacency graph,
\begin{equation}
G_{\mathcal{A}}=(V_{\mathcal{A}},E_{\mathcal{A}}),
\qquad
V_{\mathcal{A}}=\{1,\dots,M\},
\qquad
E_{\mathcal{A}}=\{(i,j):\, i\neq j,\ \Omega_{ij}\neq \emptyset\}.
\label{eq:atlas_overlap_graph}
\end{equation}
Here, $G_{\mathcal{A}}$ is the atlas overlap graph, $V_{\mathcal{A}}$ is the set of chart indices (graph vertices), and $E_{\mathcal{A}}$ is the set of edges connecting chart pairs with non-empty overlap.
This graph is not merely a bookkeeping device; it defines the neighborhood structure used later in the Schwarz sweep and in the graph-coloring step that identifies non-overlapping chart groups for parallel updates.

Figure~\ref{fig:atlas_slice} gives a schematic planar slice through this volumetric atlas. The paired arrows represent the chart maps $\varphi_i$ and their local inverses $\pi_i$, illustrating the bidirectional correspondence between reference coordinates and physical space. 
Table~\ref{tab:notation} summarizes the notation used throughout the remainder of the paper.

\begin{table}[htbp]
\centering
\footnotesize
\caption{Notation used in Section~\ref{sec:methodology}. Physical-space variables, chart-local reference variables, and global reference-ball variables are separated explicitly so that the reader can track which coordinate system is active at each step of the derivation.}
\label{tab:notation}
\begin{tabular}{p{2.35cm}p{3.15cm}p{7.05cm}}
\toprule
Group & Symbol & Role in the method \\
\midrule
\multirow{4}{*}{Physical space}
& $\Omega$, $\partial\Omega$ & physical body and its boundary \\
& $x$ & physical-space point used in the governing PDE \\
& $\Omega_i$, $\Omega_{ij}$ & chart image and chart-overlap region in physical space \\
& $u(x),\,u_h(x)$ & physical state field and the blended global field used for reporting \\
\midrule
\multirow{4}{*}{\shortstack[l]{Chart-local\\ reference}}
& $\widehat{\Omega}_i$ & reference patch attached to chart $i$ \\
& $\boldsymbol{\zeta}$ & local coordinate used to evaluate one chart-local network \\
& $\widehat{\Omega}_{ij}$ & overlap region written in the coordinates of chart $i$ \\
& $\widehat{u}_i(\boldsymbol{\zeta})$ & chart-local state field defined on $\widehat{\Omega}_i$ \\
\midrule
\multirow{2}{*}{\shortstack[l]{Global\\ reference ball}}
& $B^3=\{\boldsymbol{\xi}\in\mathbb{R}^3:\|\boldsymbol{\xi}\|<1\}$ & unit ball used only by the special global-map examples \\
& $\Psi_\theta(\boldsymbol{\xi})$ & learned or analytic global map from the reference ball to the physical domain \\
\midrule
\multirow{4}{*}{Map operators}
& $\varphi_i,\pi_i$ & chart map and its local inverse \\
& $\psi_{ij}=\pi_j\circ\varphi_i$ & transition map from chart-$i$ coordinates to chart-$j$ coordinates on an overlap \\
& $\mathbf{J}_i=D_{\boldsymbol{\zeta}}\varphi_i,\; j_i=\det\mathbf{J}_i$ & Jacobian matrix and Jacobian determinant of chart $i$ \\
& $\omega_i$ & partition-of-unity weight used in the global blend \\
\midrule
\multirow{3}{*}{\shortstack[l]{Vectors and\\boundary data}}
& $\mathbf{n},\,\widehat{\mathbf{n}}_i$ & outward unit normal on a physical or reference boundary \\
& $\mathbf{t}_N,\,\mathbf{t}^{\mathrm{obs}}$ & prescribed Neumann traction and observed traction \\
& $\mathbf{t}_{1i},\mathbf{t}_{2i},\mathbf{n}_i$ & orthonormal TNB frame at chart seed $s_i$ (distinct from boundary normal) \\
\midrule
\shortstack[l]{Shared solver\\ quantities}
& $\eta,\lambda_*$ & $\eta$ denotes global material or inverse parameters; $\lambda_*$ denotes loss weights shared across charts \\
\bottomrule
\end{tabular}
\end{table}

\begin{remark}[Post-processing]
When a single global field is needed for post-processing, we use partition-of-unity weights $\{\omega_i\}_{i=1}^{M}$ satisfying
\begin{equation}
\omega_i:\Omega\rightarrow [0,1],
\qquad
\mathrm{supp}(\omega_i)\subset \Omega_i,
\qquad
\sum_{i=1}^{M}\omega_i(x)=1.
\label{eq:atlas_pou}
\end{equation}
For any chart-local quantity $\widehat{g}_i$, the blended global field is
\begin{equation}
g_h(x)=\sum_{i=1}^{M}\omega_i(x)\,\widehat{g}_i(\pi_i(x)).
\label{eq:atlas_blend}
\end{equation}
\end{remark}

\subsection{Mapped PDE operator on a volumetric chart}
\label{sec:mapped_operator}
Let $\tensor u:\Omega\rightarrow\mathbb{R}^{q}$ be the physical state.
\[
\nabla_x \tensor u(\tensor x)\in \mathbb{R}^{q}\otimes\mathbb{R}^{3},
\qquad
(\nabla_x \tensor u)_{\alpha i}=\frac{\partial u_\alpha}{\partial x_i}.
\]
Here, $q$ denotes the number of components of the state field (e.g., $q=1$ for a scalar field and $q=3$ for 3D displacement).
We write the physical PDE in divergence form as
\begin{equation}
\nabla_x\cdot \mathcal{G}\!\left(\tensor x,\tensor u(\tensor x),\nabla_x \tensor u(\tensor x);\tensor\eta\right)
+ \mathcal{S}\!\left(\tensor x,\tensor u(\tensor x),\nabla_x \tensor u(\tensor x);\tensor\eta\right)=0
\qquad \text{in }\Omega,
\label{eq:abstract_pde}
\end{equation}
where the divergence acts on the spatial index and
\[
\mathcal{G}(\tensor x,\tensor u,\nabla \tensor u;\tensor\eta)\in \mathbb{R}^{q}\otimes\mathbb{R}^{3},
\qquad
\mathcal{S}(\tensor x,\tensor u,\nabla \tensor u;\tensor\eta)\in \mathbb{R}^{q}.
\]
Here, $\tensor\eta$ denotes the vector of model/material parameters (it may reduce to a scalar in special cases).
Physical boundary conditions are written as
\begin{equation}
\tensor u=\tensor u_D \quad \text{on }\Gamma_D,
\qquad
\mathcal{G}(\tensor x,\tensor u,\nabla_x \tensor u;\tensor\eta)\,\mathbf{n} = \mathbf{t}_N
\quad \text{on }\Gamma_N,
\label{eq:abstract_bc}
\end{equation}
with outward unit normal $\mathbf{n}$.

\begin{figure}[htbp]
\centering
\begin{tikzpicture}[
  >=Stealth,
  every node/.style={font=\small},
  formulabox/.style={draw=gray!50, fill=gray!5, rounded corners=3pt, inner sep=5pt, font=\footnotesize},
]
\definecolor{cBlue}{HTML}{4C78A8}
\definecolor{cGreen}{HTML}{59A14F}

% --- Section titles ---
\node[font=\small\bfseries, anchor=south] at (-3.2, 2.8) {Reference patch};
\node[font=\small\bfseries, anchor=south] at (3.8, 2.8) {Physical patch};

% --- Reference patch (left): circle with grid ---
\begin{scope}[shift={(-3.2, 0.8)}]
  % Grid lines inside circle to show structure
  \clip (0,0) circle (1.5cm);
  \foreach \x in {-1.2,-0.6,...,1.2} {
    \draw[cBlue!25, line width=0.3pt] (\x,-1.5) -- (\x,1.5);
  }
  \foreach \y in {-1.2,-0.6,...,1.2} {
    \draw[cBlue!25, line width=0.3pt] (-1.5,\y) -- (1.5,\y);
  }
  % Fill
  \fill[cBlue!8] (0,0) circle (1.5cm);
\end{scope}
\draw[cBlue, line width=1.1pt] (-3.2, 0.8) circle (1.5cm);
\node[font=\normalsize] at (-3.5, 0.5) {$\widehat{\Omega}_i$};
% Coordinate point
\fill[black] (-2.8, 1.0) circle (1.5pt);
\node[font=\scriptsize, anchor=north east] at (-2.85, 0.95) {$\boldsymbol{\zeta}$};
% Tangent basis vectors at zeta (orthonormal reference frame)
\definecolor{cRed}{HTML}{C44E52}
\draw[-{Stealth[length=4pt]}, cRed, line width=0.9pt] (-2.8, 1.0) -- (-2.2, 1.0);
\node[font=\scriptsize, text=cRed, anchor=south] at (-2.35, 1.02) {$\mathbf{e}_1$};
\draw[-{Stealth[length=4pt]}, cBlue, line width=0.9pt] (-2.8, 1.0) -- (-2.8, 1.6);
\node[font=\scriptsize, text=cBlue, anchor=west] at (-2.75, 1.45) {$\mathbf{e}_2$};
% Outward normal
\draw[->, line width=0.8pt] (-3.2, 2.3) -- (-3.2, 2.75);
\node[font=\scriptsize, anchor=west] at (-3.15, 2.55) {$\widehat{\mathbf{n}}_i$};

% --- Physical patch (right): smooth deformed shape with curved grid ---
\begin{scope}[shift={(3.8, 0.8)}]
  % Deformed shape
  \fill[cGreen!8]
    plot[smooth cycle, tension=0.7] coordinates {
      (-1.6, -0.9) (-1.2, 1.2) (0.2, 1.7) (1.5, 1.3)
      (1.9, 0.1) (1.6, -1.0) (0.2, -1.3)
    };
  \draw[black, line width=1.1pt]
    plot[smooth cycle, tension=0.7] coordinates {
      (-1.6, -0.9) (-1.2, 1.2) (0.2, 1.7) (1.5, 1.3)
      (1.9, 0.1) (1.6, -1.0) (0.2, -1.3)
    };
  % Deformed grid (curved lines to show Jacobian effect)
  \begin{scope}
    \clip plot[smooth cycle, tension=0.7] coordinates {
      (-1.6, -0.9) (-1.2, 1.2) (0.2, 1.7) (1.5, 1.3)
      (1.9, 0.1) (1.6, -1.0) (0.2, -1.3)
    };
    % Curved vertical lines
    \draw[cGreen!30, line width=0.3pt] plot[smooth] coordinates {(-1.0,-1.4) (-0.8,-0.2) (-0.9,0.8) (-0.7,1.6)};
    \draw[cGreen!30, line width=0.3pt] plot[smooth] coordinates {(-0.3,-1.3) (-0.1,-0.3) (-0.2,0.6) (0.0,1.7)};
    \draw[cGreen!30, line width=0.3pt] plot[smooth] coordinates {(0.4,-1.2) (0.5,-0.2) (0.3,0.7) (0.5,1.5)};
    \draw[cGreen!30, line width=0.3pt] plot[smooth] coordinates {(1.0,-1.0) (1.1,0.0) (0.9,0.8) (1.2,1.4)};
    % Curved horizontal lines
    \draw[cGreen!30, line width=0.3pt] plot[smooth] coordinates {(-1.5,-0.5) (-0.5,-0.6) (0.5,-0.7) (1.5,-0.6)};
    \draw[cGreen!30, line width=0.3pt] plot[smooth] coordinates {(-1.4,0.2) (-0.4,0.1) (0.6,0.0) (1.7,0.1)};
    \draw[cGreen!30, line width=0.3pt] plot[smooth] coordinates {(-1.3,0.9) (-0.3,0.8) (0.7,0.7) (1.6,0.8)};
  \end{scope}
\end{scope}
\node[font=\normalsize\bfseries] at (4.6, 2.6) {$\Omega_i$};
% Point
\fill[black] (4.2, 1.0) circle (1.5pt);
\node[font=\scriptsize, anchor=north west] at (4.25, 0.90) {$x{=}\varphi_i(\boldsymbol{\zeta})$};
% Pushed-forward tangent vectors (stretched/rotated by J_i)
\draw[-{Stealth[length=4pt]}, cRed, line width=0.9pt] (4.2, 1.0) -- (4.95, 1.2);
\node[font=\scriptsize, text=cRed, anchor=south] at (4.7, 1.18) {$\mathbf{J}_i \mathbf{e}_1$};
\draw[-{Stealth[length=4pt]}, cBlue, line width=0.9pt] (4.2, 1.0) -- (4.0, 1.75);
\node[font=\scriptsize, text=cBlue, anchor=east] at (4.0, 1.55) {$\mathbf{J}_i \mathbf{e}_2$};
% Normal
\draw[->, line width=0.8pt] (5.5, 1.85) -- (5.9, 2.35);
\node[font=\scriptsize, anchor=south west] at (5.85, 2.2) {$\mathbf{n}_i$};

% --- Central arrow ---
\draw[->, line width=1.4pt] (-1.2, 1.0) -- (1.4, 1.0)
  node[midway, above, font=\small] {$\varphi_i$};
\node[font=\scriptsize, anchor=north] at (0.1, 0.85) {$\mathbf{J}_i=D_{\boldsymbol{\zeta}}\varphi_i$};

% --- Formula cards (bottom) ---
\node[formulabox] (f1) at (-3.5, -1.6) {%
  \begin{tabular}{@{}c@{}}
  {\scriptsize Gradient pullback}\\[2pt]
  $\nabla_x u=\nabla_{\boldsymbol{\zeta}}\widehat{u}_i\,\mathbf{J}_i^{-1}$
  \end{tabular}
};
\node[formulabox] (f2) at (0.3, -1.6) {%
  \begin{tabular}{@{}c@{}}
  {\scriptsize Piola flux}\\[2pt]
  $\widehat{\mathcal{P}}_i=j_i\,\widehat{\mathcal{G}}_i\,\mathbf{J}_i^{-T}$
  \end{tabular}
};
\node[formulabox] (f3) at (4.2, -1.6) {%
  \begin{tabular}{@{}c@{}}
  {\scriptsize Normal map}\\[2pt]
  $\mathbf{n}_i=\mathbf{J}_i^{-T}\widehat{\mathbf{n}}_i\big/\|\mathbf{J}_i^{-T}\widehat{\mathbf{n}}_i\|$
  \end{tabular}
};
\end{tikzpicture}
\caption{Mapped-operator view on one chart. The left side shows the local reference patch (unit ball) with a regular grid; the center arrow represents the chart map $\varphi_i$ and its Jacobian $\mathbf{J}_i$; the right side shows the corresponding physical patch with the grid deformed by the map. The formula cards highlight the three identities the implementation uses directly: gradient pullback, Piola-transformed flux, and normal mapping.}
\label{fig:mapped_operator_schematic}
\end{figure}
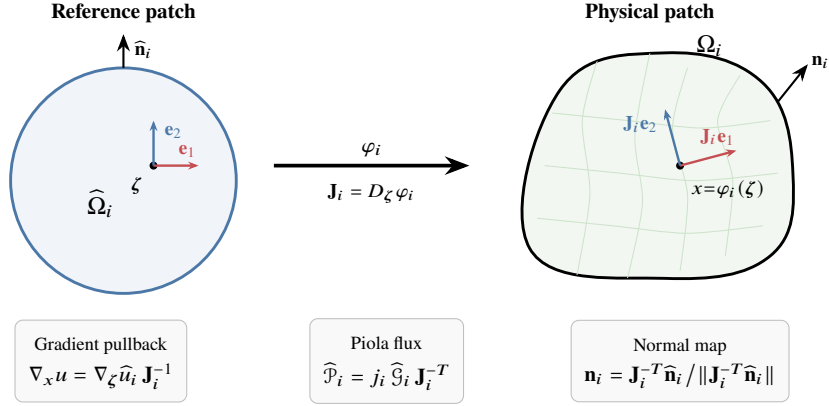

On chart $i$, let $\boldsymbol{\zeta}\in\widehat{\Omega}_i$ denote the local reference coordinate and define the pulled-back state by
\begin{equation}
\widehat{\tensor u}_i(\boldsymbol{\zeta}):=\tensor u(\varphi_i(\boldsymbol{\zeta})).
\label{eq:pullback_state}
\end{equation}
The chain rule takes the form
\begin{equation}
\nabla_x \tensor u(\varphi_i(\boldsymbol{\zeta}))
=
\nabla_{\boldsymbol{\zeta}} \widehat{\tensor u}_i(\boldsymbol{\zeta})\,\mathbf{J}_i(\boldsymbol{\zeta})^{-1}.
\label{eq:grad_pullback}
\end{equation}
The physical flux is first evaluated on the chart and then converted to a reference-space flux by the Piola transform (acting on the spatial index):
\begin{align}
\widehat{\mathcal{G}}_i(\boldsymbol{\zeta};\tensor\eta)
&:=
\mathcal{G}\!\left(\varphi_i(\boldsymbol{\zeta}),\widehat{\tensor u}_i(\boldsymbol{\zeta}),\nabla_{\boldsymbol{\zeta}} \widehat{\tensor u}_i(\boldsymbol{\zeta})\mathbf{J}_i(\boldsymbol{\zeta})^{-1};\tensor\eta\right),
\label{eq:pullback_flux}\\
\widehat{\mathcal{P}}_i(\boldsymbol{\zeta};\tensor\eta)
&:=
j_i(\boldsymbol{\zeta})\,\widehat{\mathcal{G}}_i(\boldsymbol{\zeta};\tensor\eta)\,\mathbf{J}_i(\boldsymbol{\zeta})^{-T}.
\label{eq:piola_flux}
\end{align}
The pulled-back source term is
\begin{equation}
\widehat{\mathcal{S}}_i(\boldsymbol{\zeta};\tensor\eta)
:=
\mathcal{S}\!\left(\varphi_i(\boldsymbol{\zeta}),\widehat{\tensor u}_i(\boldsymbol{\zeta}),\nabla_{\boldsymbol{\zeta}} \widehat{\tensor u}_i(\boldsymbol{\zeta})\mathbf{J}_i(\boldsymbol{\zeta})^{-1};\tensor\eta\right).
\label{eq:pullback_source}
\end{equation}
We reserve the notation $\widehat{\mathcal{P}}_i$ for the Piola-transformed flux entering the reference-space divergence and $\widehat{\mathcal{S}}_i$ for the pulled-back source term, so the flux and source roles remain distinct throughout the derivation.
With these definitions, the Piola identity $\nabla_x\cdot\mathcal{G} = j_i^{-1}\,\nabla_{\boldsymbol{\zeta}}\cdot(j_i\,\mathcal{G}\,\mathbf{J}_i^{-T}) = j_i^{-1}\,\nabla_{\boldsymbol{\zeta}}\cdot\widehat{\mathcal{P}}_i$ applied to \eqref{eq:abstract_pde} and multiplied through by $j_i$ yields the mapped residual on the reference chart:
\begin{equation}
\widehat{\mathcal{R}}_i(\boldsymbol{\zeta};\tensor\eta)
:=
\nabla_{\boldsymbol{\zeta}}\cdot \widehat{\mathcal{P}}_i(\boldsymbol{\zeta};\tensor\eta)
+
j_i(\boldsymbol{\zeta})\,\widehat{\mathcal{S}}_i(\boldsymbol{\zeta};\tensor\eta),
\qquad \boldsymbol{\zeta}\in \widehat{\Omega}_i.
\label{eq:mapped_residual}
\end{equation}

For the Dirichlet boundary we use
\begin{equation}
\widehat{\mathcal{B}}_{i,D}(\boldsymbol{\zeta}):=\widehat{\tensor u}_i(\boldsymbol{\zeta})-\tensor u_D(\varphi_i(\boldsymbol{\zeta})),
\qquad \boldsymbol{\zeta}\in \widehat{\Gamma}_{i,D}:=\pi_i(\Gamma_D\cap \Omega_i).
\label{eq:mapped_dirichlet}
\end{equation}
For Neumann or traction conditions, if $\widehat{\mathbf{n}}_i(\boldsymbol{\zeta})$ is a reference outward normal on the relevant portion of $\partial\widehat{\Omega}_i$, the physical normal is
\begin{equation}
\mathbf{n}_i(\varphi_i(\boldsymbol{\zeta}))
=
\frac{\mathbf{J}_i(\boldsymbol{\zeta})^{-T}\widehat{\mathbf{n}}_i(\boldsymbol{\zeta})}
{\|\mathbf{J}_i(\boldsymbol{\zeta})^{-T}\widehat{\mathbf{n}}_i(\boldsymbol{\zeta})\|},
\label{eq:mapped_normal}
\end{equation}
and the pulled-back Neumann residual is
\begin{equation}
\widehat{\mathcal{B}}_{i,N}(\boldsymbol{\zeta};\tensor\eta)
:=
\widehat{\mathcal{G}}_i(\boldsymbol{\zeta};\tensor\eta)\,\mathbf{n}_i(\varphi_i(\boldsymbol{\zeta})) - \mathbf{t}_N(\varphi_i(\boldsymbol{\zeta})),
\label{eq:mapped_neumann}
\end{equation}
where $\widehat{\mathcal{G}}_i$ is the physical flux from \eqref{eq:pullback_flux} and $\mathbf{n}_i$ is the physical normal from \eqref{eq:mapped_normal}, both evaluated at the chart-mapped point $\varphi_i(\boldsymbol{\zeta})$. Table~\ref{tab:operator_map} summarizes the correspondence between physical-space operators and their chart-local counterparts.

\begin{table}[htbp]
\centering
\caption{Operator pullback summary. Each physical-space quantity on the left is expressed in chart-local reference coordinates on the right via the decoder map $\varphi_i$ and its Jacobian $\mathbf{J}_i$.}
\label{tab:operator_map}
\smallskip
\small
\begin{tabular}{lll}
\toprule
Physical operator & Reference-chart form & Eq. \\
\midrule
State field $u(x)$ & $\widehat{u}_i(\boldsymbol{\zeta}) = u(\varphi_i(\boldsymbol{\zeta}))$ & \eqref{eq:pullback_state} \\
Gradient $\nabla_x u$ & $\nabla_{\boldsymbol{\zeta}}\widehat{u}_i\,\mathbf{J}_i^{-1}$ & \eqref{eq:grad_pullback} \\
Flux $\mathcal{G}(x,u,\nabla u)$ & $\widehat{\mathcal{G}}_i = \mathcal{G}(\varphi_i,\widehat{u}_i,\nabla_{\boldsymbol{\zeta}}\widehat{u}_i \mathbf{J}_i^{-1})$ & \eqref{eq:pullback_flux} \\
Piola flux $\widehat{\mathcal{P}}_i$ & $j_i\,\widehat{\mathcal{G}}_i\,\mathbf{J}_i^{-T}$ & \eqref{eq:piola_flux} \\
Source $\mathcal{S}(x,u,\nabla u)$ & $\widehat{\mathcal{S}}_i = \mathcal{S}(\varphi_i,\widehat{u}_i,\nabla_{\boldsymbol{\zeta}}\widehat{u}_i \mathbf{J}_i^{-1})$ & \eqref{eq:pullback_source} \\
PDE residual & $\widehat{\mathcal{R}}_i = \nabla_{\boldsymbol{\zeta}}\cdot\widehat{\mathcal{P}}_i + j_i\,\widehat{\mathcal{S}}_i$ & \eqref{eq:mapped_residual} \\
Outward normal $\mathbf{n}$ & $\mathbf{J}_i^{-T}\widehat{\mathbf{n}}_i / \|\mathbf{J}_i^{-T}\widehat{\mathbf{n}}_i\|$ & \eqref{eq:mapped_normal} \\
Dirichlet BC & $\widehat{u}_i(\boldsymbol{\zeta}) - u_D(\varphi_i(\boldsymbol{\zeta}))$ & \eqref{eq:mapped_dirichlet} \\
Neumann BC & $\widehat{\mathcal{G}}_i\,\mathbf{n}_i - \mathbf{t}_N$ & \eqref{eq:mapped_neumann} \\
\bottomrule
\end{tabular}
\end{table}

\begin{remark}[Global maps for simply connected domains]
In the special case where a single low-distortion volumetric parametrization $\Psi_\theta:B^3\to\Omega$ is feasible, the atlas reduces to one chart. The ellipsoid verification (Example~1) uses this special case with an analytic map; the rabbit and torus benchmarks do not. Details of the learned global-map training objective and algorithm are given in Appendix~\ref{sec:appendix_global_map}.
\end{remark}

\section{Solvers on a fixed atlas}
\label{sec:solver_method}
This section explains how a fixed neural atlas and the corresponding mapped operators defined in Section \ref{sec:methodology} is used to solve PDEs while bypassing the meshing procedure that renders the geometry. 
We use a physics-informed neural network (Section~\ref{sec:schwarz_pinn}) and a finite-element method (Section~\ref{sec:schwarz_fem}) to demonstrate how the neural atlas can be used to solve PDEs in a domain-decomposed setting. In principle, the atlas framework may also be extended to other PDE solvers, such as smoothed particle hydrodynamics (SPH) or the material point method, but this is outside the scope of the present study. 

\subsection{Schwarz alternating physics-informed neural network on a fixed atlas}
\label{sec:schwarz_pinn}
After the atlas is obtained, its coordinate charts can be used to decompose the spatial domain for the PDE solver. In this setting, each chart becomes a local subproblem, each overlap $\Omega_{ij}$ acts as an artificial interface, and each transition map $\psi_{ij}$ provides a consistent way to compare neighboring chart fields \citep{quarteroni1999domain}. For coordinate-chart PINNs, this setting enables us to formulate residual minimization in reference coordinates for each coordinate chart. The overlap graph $G_{\mathcal A}$ identifies which neighboring charts must exchange traces. Compared with a single global PINN, this structure improves locality in approximation, conditioning, and sampling. 

Let the state network on chart $i$ be
\begin{equation}
\widehat{u}_i(\cdot;\theta_i):\widehat{\Omega}_i\rightarrow \mathbb{R}^{q}
\label{eq:local_network}
\end{equation}
where $\theta_i$ denotes the trainable parameters of the chart-$i$ network. 
The local physical representation reads,
\begin{equation}
u_i(\tensor x;\theta_i):=\widehat{u}_i(\pi_i(\tensor x);\theta_i),
\qquad x\in\Omega_i.
\label{eq:local_physical_field}
\end{equation}

For overlaps, the artificial Schwarz boundary of chart $i$ is
\begin{equation}
\Gamma_i:=\partial\Omega_i\setminus \partial\Omega,
\qquad
\Gamma_{ij}:=\Gamma_i\cap \Omega_j,
\qquad
\mathcal{N}(i):=\{j\neq i:\Gamma_{ij}\neq\emptyset\}.
\label{eq:schwarz_sets}
\end{equation}
In reference coordinates, the corresponding sets are $\widehat{\Gamma}_{ij}:=\pi_i(\Gamma_{ij})\subset\widehat{\Omega}_i$ and $\widehat{\Gamma}_{i,\partial\Omega}:=\pi_i(\partial\Omega\cap\partial\Omega_i)$, which are the domains on which interface and physical-boundary penalties are sampled in the loss below. The sets $\Gamma_{ij}$ play the same conceptual role as Schwarz interfaces in classical overlapping domain decomposition: they are not physical boundaries of the PDE, but the internal regions where chart-local solutions must agree strongly enough to define a coherent global field. In the chart setting the same overlap also provides the transition map $\psi_{ij}$ needed to compare chart-local predictions in compatible coordinates. In the rabbit implementation, interface samples are drawn from points whose stored atlas membership includes both charts, and the rigid local coordinates \eqref{eq:atlas_rigid_local_coords} are evaluated in both neighboring frames before the chart-to-chart penalties are formed.

In Schwarz alternating method \citep{lions1988schwarz}, one would impose Dirichlet, Neumann, or Robin data on these artificial interfaces and solve each local problem to completion before moving to the next subdomain. In the Schwarz PINN model, 
 neighbor networks are frozen while chart $i$ is updated, and compatibility is enforced by sampled penalties on the overlap rather than by exact boundary imposition. Writing $\theta_j^{(k,\mathrm{fr})}$ for the frozen neighbor state available during sweep $k$, the interface-value term is
sampled directly from the neighboring local field $u_j$, not from the globally blended field $u_h$. This distinction is important for reproducibility: in the released solver, Schwarz transmission is chart-to-chart, while the global blend is used only for diagnostics and post-processing after a proposed sweep.
\begin{equation}
\mathcal{J}_{i,\mathrm{iv}}^{(k)}
=
\sum_{j\in \mathcal{N}(i)}
\mathbb{E}_{\boldsymbol{\zeta}\sim \widehat{\Gamma}_{ij}}
\left[
\left|
\widehat{u}_i(\boldsymbol{\zeta};\theta_i)
-
\widehat{u}_j\!\left(\psi_{ij}(\boldsymbol{\zeta});\theta_j^{(k,\mathrm{fr})}\right)
\right|^2
\right].
\label{eq:schwarz_iv_loss}
\end{equation}
To ensure continuity, one may also penalizes projected normal-flux mismatch. Let $\mathbf{n}_{ij}(\boldsymbol{\zeta})$ denote the outward unit normal of $\partial\Omega_i$ at the physical point $\varphi_i(\boldsymbol{\zeta})$, restricted to the interface $\Gamma_{ij}$. Define
\begin{equation}
q_i(\boldsymbol{\zeta};\theta_i)
:=
\mathbf{n}_{ij}(\boldsymbol{\zeta})\cdot
\nabla_x u_i\!\left(\varphi_i(\boldsymbol{\zeta});\theta_i\right),
\label{eq:schwarz_flux_trace}
\end{equation}
and
\begin{equation}
\mathcal{J}_{i,\mathrm{if}}^{(k)}
=
\sum_{j\in \mathcal{N}(i)}
\mathbb{E}_{\boldsymbol{\zeta}\sim \widehat{\Gamma}_{ij}}
\left[
\left|
q_i(\boldsymbol{\zeta};\theta_i)
-
q_j\!\left(\psi_{ij}(\boldsymbol{\zeta});\theta_j^{(k,\mathrm{fr})}\right)
\right|^2
\right].
\label{eq:schwarz_if_loss}
\end{equation}
For completeness, one may also expose a Robin-style variant,i.e.,
\begin{equation}
\mathcal{J}_{i,\mathrm{rb}}^{(k)}
=
\sum_{j\in \mathcal{N}(i)}
\mathbb{E}_{\boldsymbol{\zeta}\sim \widehat{\Gamma}_{ij}}
\left[
\left|
\lambda_R
\left(
\widehat{u}_i(\boldsymbol{\zeta};\theta_i)
-
\widehat{u}_j\!\left(\psi_{ij}(\boldsymbol{\zeta});\theta_j^{(k,\mathrm{fr})}\right)
\right)
+
q_i(\boldsymbol{\zeta};\theta_i)
-
q_j\!\left(\psi_{ij}(\boldsymbol{\zeta});\theta_j^{(k,\mathrm{fr})}\right)
\right|^2
\right],
\label{eq:schwarz_robin_loss}
\end{equation}
This option is nevertheless not exercised in our numerical experiments. 
%In volumetric-atlas mode, the code can further add an overlap-interior value penalty on sampled points from $\widehat{\Omega}_{ij}$; despite the legacy variable name \texttt{overlap\_h1}, the current implementation penalizes value mismatch over the sampled overlap volume rather than the full $H^1$ seminorm.
At Schwarz sweep $k$, the local chart objective is therefore
\begin{align}
\mathcal{J}_i^{(k)}(\theta_i;\eta)
=&
\lambda_{\mathrm{pde}}\,
\mathbb{E}_{\boldsymbol{\zeta}\sim \widehat{\Omega}_i}\!\left[\|\widehat{\mathcal{R}}_i(\boldsymbol{\zeta};\theta_i,\eta)\|_2^2\right]
+
\lambda_{\mathrm{bc}}\,
\mathbb{E}_{\boldsymbol{\zeta}\sim \widehat{\Gamma}_{i,\partial\Omega}}\!\left[\|\widehat{\mathcal{B}}_i(\boldsymbol{\zeta};\theta_i,\eta)\|_2^2\right]
\nonumber\\
&+
\lambda_{\mathrm{iv}}\,\mathcal{J}_{i,\mathrm{iv}}^{(k)}
+
\lambda_{\mathrm{if}}\,\widetilde{\mathcal{J}}_{i,\mathrm{if}}^{(k)}
+
\lambda_{\mathrm{ov}}\,\mathcal{J}_{i,\mathrm{ov}}^{(k)}
+
\lambda_{\mathrm{data}}\,\mathcal{J}_{i,\mathrm{data}}(\theta_i,\eta),
\label{eq:schwarz_local_loss}
\end{align}
where $\widetilde{\mathcal{J}}_{i,\mathrm{if}}^{(k)}$ denotes either the projected-flux penalty \eqref{eq:schwarz_if_loss} or the Robin residual \eqref{eq:schwarz_robin_loss}, and $\mathcal{J}_{i,\mathrm{ov}}^{(k)}$ denotes the optional volumetric overlap penalty. The practical default is the projected-flux penalty with frozen-neighbor traces. This distinction matters: the present solver is Schwarz-inspired in its transmission logic, but each chart update is still carried out by optimizer steps on a neural loss, not by an exact local PDE solve.

One sweep of the atlas solver can be written abstractly as
\begin{equation}
\Theta^{(k+1)}=\mathfrak{S}\!\left(\Theta^{(k)};\eta\right),
\qquad
\Theta^{(k)}:=(\theta_1^{(k)},\dots,\theta_M^{(k)}),
\label{eq:schwarz_sweep}
\end{equation}
where $\mathfrak{S}$ denotes one multiplicative Schwarz pass over the chart set. In the PINN solver, multiplicativity is enforced on every overlapping pair: charts updated later in the sweep use the most recently accepted states of previously updated neighbors. Charts that do not overlap are assigned to the same graph color and can be updated concurrently on available accelerator backends without violating the frozen-neighbor assumption. The resulting algorithm is therefore a multiplicative Schwarz scheme with restricted within-color parallelism, rather than a fully additive update.
The color groups used in Algorithm~\ref{alg:atlas_sapinn_training} are derived from a graph-coloring of the chart-overlap adjacency matrix: two charts may share a color only if they do not overlap, which makes within-color updates independent for the purposes of the frozen-neighbor Schwarz step.

After an accepted sweep, the global blended field is
\begin{equation}
u_h^{(k)}(x)=\sum_{i=1}^{M}\omega_i(x)\,u_i(x;\theta_i^{(k)}).
\label{eq:schwarz_global_field}
\end{equation}
where $M$ is the total number of charts in the atlas.

\begin{figure}[htbp]
\centering
\resizebox{\textwidth}{!}{%
\begin{tikzpicture}[
  >=Stealth,
  every node/.style={font=\small},
  stagebox/.style={draw, rounded corners=4pt, line width=0.9pt, minimum width=1.65cm, minimum height=3.2cm, inner sep=3pt, align=center},
  smallbox/.style={draw, rounded corners=2pt, line width=0.7pt, minimum width=1.3cm, inner sep=2.5pt, font=\scriptsize, align=center},
]
\definecolor{cBlue}{HTML}{4C78A8}
\definecolor{cGreen}{HTML}{59A14F}
\definecolor{cOrange}{HTML}{F28E2B}
\definecolor{cRed}{HTML}{C44E52}

% ============================================================
%  Panel (a): Overlapping chart geometry
% ============================================================
\begin{scope}[shift={(-3.8, 0)}]
  \node[font=\small\bfseries, anchor=south] at (0, 2.8) {(a)\; Overlapping chart images};

  % Body outline
  \draw[line width=1.2pt, fill=gray!5]
    plot[smooth cycle, tension=0.72] coordinates {
      (-2.0, 0.5) (-1.3, 2.2) (0.4, 2.5) (1.6, 1.8)
      (2.0, 0.5) (1.4,-1.1) (0.0,-1.6) (-1.4,-0.9)
    };
  \node[font=\small\bfseries] at (1.6, 2.3) {$\Omega$};

  % Chart i (blue)
  \draw[cBlue, line width=0.9pt, fill=cBlue!15] (-0.35, 0.5) ellipse (1.3cm and 1.25cm);
  \node[font=\small\bfseries, text=cBlue] at (-1.1, 1.65) {$\Omega_i$};

  % Chart j (green)
  \draw[cGreen, line width=0.9pt, fill=cGreen!15] (0.7, 0.45) ellipse (1.25cm and 1.2cm);
  \node[font=\small\bfseries, text=cGreen] at (1.4, 1.55) {$\Omega_j$};

  % Overlap label
  \node[fill=white, inner sep=1.5pt, font=\scriptsize] at (0.22, 0.48) {$\Omega_{ij}$};

  % Interface lines
  \draw[dashed, gray!70, line width=0.7pt] (0.22, -0.7) -- (0.22, 1.75);
  \node[font=\scriptsize, text=cBlue, rotate=90] at (0.04, 1.35) {$\Gamma_{ij}$};
  \node[font=\scriptsize, text=cGreen, rotate=90] at (0.40, -0.25) {$\Gamma_{ji}$};

  % Data exchange arrows
  \draw[->, cGreen, line width=0.8pt] (1.1, -0.35) to[bend left=15]
    node[font=\scriptsize, right, text=cGreen] {neighbor trace} (0.45, 0.05);
  \draw[->, cBlue, line width=0.8pt] (-0.15, -0.05) to[bend left=15]
    node[font=\scriptsize, left, text=cBlue] {interface loss} (0.1, -0.55);

  % Legend boxes
  \node[smallbox, draw=cBlue, fill=cBlue!8] at (-0.9, -2.2) {active chart $i$};
  \node[smallbox, draw=cGreen, fill=cGreen!8] at (0.9, -2.2) {frozen neighbor $j$};
\end{scope}

% ============================================================
%  Panel (b): One multiplicative sweep
% ============================================================
\begin{scope}[shift={(2.6, 0)}]
  \node[font=\small\bfseries, anchor=south] at (1.2, 2.8) {(b)\; One multiplicative sweep};

  % Stage 1: Snapshot
  \node[stagebox, draw=cBlue, fill=cBlue!6] (S1) at (0, 0) {};
  \node[font=\scriptsize\bfseries, anchor=north] at (0, 1.4) {1.\,Snapshot};
  \node[font=\scriptsize, text=black] at (0, 0.75) {save $\Theta^{(k)}$};
  \node[font=\scriptsize, text=gray!70] at (0, 0.28) {rollback state};
  \node[font=\scriptsize, text=gray!70] at (0, -0.18) {refresh pools};

  % Stage 2: Local updates
  \node[stagebox, draw=cGreen, fill=cGreen!6, minimum width=1.85cm] (S2) at (2.3, 0) {};
  \node[font=\scriptsize\bfseries, anchor=north] at (2.3, 1.4) {2.\,Local updates};
  % Mini chart overlaps
  \draw[cBlue, fill=cBlue!15, line width=0.6pt] (2.0, 0.45) ellipse (0.3cm and 0.42cm);
  \draw[cGreen, fill=cGreen!15, line width=0.6pt] (2.35, 0.45) ellipse (0.3cm and 0.42cm);
  \draw[cOrange, fill=cOrange!15, line width=0.6pt] (2.7, 0.45) ellipse (0.3cm and 0.42cm);
  \draw[cRed, line width=1.0pt, rounded corners=2pt] (2.08, -0.02) rectangle (2.62, 0.92);
  \node[font=\scriptsize, text=gray!70] at (2.3, -0.35) {update one chart};
  \node[font=\scriptsize, text=gray!70] at (2.3, -0.72) {or color group};
  \node[font=\scriptsize, text=gray!70] at (2.3, -1.15) {neighbors frozen};

  % Stage 3: Global check
  \node[stagebox, draw=cRed, fill=cRed!6] (S3) at (4.4, 0) {};
  \node[font=\scriptsize\bfseries, anchor=north] at (4.4, 1.4) {3.\,Global check};
  \node[smallbox, draw=cGreen, fill=white] (acc) at (4.4, 0.6) {accept};
  \node[smallbox, draw=cRed, fill=white] (rej) at (4.4, -0.15) {rollback};
  \node[smallbox, draw=gray!50, fill=white, font=\scriptsize] at (4.4, -0.9) {PDE\\interface\\rel-$L^2$};
  \draw[->, cGreen, line width=0.7pt] (4.4, 0.28) -- (4.4, 0.42);
  \draw[->, cRed, line width=0.7pt] (4.4, 0.10) -- (4.4, 0.0);

  % Stage 4: Checkpoint
  \node[stagebox, draw=cOrange, fill=cOrange!6] (S4) at (6.2, 0) {};
  \node[font=\scriptsize\bfseries, anchor=north] at (6.2, 1.4) {4.\,Checkpoint};
  \node[smallbox, draw=black!60, fill=white] at (6.2, 0.55) {blend field};
  \node[smallbox, draw=black!60, fill=white] at (6.2, -0.15) {score sweep};
  \node[smallbox, draw=black!60, fill=white] at (6.2, -0.80) {continue};

  % Arrows between stages
  \draw[->, gray!70, line width=1.0pt] (S1.east) -- (S2.west);
  \draw[->, gray!70, line width=1.0pt] (S2.east) -- (S3.west);
  \draw[->, gray!70, line width=1.0pt] (S3.east) -- (S4.west);

  % Bottom annotation
  \node[font=\scriptsize, text=gray!60, anchor=north] at (3.1, -2.05) {multiplicative on true overlaps; parallel within non-overlapping color groups};
\end{scope}
\end{tikzpicture}%
}% end resizebox
\caption{Conceptual and algorithmic view of fixed-atlas SA-PINN coupling. Panel~(a) shows two overlapping chart images inside the physical body. The overlap $\Omega_{ij}$ induces the artificial interfaces $\Gamma_{ij}$ and $\Gamma_{ji}$ on which neighboring chart solutions exchange temporary transmission data during training. Panel~(b) shows one multiplicative Schwarz sweep: the solver snapshots the full chart state, updates one active chart or non-overlapping color group while neighbor states are detached, evaluates global residual and interface diagnostics, and then accepts the sweep or restores the previous state.}
\label{fig:schwarz_schematic}
\end{figure}
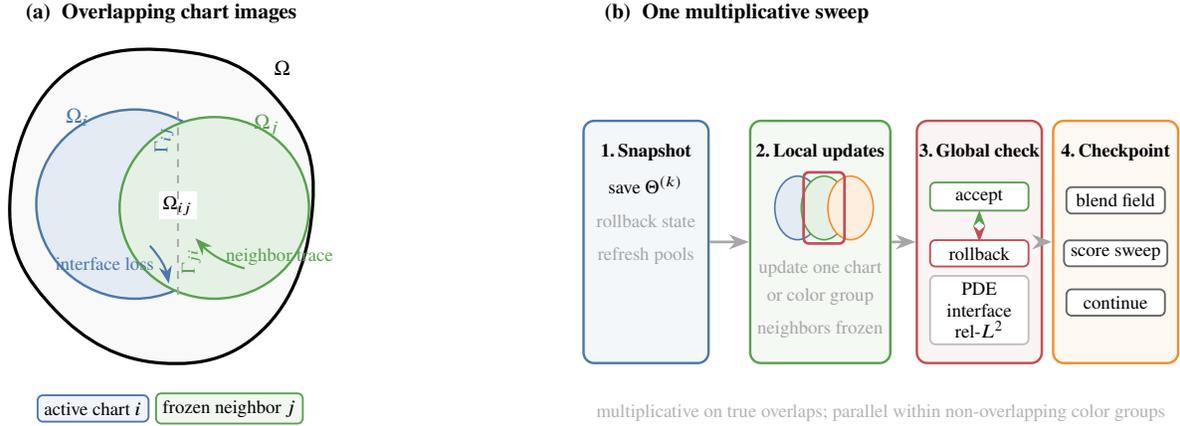

\begin{algorithm}[htbp]
\caption{Fixed-atlas SA-PINN training by multiplicative Schwarz sweeps}
\label{alg:atlas_sapinn_training}
\small
\begin{algorithmic}[1]
\Statex \textbf{Inputs:} frozen atlas $\mathcal{A}$; chart-local networks $\{\widehat{u}_i(\cdot;\theta_i)\}_{i=1}^M$;
\StateX optional global parameters $\eta$; local optimizers; sweep budget $K$; local-step budget $N_{\mathrm{loc}}$
\Statex \textbf{Outputs:} selected checkpoint $\{\theta_i\}_{i=1}^M$ and blended field $u_h$
\Statex \textit{Initialization}
\State Initialize chart parameters $\theta_i$ and any trainable global parameters $\eta$
\State Evaluate the initial blended field and initialize checkpoint metrics
\For{$k=0,\dots,K-1$}
    \Statex \textit{Pre-sweep setup}
    \State Store a snapshot of the full chart state $\Theta^{(k)}$
    \State Refresh sweep-level sampling pools and effective interface weights
        \For{each color group $g$ from the graph-coloring of the chart-overlap adjacency matrix}
        \Statex \textit{Local chart updates}
        \For{each active chart $i\in g$ sequentially, or concurrently when charts in $g$ do not overlap}
            \For{$n=1,\dots,N_{\mathrm{loc}}$}
                \State Sample PDE, physical-boundary, interface, and optional supervision points on chart $i$
                \State Query neighbors only through frozen predictions on sampled overlap points
                \State Assemble the local loss $\mathcal{J}_i^{(k)}$ in \eqref{eq:schwarz_local_loss}
                \State Update $\theta_i$ while all neighbor contributions remain detached
            \EndFor
        \EndFor
    \EndFor
    \Statex \textit{Global acceptance test}
    \State Form the proposed blended field by \eqref{eq:schwarz_global_field}
    \State Evaluate global PDE, boundary, interface, and relative-$L^2$ diagnostics
    \If{the sweep violates the trust-region admissibility test}
        \State Restore the pre-sweep snapshot $\Theta^{(k)}$ and reduce the learning rate
    \Else
        \State Accept the sweep and refresh checkpoint statistics
        \State Update adaptive weights if enabled
    \EndIf
\EndFor
\State \Return the selected checkpoint and its blended field $u_h$
\end{algorithmic}
\end{algorithm}

Algorithm~\ref{alg:atlas_sapinn_training} records the core training loop. The key point is that acceptance is evaluated at the sweep level, not chart by chart. As such, a candidate sweep is formed first, then accepted or rejected using global diagnostics. Consequently, later chart updates within the same sweep are interpreted in the context of the current provisional state, and the solver commits a new iterate only after the sweep-level decision is made. Parallel updates are permitted only within non-overlapping graph-color groups; no two charts in the same color share an overlap interface. This coloring preserves independent within-color updates while retaining multiplicative information transfer across true overlap pairs through the sweep ordering.

\subsection{Chart-local finite-element solver on a fixed atlas}
\label{sec:schwarz_fem}
This section shows that Galerkin finite element can also be cast as interchangeable chart-local solvers that consume the same geometric data structures and Schwarz coupling protocol, differing only in the choice of trial space and the mechanism by which the local subproblem is solved.
Our starting point is the observation that the mapped residual \eqref{eq:mapped_residual} and the Piola-transformed flux \eqref{eq:piola_flux} are coordinate-level identities: they hold for \emph{any} admissible trial function on the reference chart, regardless of whether that trial function is a neural network $\widehat{u}_i(\boldsymbol{\zeta};\theta_i)$ or a piecewise-linear finite-element expansion $\widehat{u}_i^h(\boldsymbol{\zeta}) = \sum_a N_a(\boldsymbol{\zeta})\,d_{i,a}$ with nodal coefficients $\mathbf{d}_i\in\mathbb{R}^{n_i\times q}$ and standard Lagrange shape functions $N_a$. The Galerkin weak form on chart~$i$ reads: find $\widehat{u}_i^h\in V_i^h$ such that
\begin{equation}
\int_{\widehat{\Omega}_i}
(\nabla_{\boldsymbol{\zeta}} N_a)^T\,
\widehat{\mathcal{P}}_i(\boldsymbol{\zeta};\widehat{u}_i^h)\,
d\boldsymbol{\zeta}
\;=\;
\int_{\widehat{\Omega}_i}
j_i(\boldsymbol{\zeta})\,\widehat{\mathcal{S}}_i(\boldsymbol{\zeta};\widehat{u}_i^h)\,N_a\,
d\boldsymbol{\zeta}
\qquad \forall\,N_a\in V_i^h,
\label{eq:fem_weak_form}
\end{equation}
where $\widehat{\mathcal{P}}_i$ is the same Piola flux \eqref{eq:piola_flux} that the PINN evaluates pointwise, and the reference-domain integrals are evaluated by standard Gauss quadrature on each tetrahedron. This is the standard Bubnov--Galerkin statement pulled back to the reference chart through the map $\varphi_i$; the only non-standard ingredient is that the metric coefficients $\mathbf{J}_i$ and $j_i$ are supplied by the frozen atlas rather than by a conventional body-fitted mesh.
For linear elliptic operators, \eqref{eq:fem_weak_form} yields a sparse symmetric positive-definite system whose assembly and solution follow the standard finite-element workflow. For nonlinear constitutive laws---hyperelasticity, rate-independent plasticity, or coupled multi-physics---the weak form yields a nonlinear residual
\begin{equation}
\mathbf{R}_i(\mathbf{d}_i) \;=\; \mathbf{f}_{\mathrm{int},i}(\mathbf{d}_i) - \mathbf{f}_{\mathrm{ext},i} \;=\; \mathbf{0},
\label{eq:fem_residual}
\end{equation}
which is solved by Newton--Raphson iteration. The consistent tangent stiffness $\mathbf{K}_i = \partial\mathbf{R}_i/\partial\mathbf{d}_i$ is assembled through the same automatic-differentiation infrastructure used by the PINN: PyTorch's \texttt{autograd} differentiates the mapped constitutive operator $\widehat{\mathcal{P}}_i$ with respect to the deformation gradient, and the resulting material tangent modulus $\mathbb{C}_i = \partial\widehat{\mathcal{P}}_i/\partial\mathbf{F}$ is contracted with shape-function gradients at each quadrature point in the standard fashion \citep{simo_computational_1998,de_souza_neto_computational_2011}. For history-dependent constitutive laws (e.g., the smooth return mapping of Appendix~\ref{sec:appendix_constitutive}), the autograd tape records the full algorithmic path through the return-mapping corrector, so the consistent algorithmic tangent is obtained exactly without manual derivation of the elasto-plastic modulus.

The reference domain $\widehat{\Omega}_i \subset [-r_i, r_i]^3$ is discretized by a structured hexahedral grid with Freudenthal six-tetrahedron decomposition, yielding $(n_{\mathrm{cells}}+1)^3$ nodes and $6\,n_{\mathrm{cells}}^3$ P1 tetrahedra per chart. This structured mesh is generated once per chart from the support radius $r_i$ alone, with no surface-conforming mesh generation step---the geometry is encoded entirely through the chart map $\varphi_i$ and its Jacobian $\mathbf{J}_i$, which enter the integrands of \eqref{eq:fem_weak_form}. Nodes whose reference-domain images fall outside the physical body (detected by the frozen SDF) are excluded from the assembly: elements with all nodes outside the body contribute zero to the chart-local stiffness and load vectors, and partially exterior elements are retained with their interior quadrature points contributing through the standard mapped integrands. The mesh generation cost is therefore $O(n_{\mathrm{cells}}^3)$ per chart and is independent of the geometric complexity of $\partial\Omega$; all geometric information enters through the frozen decoder and SDF evaluations at quadrature points.

\subsubsection{Schwarz coupling for the FEM solver}

Inter-chart coupling follows the same multiplicative Schwarz procedure as that of the PINN (Section~\ref{sec:schwarz_pinn}). The major departure is that the admissible solution of each chart is obtained through a gradient-based solver rather than searched in a non-convex landscape. This simplification leads to the classical multiplicative Schwarz method \citep{quarteroni1999domain} rather than its PINN-relaxed variant. This setting is helpful for problems with history-dependent constitutive laws, where each local solve must reach equilibrium before internal variables are updated. Because approximate gradient-descent solutions cannot guarantee this, the Schwarz PINN used in this study could potentially corrupt the stored plastic state. At Schwarz sweep~$k$, the solver visits charts by color groups from the overlap adjacency graph. For each active chart~$i$:
\begin{enumerate}
\item Schwarz interface nodes $\boldsymbol{\zeta}_a \in \widehat{\Gamma}_{ij}$ receive Dirichlet values from the current solution on neighboring chart~$j$:
\begin{equation}
d_{i,a} \;\leftarrow\; \widehat{u}_j^h\!\left(\psi_{ij}(\boldsymbol{\zeta}_a);\,\mathbf{d}_j^{(k)}\right),
\qquad a \in \mathcal{I}_{ij},
\label{eq:fem_schwarz_bc}
\end{equation}
where $\psi_{ij}$ is the chart transition map and $\mathcal{I}_{ij}$ denotes the index set of chart-$i$ boundary nodes that lie within chart~$j$'s support.
\item Physical boundary conditions (Dirichlet or Neumann) are imposed on nodes $\boldsymbol{\zeta}_a \in \widehat{\Gamma}_{i,\partial\Omega}$ via the mapped boundary operators \eqref{eq:mapped_dirichlet}--\eqref{eq:mapped_neumann}.
\item The chart-local residual \eqref{eq:fem_residual} is assembled and solved by Newton--Raphson for $\mathbf{d}_i^{(k+1)}$.
\end{enumerate}
For problems with history-dependent constitutive laws (e.g., the elastoplastic forward BVP of Example~3), internal variables---accumulated plastic strain, back-stress tensors---are stored at each quadrature point within each chart and updated locally during the Newton iteration. Only displacement traces are exchanged across chart boundaries; internal-variable fields remain private to each chart. This locality is a direct consequence of the multiplicative decomposition $\mathbf{F}=\mathbf{F}^e\mathbf{F}^p$: the plastic state is determined by the local deformation history, and Schwarz coupling need enforce only displacement continuity at interfaces for the global equilibrium problem to be well-posed.

\begin{remark}[Tangent computation and operator modularity]
\label{rem:tangent_modularity}
The autograd-based tangent assembly deserves emphasis because it decouples the solver infrastructure from the constitutive law. In a conventional finite-element code, each new material model requires a hand-derived consistent tangent modulus; errors in this derivation are a well-known source of convergence failures \citep{simo_computational_1998}. Here, the chart-local residual $\mathbf{R}_i(\mathbf{d}_i)$ is assembled as a standard PyTorch computation graph, and the tangent $\mathbf{K}_i = \partial\mathbf{R}_i/\partial\mathbf{d}_i$ is obtained by reverse-mode differentiation through the entire constitutive path---including, for plasticity, the smooth return-mapping corrector. Adding a new constitutive law therefore requires only a forward evaluation of the stress; the consistent tangent follows automatically. This is the same automatic-differentiation principle exploited in differentiable simulation \citep{hu_difftaichi_2020}, applied here within the atlas-based Schwarz framework.
\end{remark}

\begin{algorithm}[htbp]
\caption{Chart-local P1 FEM solver with multiplicative Schwarz coupling}
\label{alg:atlas_fem_schwarz}
\small
\begin{algorithmic}[1]
\Statex \textbf{Inputs:} frozen atlas $\mathcal{A}=\{(\varphi_i,\mathbf{J}_i,r_i)\}_{i=1}^M$; overlap graph $G_{\mathcal{A}}$ with coloring;
\StateX mesh parameter $n_{\mathrm{cells}}$; loading schedule $\{\delta^{(t)}\}_{t=1}^{N_{\mathrm{step}}}$; Schwarz sweep budget $K$;
\StateX Newton tolerance $\epsilon_{\mathrm{NR}}$; constitutive law $\widehat{\mathcal{P}}_i(\cdot)$
\Statex \textbf{Outputs:} chart-local nodal solutions $\{\mathbf{d}_i\}_{i=1}^M$ and internal-variable fields
\Statex
\State Generate structured P1 tetrahedral meshes on each $\widehat{\Omega}_i\subset[-r_i,r_i]^3$
\State Precompute shape-function gradients, quadrature weights, $\mathbf{J}_i$, $j_i$ at each Gauss point
\State Identify boundary node sets: $\widehat{\Gamma}_{i,\partial\Omega}$ (physical BC) and $\widehat{\Gamma}_{ij}$ (Schwarz interface) for each chart pair
\State Initialize $\mathbf{d}_i^{(0)}=\mathbf{0}$, internal variables $\leftarrow$ elastic reference state
\For{$t=1,\dots,N_{\mathrm{step}}$} \Comment{Incremental load stepping}
    \State Update external load $\delta^{(t)}$ and corresponding physical boundary data
    \For{$k=1,\dots,K$} \Comment{Multiplicative Schwarz sweeps}
        \For{each color group $g$ from the overlap graph coloring}
            \For{each chart $i\in g$}
                \State Impose physical BCs: $d_{i,a}\leftarrow u_D(\varphi_i(\boldsymbol{\zeta}_a))$ for $a\in\widehat{\Gamma}_{i,\partial\Omega}$
                \State Impose Schwarz BCs: $d_{i,a}\leftarrow \widehat{u}_j^h(\psi_{ij}(\boldsymbol{\zeta}_a);\mathbf{d}_j^{(k)})$ for $a\in\mathcal{I}_{ij}$, $j\in\mathcal{N}(i)$
                \State \textbf{Newton--Raphson:}
                \Indent
                    \State Assemble $\mathbf{R}_i(\mathbf{d}_i)$ and tangent $\mathbf{K}_i$ from mapped weak form \eqref{eq:fem_weak_form}
                    \State Solve $\mathbf{K}_i\,\Delta\mathbf{d}_i=-\mathbf{R}_i$; update $\mathbf{d}_i\leftarrow\mathbf{d}_i+\Delta\mathbf{d}_i$
                    \State Repeat until $\|\mathbf{R}_i\|/\|\mathbf{f}_{\mathrm{ext},i}\|<\epsilon_{\mathrm{NR}}$
                \EndIndent
            \EndFor
        \EndFor
        \State Evaluate inter-chart displacement jump $\max_{(i,j)\in G_{\mathcal{A}}}\|\mathbf{d}_i-\mathbf{d}_j\|_{\Gamma_{ij}}$
    \EndFor
    \State Update history-dependent internal variables (plastic strain, back-stress)
    \State Store checkpoint if requested
\EndFor
\State Form global blended field $u_h(x)=\sum_i\omega_i(x)\,\widehat{u}_i^h(\pi_i(x);\mathbf{d}_i)$ via partition of unity
\end{algorithmic}
\end{algorithm}

\subsubsection{Comparison of PINN and FEM algorithms}
\label{sec:pinn_vs_fem}

Table~\ref{tab:pinn_vs_fem} compares the two chart-local solver families on atlases built by the same pipeline. A important key is that that the atlas construction pipeline (SDF learning, chart seeding, decoder training, quality gating) and the Schwarz coupling logic (overlap graph, color groups, interface transmission) are \emph{shared infrastructure} that both solvers consume without modification. 
Once the atlas passes its quality gates, either Algorithm~\ref{alg:atlas_sapinn_training} or Algorithm~\ref{alg:atlas_fem_schwarz} may be applied, and new chart-local discretization families---spectral elements, isogeometric patches, or neural operators---could be introduced by supplying only a local trial-function class and the corresponding assembly or evaluation routine. This modularity is a direct consequence of the separation between geometry (encoded in the atlas) and physics (encoded in the chart-local solver), and it is the primary architectural contribution of the present work.

\begin{table}[htbp]
\centering
\small
\setlength{\tabcolsep}{4pt}
\caption{Comparisons between the two chart-local solver families on the atlas pipeline. Both solvers share the same decoder architecture, overlap graph structure $G_{\mathcal{A}}$, Schwarz transmission logic, and partition-of-unity blending \eqref{eq:schwarz_global_field}; the chart count is a user parameter.}
\label{tab:pinn_vs_fem}
\begin{tabular}{p{3.5cm}p{5.2cm}p{5.2cm}}
\toprule
& \textbf{Chart-local PINN} (Alg.~\ref{alg:atlas_sapinn_training}) & \textbf{Chart-local FEM} (Alg.~\ref{alg:atlas_fem_schwarz}) \\
\midrule
Trial function & Neural network $\widehat{u}_i(\boldsymbol{\zeta};\theta_i)$ & P1 expansion $\sum_a N_a\,d_{i,a}$ \\
Local solve & Approximate (gradient descent) & Exact (Newton--Raphson to tolerance) \\
Mesh generation & None (meshfree collocation) & Structured Freudenthal tets on $[-r_i,r_i]^3$ \\
Convergence guarantee & Empirical & $O(h^2)$ for P1 on smooth solutions \\
Nonlinear operators & Pointwise autograd residual & Requires tangent assembly per Newton step \\
Schwarz coupling & Penalty-based on sampled overlaps & Dirichlet trace injection at interface nodes \\
%History-dependent laws & Not yet demonstrated & Demonstrated (Example~3, $J_2$ plasticity) \\
Gradient w.r.t.\ parameters & Direct (backpropagation) & Implicit function theorem \\
\bottomrule
\end{tabular}
\end{table}

%\paragraph{Meshfree neural solver.}
%The PINN solver avoids reference-domain mesh generation and stiffness-matrix assembly entirely, and it extends naturally to nonlinear operators because the PDE residual is evaluated pointwise through autograd. These are practical advantages when the constitutive law is complex or when the user wishes to explore different PDE operators on the same geometry without regenerating element connectivity. However, the local solve is approximate (driven by stochastic gradient descent rather than Newton iteration), and no mesh-refinement convergence guarantee is available.

%\paragraph{Chart-local Galerkin solver.}
%The FEM solver provides a classical mesh-refinement convergence guarantee ($O(h^2)$ for P1 elements on sufficiently smooth solutions) and solves each chart-local subproblem to machine-precision Newton tolerance. For incremental problems with history-dependent internal variables---where each load step must equilibrate before advancing to the next---the exactness of the Newton solve is essential, because approximate gradient-descent updates cannot guarantee equilibrium at every step and would corrupt the stored plastic state. The trade-off is that each Newton step requires tangent stiffness assembly, which in the current dense implementation scales as $O(n_i^2)$ in memory and $O(n_i^3)$ in solve time, where $n_i$ is the chart-local DOF count. Sparse or iterative solvers would alleviate this cost for finer chart-local meshes.

%\paragraph{Shared infrastructure and extensibility.}

\section{Atlas constructions}
\label{sec:implementation}

This section details the workflow that produces a frozen atlas. It also describes the quality gates that certify the atlas before any PDE solver is invoked and the chart-local solver architectures. Architectural specifications and the loss functions used to train the coordinate charts are collected in Appendix~\ref{sec:appendix_architectures}--\ref{sec:appendix_pipeline}. Solver stabilization heuristics are given in Appendix~\ref{sec:appendix_stabilization} to ensure reproducibility. 
All neural networks are implemented in PyTorch \citep{paszke2019pytorch} with automatic differentiation for both PDE residuals and chart Jacobians. The atlas, including the signed distance function, coordinate charts, decoder maps, mask weights, and overlap structure, is constructed and frozen before any PDE solver is used. Geometric quality gates ensure that if a benchmark fails after passing geometry quality control, the failure can be attributed to the PDE solver or coupling logic.
We use the Stanford Bunny geometry as a representative test bed. 

\subsection{Atlas construction pipeline}
\label{sec:impl_pipeline}

The geometry stage follows a four-step pipeline, each producing a mathematical object required by the downstream solver. The pipeline is designed so that every stage can be verified independently: the SDF defines the domain, the seeds and frames initialize the charts, the decoders and masks refine them, and the quality gates accept or reject the result before any PDE solver is invoked.

\subsubsection{Volumetric signed-distance modeling}
\label{sec:impl_sdf}
The construction of the atlas requires a representation of geometry to parametrize the coordinate charts. In this work, we leverage the signed distance function as the input to construct the atlas. This signed distance function can be obtained externally or can be recovered from other forms of representations, such as meshes, point clouds, or parametric surfaces. In this work, we train a neural signed distance function $s_\vartheta:\mathbb{R}^3\to\mathbb{R}$ from surface samples with oriented normals. The learned level set defines the computational domain and its boundary:
\begin{equation}
s_{\vartheta}:\mathbb{R}^3\rightarrow \mathbb{R},
\qquad
\Omega \approx \{x\in\mathbb{R}^3:s_{\vartheta}(x)<0\},
\qquad
\partial\Omega \approx \{x\in\mathbb{R}^3:s_{\vartheta}(x)=0\}.
\label{eq:impl_sdf_levelset}
\end{equation}
The SDF serves two roles that a surface mesh alone cannot provide: (i)~it furnishes a continuous inside/outside classifier for chart-seed acceptance and domain-containment losses, and (ii)~its gradient field yields approximate surface normals for chart-frame initialization, both without the combinatorial complexity of watertight mesh repair. The training loss combines four objectives---surface fit, normal alignment, Eikonal regularization $(\|\nabla s_\vartheta\|-1)^2$, and sign supervision via offset probes---whose explicit forms are given in Appendix~\ref{sec:appendix_pipeline} (Eqs.~\eqref{eq:impl_sdf_loss}--\eqref{eq:impl_sdf_sign}).

\subsubsection{Seed placement and local frames}
\label{sec:impl_seeds}

The domain is covered by $M$ overlapping ball charts, each anchored at a seed $c_i$ with an orthonormal frame $\mathbf{Q}_i=[\mathbf{t}_{1,i}\;\mathbf{t}_{2,i}\;\mathbf{n}_i]$ and support radius $r_i$. Seeds are accepted only where $s_\vartheta(c_i)<0$. The choice of ball-shaped reference domains $\widehat{\Omega}_i\subset\mathbb{R}^3$ provides an isotropic reference metric that avoids directional bias in collocation sampling and stabilizes Jacobian conditioning.

Each physical point $x$ is expressed in chart-local coordinates via the rigid frame projection defined in~\eqref{eq:atlas_rigid_local_coords}. A point belongs to chart $i$ when it satisfies the distance-based membership criterion
\begin{equation}
d_i(x)=\|x-c_i\|_2,
\qquad
\chi_i(x)=\mathbf{1}\!\left\{
d_i(x)\le (1+\alpha)\min_{j} d_j(x)
\right\},
\label{eq:impl_membership}
\end{equation}
where $\alpha>0$ controls the overlap fraction: larger $\alpha$ increases the communication bandwidth of the Schwarz iteration at the cost of additional per-chart evaluations, while smaller $\alpha$ reduces overlap but may leave gaps in the partition of unity. Bijectivity is not guaranteed by construction; it is monitored a~posteriori through the determinant barrier and foldover gates described below.

\subsubsection{Decoder and mask training}
\label{sec:impl_decoder}

The abstract chart map $\varphi_i$ introduced in Section~\ref{sec:methodology} is realized as a rigid affine frame augmented by a learned residual:
\begin{align}
\Phi_i(\boldsymbol{\zeta})
=&\;
c_i
+
\zeta_1 \mathbf{t}_{1,i}
+
\zeta_2 \mathbf{t}_{2,i}
+
\zeta_3 \mathbf{n}_i
+
a_i r_i\,\mathcal{N}_i(\boldsymbol{\zeta}/r_i),
\label{eq:impl_decoder_map}
\end{align}
where $\mathcal{N}_i$ is a tanh MLP and $a_i=0.20\,\tanh(\texttt{raw\_scale})$ is a learned residual amplitude. This design is central to the atlas construction: at initialization the residual vanishes, so each chart starts as a rigid ball with $\det\mathbf{J}_i>0$ everywhere. Training then deforms the chart to conform to the local geometry, while the bounded amplitude $a_i r_i$ prevents the residual from overwhelming the affine frame and introducing foldover. The decoder is paired with a mask network $M_i$ that produces validity logits and, through them, both a hard membership probability and smooth partition-of-unity weights:
\begin{align}
\ell_i(\boldsymbol{\zeta})
=&\;
M_i(\boldsymbol{\zeta}/r_i),
\qquad
p_i(x)=\sigma\!\big(\ell_i(\boldsymbol{\zeta}_i(x))\big),
\qquad
\omega_i(x)=\frac{\exp(\ell_i(\boldsymbol{\zeta}_i(x)))}{\sum_j \exp(\ell_j(\boldsymbol{\zeta}_j(x)))}.
\label{eq:impl_mask_weights}
\end{align}
The sigmoid output $p_i$ serves as a differentiable chart-membership classifier for collocation-point assignment, while the softmax weights $\omega_i$ provide the smooth blending coefficients needed by the Schwarz coupling and the global partition of unity.

The decoder and mask networks are trained jointly with a composite atlas loss
\begin{equation}
\mathcal{L}_{\mathrm{atlas}}
=
\lambda_{\mathrm{rec/dom}}\mathcal{L}_{\mathrm{rec/dom}}
+
\lambda_{\mathrm{mask}}\mathcal{L}_{\mathrm{mask}}
+
\lambda_{\mathrm{ov}}\mathcal{L}_{\mathrm{ov}}
+
\lambda_{\mathrm{jac}}\mathcal{L}_{\mathrm{jac}}
+
\lambda_{\mathrm{cov}}\mathcal{L}_{\mathrm{cov}},
\label{eq:impl_atlas_total}
\end{equation}
whose five terms penalize reconstruction error ($\mathcal{L}_{\mathrm{rec}}$) or domain-containment violation ($\mathcal{L}_{\mathrm{dom}}$), mask misclassification ($\mathcal{L}_{\mathrm{mask}}$), overlap inconsistency between neighboring charts ($\mathcal{L}_{\mathrm{ov}}$), non-positive Jacobian determinants ($\mathcal{L}_{\mathrm{jac}}$), and insufficient coverage ($\mathcal{L}_{\mathrm{cov}}$), respectively. The surface-based term $\mathcal{L}_{\mathrm{rec}}$ is used when point-cloud data are available; the volumetric variant $\mathcal{L}_{\mathrm{dom}}$ extends the decoder into the interior using the SDF as a containment oracle. Explicit functional forms for each component are given in Appendix~\ref{sec:appendix_pipeline} (Eqs.~\eqref{eq:impl_atlas_recon}--\eqref{eq:impl_atlas_cov}).

\subsubsection{Quality gates}
\label{sec:impl_gates}

The frozen-atlas philosophy requires a binary accept/reject decision before any PDE solver is invoked, rather than soft regularization that could leave residual geometric defects for the solver to absorb. The atlas is accepted only if the following metrics lie within prescribed bands:
\begin{align}
g_{\mathrm{cov}}
=&\;
\mathbb{P}_{x\sim X}
\left[
\max_i p_i(x)>\tau_{\mathrm{cov}}
\right],
\qquad
g_{\mathrm{ov}}
=
\mathbb{E}_{x\in X_{\mathrm{ov}}}
\left[
\|\Phi_i(\boldsymbol{\zeta}_i(x))-\Phi_j(\boldsymbol{\zeta}_j(x))\|_2
\right],
\nonumber\\
g_{\mathrm{fold}}
=&\;
\frac{1}{N_J}
\sum_{i,\boldsymbol{\zeta}}
\mathbf{1}_{\{\det D\Phi_i(\boldsymbol{\zeta})\le 0\}},
\label{eq:impl_gate_metrics}\\
g_{\mathrm{rmse}}
=&\;
\left(
\mathbb{E}_{x\sim X}
\left[
\left\|
\sum_{i=1}^{M}\omega_i(x)\Phi_i(\boldsymbol{\zeta}_i(x))-x
\right\|_2^2
\right]
\right)^{1/2}.
\label{eq:impl_gate_rmse}
\end{align}
Here $g_{\mathrm{cov}}$ measures the fraction of domain points covered by at least one chart, $g_{\mathrm{ov}}$ quantifies the mean pairwise transition-map discrepancy in overlap regions, $g_{\mathrm{fold}}$ records the fraction of Jacobian-evaluation points with non-positive determinant, and $g_{\mathrm{rmse}}$ gives the partition-of-unity-weighted reconstruction accuracy. Failed gates trigger iterative refinement of chart seeds, support radii, decoder states, and mask parameters.

\subsubsection{Benchmark performance on Stanford Bunny atlas}
\label{sec:atlas_diagnostics}

\begin{figure}[htbp]
\centering
\includegraphics[width=\textwidth]{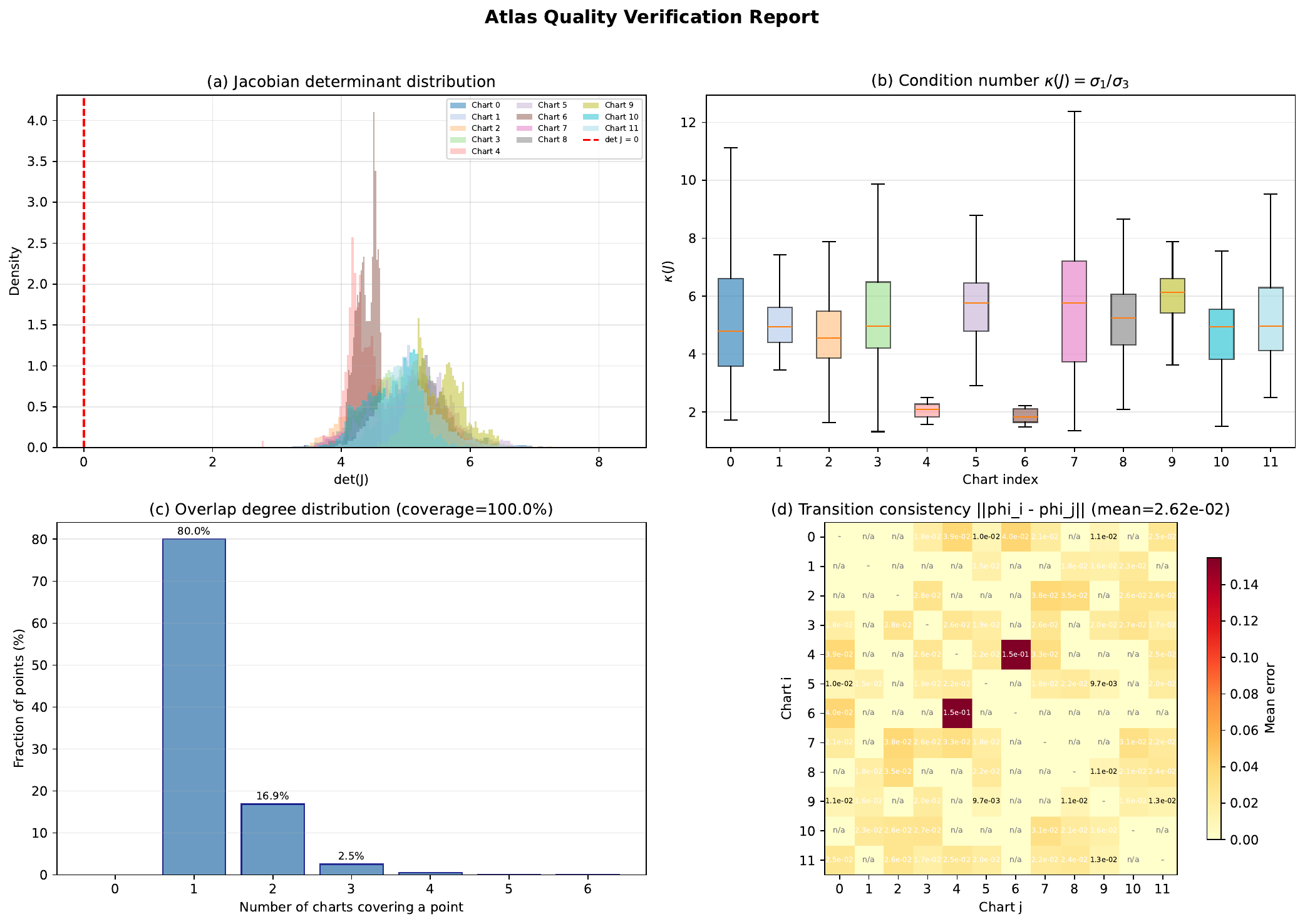}
\caption{Atlas quality verification for the 12-chart rabbit atlas.
(a)~Jacobian determinant distributions per chart evaluated at membership-filtered points; all charts maintain $\det\mathbf{J}_i>0$.
(b)~Condition number $\kappa(\mathbf{J}_i)$ box plots per chart; median values range from 1.8 to 6.1.
(c)~Overlap degree distribution: 80\% of surface points lie in exactly one chart.
(d)~Pairwise transition-map consistency error $\|\varphi_i-\varphi_j\|$ (mean across 37 overlapping pairs: $2.62\times10^{-2}$).}
\label{fig:atlas_quality}
\end{figure}

To verify that the frozen atlas provides well-conditioned coordinate charts, we conduct a numerical experiment on the Stanford Bunny geometry \citep{turk1994zippered}, a standard benchmark used for meshing. In this example, the unit ball is used as the reference domain to avoid edges and corners of the 3D cube, which are only $C^0$-continuous. 
Table~\ref{tab:atlas_diagnostics} and Figure~\ref{fig:atlas_quality} report four categories of quantitative diagnostics for the canonical 12-chart rabbit atlas, computed on two sample populations per chart: (i)~membership-filtered points (the physical-domain samples actually used by the solver, 1{,}500--5{,}000 per chart) and (ii)~5{,}000 dense uniform samples over the full reference ball (including regions outside the body).

\paragraph{Jacobian determinant histograms.}
On the physical domain, all 12 charts maintain strictly positive Jacobian determinants ($\min_i\min_{\boldsymbol{\zeta}} j_i>0$; Table~\ref{tab:atlas_diagnostics}, columns~2--4), confirming zero foldover within the region where the PDE is evaluated.
The per-chart distributions (Figure~\ref{fig:atlas_quality}a) cluster between 3 and 8, reflecting the volume expansion from the reference ball to the irregularly shaped physical patches.
On the full reference ball, five charts (0, 1, 2, 7, 11) exhibit negative determinants at 0.3\%--5.6\% of samples (Table~\ref{tab:atlas_diagnostics}, columns~5--7); these degenerate regions lie outside the body surface and are filtered by the SDF rejection step before any PDE collocation or finite-element assembly.

\paragraph{Distortion and conditioning.}
The 95th-percentile condition numbers $\kappa(\mathbf{J}_i)=\sigma_1/\sigma_3$ range from 2.2 (Chart~6) to 9.5 (Chart~7) on the physical domain, with a chart-averaged mean of 7.0 (Figure~\ref{fig:atlas_quality}b).
Charts 4 and 6, whose seeds lie in geometrically simple regions (torso), have $\kappa_{95}<2.5$; charts near thin features (ears, tail) exhibit moderately higher distortion.
No chart reaches $\kappa_{95}>15$ on the physical domain, confirming that the rigid-frame-plus-residual decoder design (Eq.~\eqref{eq:impl_decoder_map}) controls distortion without explicit isotropy regularization.

\paragraph{Coverage and overlap volume.}
Coverage is 100\%: every surface point belongs to at least one chart.
The overlap degree distribution (Figure~\ref{fig:atlas_quality}c) shows that 80\% of surface points belong to exactly one chart, 16.9\% to two charts, 2.5\% to three, and 0.5\% to four or more.
This means ${\sim}20\%$ of the body volume participates in Schwarz transmission, providing substantial overlap for interface coupling while keeping the majority of the domain single-chart (and therefore computationally inexpensive).
The partition-of-unity entropy $H(\omega)=-\sum_i\omega_i\ln\omega_i$ averages 0.048 across surface points, indicating that most points are dominated by a single chart weight close to one.

\paragraph{Transition-map consistency.}
For each of the 37 overlapping chart pairs $(i,j)$, the round-trip error $\|\varphi_i(\boldsymbol{\zeta}_i(x))-\varphi_j(\boldsymbol{\zeta}_j(x))\|$ is evaluated at shared membership points (Figure~\ref{fig:atlas_quality}d).
The overall mean is $2.62\times10^{-2}$, and 35 of 37 pairs have mean error below $4\times10^{-2}$.
The single outlier pair $(4,6)$ has only one shared point and a consistency error of $1.55\times10^{-1}$; this pair contributes negligibly to the global solution because its overlap region is vanishingly small.
These transition errors are the atlas-level analogue of patch-interface mismatch in multi-patch IGA, and their magnitude is consistent with the downstream Schwarz coupling successfully driving inter-chart agreement to the $10^{-2}$ level observed in the PDE solution (Table~\ref{tab:rabbit_solver_comparison}).

\begin{table}[htbp]
\centering
\caption{Atlas quality diagnostics for the canonical 12-chart rabbit atlas.
\emph{Membership-filtered} evaluates at surface points assigned to each chart;
\emph{full ball} samples the entire reference domain including regions outside the body.
Foldover is the fraction of samples with $\det\mathbf{J}_i \leq 0$.}
\label{tab:atlas_diagnostics}
\smallskip
\small
\begin{tabular}{lcccccc}
\toprule
 & \multicolumn{3}{c}{Membership-filtered} & \multicolumn{3}{c}{Full reference ball} \\
\cmidrule(lr){2-4}\cmidrule(lr){5-7}
Chart & $\bar{j}_i$ & $\kappa_{95}$ & fold\% & $\min j_i$ & $\kappa_{95}$ & fold\% \\
\midrule
0  & 4.95 & 9.16  & 0.00 & $-3.24$ & 14.09 & 5.60 \\
1  & 4.97 & 6.72  & 0.00 & $-0.40$ &  6.74 & 0.74 \\
2  & 4.73 & 7.71  & 0.00 & $-6.38$ &  6.69 & 2.18 \\
3  & 4.87 & 8.90  & 0.00 &  1.40   &  6.78 & 0.00 \\
4  & 4.34 & 2.37  & 0.00 &  1.74   &  2.42 & 0.00 \\
5  & 5.30 & 7.61  & 0.00 &  0.70   &  8.49 & 0.00 \\
6  & 4.38 & 2.21  & 0.00 &  1.82   &  2.13 & 0.00 \\
7  & 5.00 & 9.49  & 0.00 & $-0.55$ &  9.68 & 0.34 \\
8  & 5.08 & 7.22  & 0.00 &  0.48   &  9.31 & 0.00 \\
9  & 5.47 & 7.29  & 0.00 &  0.86   &  8.86 & 0.00 \\
10 & 4.78 & 6.65  & 0.00 &  0.35   &  6.44 & 0.00 \\
11 & 5.04 & 8.98  & 0.00 & $-1.34$ & 13.51 & 2.18 \\
\midrule
Avg & 4.91 & 7.03 & 0.00 & $-0.38$ &  7.93 & 0.92 \\
\bottomrule
\end{tabular}
\end{table}

\begin{remark} (Chart-local solvers and CompactChartNet)
%\subsection{Chart-local solvers and CompactChartNet}
\label{sec:impl_compact}

Two chart-local architectures are used for the rabbit Poisson benchmark. The dense baseline is a $64\times 4$ tanh MLP per chart. The compact alternative, \texttt{CompactChartNet}, replaces each monolithic chart network with $9$ sub-seeds, one $32\times 2$ tanh subnet per sub-seed, and a softmax partition of unity with scale $\tau=0.125\,r_i$:
\begin{equation}
u_{\mathrm{chart}}(\boldsymbol{\zeta})
=
\sum_{m=0}^{8}\phi_m(\boldsymbol{\zeta})\,u_m(\boldsymbol{\zeta}-\boldsymbol{s}_m),
\qquad
\phi_m(\boldsymbol{\zeta})=\frac{\exp(-\|\boldsymbol{\zeta}-\boldsymbol{s}_m\|^2/2\tau^2)}{\sum_{m'}\exp(-\|\boldsymbol{\zeta}-\boldsymbol{s}_{m'}\|^2/2\tau^2)}.
\label{eq:compact_pou_impl}
\end{equation}
This localized construction produces smoother Schwarz convergence histories than the dense baseline. The same atlas pipeline also hosts a chart-local P1 FEM variant that reuses the decoder, mask, overlap graph, and Schwarz infrastructure, confirming that the atlas serves as a solver-independent geometric substrate. Solver stabilization employs staged PDE-weight ramping, PCGrad \citep{yu2020gradient} for gradient-conflict mitigation, and a trust-region filter that rejects Schwarz updates causing unacceptable error increases. Adaptive regularization heuristics for the overlap weight and CompactChartNet bandwidth are detailed in Appendix~\ref{sec:appendix_stabilization}.
\end{remark}

\section{Numerical experiments on partial differential equations}
\label{sec:benchmark_formulations}

The five examples are sequence of verification exercises and capability tests. Example~1 (ellipsoid Poisson) verifies that the mapped operator is mathematically correct on analytic geometry with a known closed-form solution. Example~2 (rabbit Poisson) asks whether the framework can handle an arbitrary 3D shape---imported from scan data---without a unified volumetric mesh, and cross-validates the atlas by running both a PINN and a classical FEM on the same pipeline. Example~3 (torus forward BVP) tests whether the atlas and Schwarz coupling survive a nonlinear equilibrium solve with history-dependent $J_2$ elastoplasticity. Example~4 (torus inverse Neo-Hookean) reverses the information flow, using observed boundary data to recover constitutive parameters across multiple charts. Example~5 (elastoplastic inverse) extends the identification to cyclic loading with plasticity, recovering yield stress and kinematic hardening from reaction-force data. Each example therefore adds one layer of complexity---geometry, nonlinearity, inverse capability, or history dependence---so that a failure at any stage isolates the source of the difficulty.

% Reporting protocol and traceability table moved to supplementary artifact package.

% Example 1 source traceability (comment only; not visible in the compiled PDF).
% Paths below are relative to the PINN_coordinate_chart_3Dgeometry/ project root.
% Code:
%   - src/pinn_3d_ellipsoid_mapped_sphere.py
%     Analytic mapped-Poisson verification script for the ellipsoid benchmark.
% Data / run artifacts:
%   - runs/examples/ellipsoid3d_20260212_180204/
%     Canonical archived run directory used for the reported verification metrics and figure export.
% Notes:
%   - This benchmark has no external raw geometry directory because the ellipsoid map, forcing,
%     and exact solution are generated analytically inside the script.
% Code location: manuscript_experiments/example1_forward_poisson/
% Figure scripts: manuscript/scripts_figures/example1_forward_poisson.py
%                 manuscript/scripts_figures/example1_forward_poisson_3d.py
\subsection{Example 1: Verifcation with Laplac's equation on the ellipsoid}
\label{sec:example1}
\begin{figure}[htbp]
\centering
\includegraphics[width=0.97\textwidth]{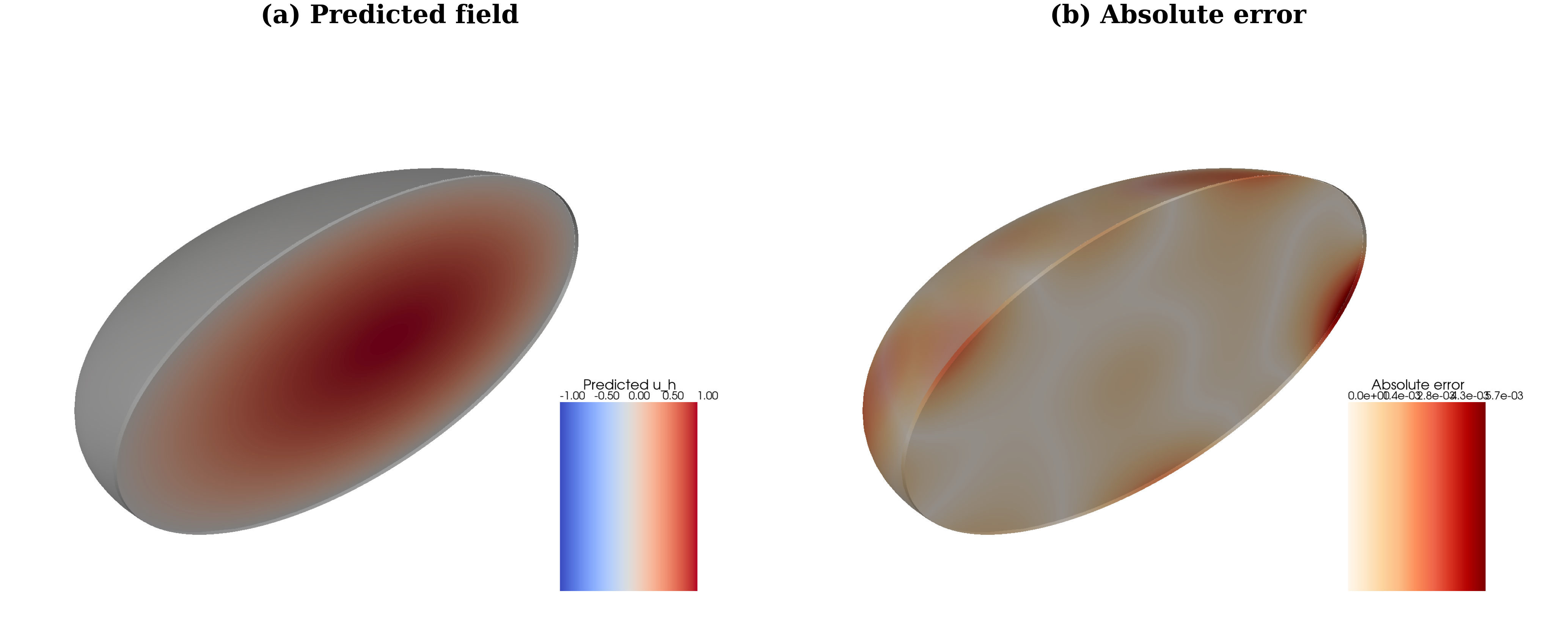}
\caption{Ellipsoidal verification benchmark rendered in 3D. The ellipsoid is clipped along a symmetry plane to expose the interior solution field. Panel~(a) shows the predicted field $u_h$, which peaks at the center and vanishes on the boundary as expected for the quadratic exact solution $u^*=1-\|\boldsymbol{\xi}\|^2$. Panel~(b) shows the absolute error $|u_h-u^*|$, confirming that the operator pullback and the analytic geometric map are implemented correctly (relative $L^2$ error $=2.26\times10^{-3}$, maximum error $=5.23\times10^{-3}$).}
\label{fig:forward_poisson_ellipsoid}
\end{figure}

We use the Laplace's equation with known analytical solution to verify our implementation. 
The corresponding boundary value problem reads,
\begin{equation}
-\Delta_x u = f \quad \text{in }\Omega,
\qquad
u=g \quad \text{on }\partial\Omega.
\label{eq:ex1_phys}
\end{equation}
For a chart-local scalar field, the mapped residual is
\begin{equation}
\widehat{\mathcal{R}}^{(1)}_i(\boldsymbol{\zeta})
:=
-\nabla_{\boldsymbol{\zeta}}\cdot\left(
\mathbf{A}_i(\boldsymbol{\zeta})\nabla_{\boldsymbol{\zeta}} \widehat{u}_i(\boldsymbol{\zeta})
\right)
- j_i(\boldsymbol{\zeta})\,f(\varphi_i(\boldsymbol{\zeta})),
\qquad
\mathbf{A}_i(\boldsymbol{\zeta}):=j_i(\boldsymbol{\zeta})\mathbf{J}_i(\boldsymbol{\zeta})^{-1}\mathbf{J}_i(\boldsymbol{\zeta})^{-T}.
\label{eq:ex1_residual}
\end{equation}
The boundary loss is
\begin{equation}
\widehat{\mathcal{B}}^{(1)}_{i,D}(\boldsymbol{\zeta}):=\widehat{u}_i(\boldsymbol{\zeta})-g(\varphi_i(\boldsymbol{\zeta})).
\label{eq:ex1_bc}
\end{equation}
The verification setup uses a single analytic map
\[
\varphi(\xi_1,\xi_2,\xi_3)=(a\xi_1,b\xi_2,c\xi_3),
\qquad
\boldsymbol{\xi}\in B^3,
\qquad
(a,b,c)=(2.0,\,1.4,\,0.9),
\]
which pulls the ellipsoid
$\Omega=\{{\bf x}\in\mathbb{R}^3: x_1^2/a^2+x_2^2/b^2+x_3^2/c^2<1\}$
back to the unit ball.  The Jacobian is constant, $\mathbf{J}=\diag(a,b,c)$, with determinant $j=abc$.  Homogeneous Dirichlet data $g=0$ are prescribed on $\partial\Omega$.  The exact solution in physical coordinates is
\begin{equation}
u^{\ast}_{\mathrm{ell}}(\mathbf{x})
=1-\frac{x_1^2}{a^2}-\frac{x_2^2}{b^2}-\frac{x_3^2}{c^2},
\label{eq:ellipsoid_exact_phys}
\end{equation}
and its pull-back to the reference ball is
\begin{equation}
\widehat{u}^{\ast}_{\mathrm{ell}}(\boldsymbol{\xi})=1-\|\boldsymbol{\xi}\|_2^2.
\label{eq:ellipsoid_exact}
\end{equation}
Substituting~\eqref{eq:ellipsoid_exact_phys} into~\eqref{eq:ex1_phys} gives the constant source term
\begin{equation}
f = 2\!\left(\frac{1}{a^2}+\frac{1}{b^2}+\frac{1}{c^2}\right)
\approx 4.039.
\label{eq:ellipsoid_forcing}
\end{equation}
Because the geometry, the exact solution, and the forcing are all analytic, this example isolates the operator pullback from any geometry-learning uncertainty or overlap coupling. The canonical run reaches relative $L^2$ error $2.26\times 10^{-3}$, absolute $L^2$ error $1.08\times 10^{-3}$, and maximum pointwise error $5.23\times 10^{-3}$ in $106.3$\,s, confirming that the mapped operators and neural approximation are implemented correctly before atlas coupling and learned geometry are introduced, as shown in Figure \ref{fig:forward_poisson_ellipsoid}.

\subsection{Example 2: rabbit Poisson on a fixed atlas}
\label{sec:example2}
The rabbit Poisson benchmark investigates whether the atlas pipeline can support a complex-geometry forward solve without being tied to a single local solver family. Two atlases built by the same four-stage pipeline (Section~\ref{sec:implementation})---a twelve-chart atlas for the PINN and an eight-chart atlas for the FEM---support two realizations of the mapped Poisson problem: chart-local PINN surrogates and chart-local P1 FEM solves. Both atlases satisfy the same quality gates; the chart count is a user parameter that trades per-chart complexity against the number of Schwarz interfaces, analogous to choosing a mesh resolution. The physical PDE is again \eqref{eq:ex1_phys}, with the manufactured solution given by
\begin{equation}
u^{\ast}_{\mathrm{rabbit}}(x)=\sin(\pi x_1)\sin(\pi x_2)\sin(\pi x_3),
\qquad
f=3\pi^2 u^{\ast}_{\mathrm{rabbit}},
\label{eq:rabbit_poisson_exact}
\end{equation}
Dirichlet boundary condition is applied on the surface by substituting the position vectors into \ref{eq:rabbit_poisson_exact}, while the interier $u$ is obtained from the solver.  

\subsubsection{Schwarz PINN}
The interior and boundary residuals are the same as in Example~1, but the loss adds overlap coupling:
\begin{align}
\mathcal{L}^{(2)}
=&
\lambda_{\mathrm{pde}}\sum_{i=1}^{M}\mathbb{E}_{\widehat{\Omega}_i}
\left[\left|\widehat{\mathcal{R}}^{(1)}_i\right|^2\right]
+
\lambda_{\mathrm{bc}}\sum_{i=1}^{M}\mathbb{E}_{\widehat{\Gamma}_{i,D}}
\left[\left|\widehat{\mathcal{B}}^{(1)}_{i,D}\right|^2\right]
\nonumber\\
&+
\lambda_{\mathrm{if,val}}\sum_{i<j}\mathbb{E}_{\widehat{\Omega}_{ij}}
\left[
\left|
\widehat{u}_i(\boldsymbol{\zeta})-\widehat{u}_j(\psi_{ij}(\boldsymbol{\zeta}))
\right|^2
\right]
\nonumber\\
&+
\lambda_{\mathrm{if,flux}}\sum_{i<j}\mathbb{E}_{\widehat{\Omega}_{ij}}
\left[
\left|
\mathbf{n}_{ij}\cdot \nabla_x u_i - \mathbf{n}_{ij}\cdot \nabla_x u_j
\right|^2
\right]
+
\lambda_{\mathrm{sup}}\mathcal{L}_{\mathrm{sup}}.
\label{eq:rabbit_poisson_loss}
\end{align}
Here $\mathcal{L}_{\mathrm{sup}}$ denotes an optional manufactured-solution anchor used in the rabbit PINN runs to stabilize Schwarz iteration. Each chart-local neural field can be represented by a dense MLP, a residual MLP, or a compact chart-partitioned network. The flagship run uses the compact architecture.
This numerical example is designed to examine whether the geometry handling, chart-local approximation, and Schwarz transmission are functioning simultaneously in different solver settings. 
 
In principle, one may also construct a single global mapping to solve the Poisson problem on the Bunny geometry. In practice, however, obtaining such a mapping with sufficiently low distortion is difficult because of thin features (e.g., the ears). Resolving these details would also require a larger neural network to achieve sufficient expressivity and resolution, further increasing computational cost. By contrast, the atlas approach uses a divide-and-conquer strategy, reducing per-chart expressivity demands by partitioning the geometry into coordinate charts.  
 Once the atlas is frozen, the PINN solver generates interior collocation by SDF rejection in chart coordinates, trains one scalar field network per chart, and enforces compatibility by interface-value and interface-flux penalties inside multiplicative Schwarz sweeps with a trust-region acceptance filter.

\subsubsection{Schwarz FEM}
 In the atlas-based FEM, each chart solves the same mapped Poisson operator in weak form,
\[
\int_{\widehat{\Omega}_i} (\nabla_{\boldsymbol{\zeta}} v)^T
\mathbf{A}_i(\boldsymbol{\zeta})\,
\nabla_{\boldsymbol{\zeta}} \widehat{u}_i
\,d\boldsymbol{\zeta}
=
\int_{\widehat{\Omega}_i}
j_i(\boldsymbol{\zeta})\,f(\varphi_i(\boldsymbol{\zeta}))\,v
\,d\boldsymbol{\zeta},
\]
with Dirichlet data supplied by the physical boundary and by Schwarz traces on artificial chart boundaries. Each chart uses a P1 tetrahedral discretization of the structured cube $[-r_i,r_i]^3$ split by Freudenthal tetrahedra. The implementation reuses the same decoder--mask atlas machinery, overlap coloring, and partition-of-unity blending as the PINN solver; Schwarz boundary values are again obtained from neighboring chart solutions through the atlas, and the FEM runs disable SDF filtering by default for these volumetric chart meshes.

\subsubsection{Numerical results}
\begin{figure}[htbp]
\centering
\includegraphics[width=0.98\textwidth]{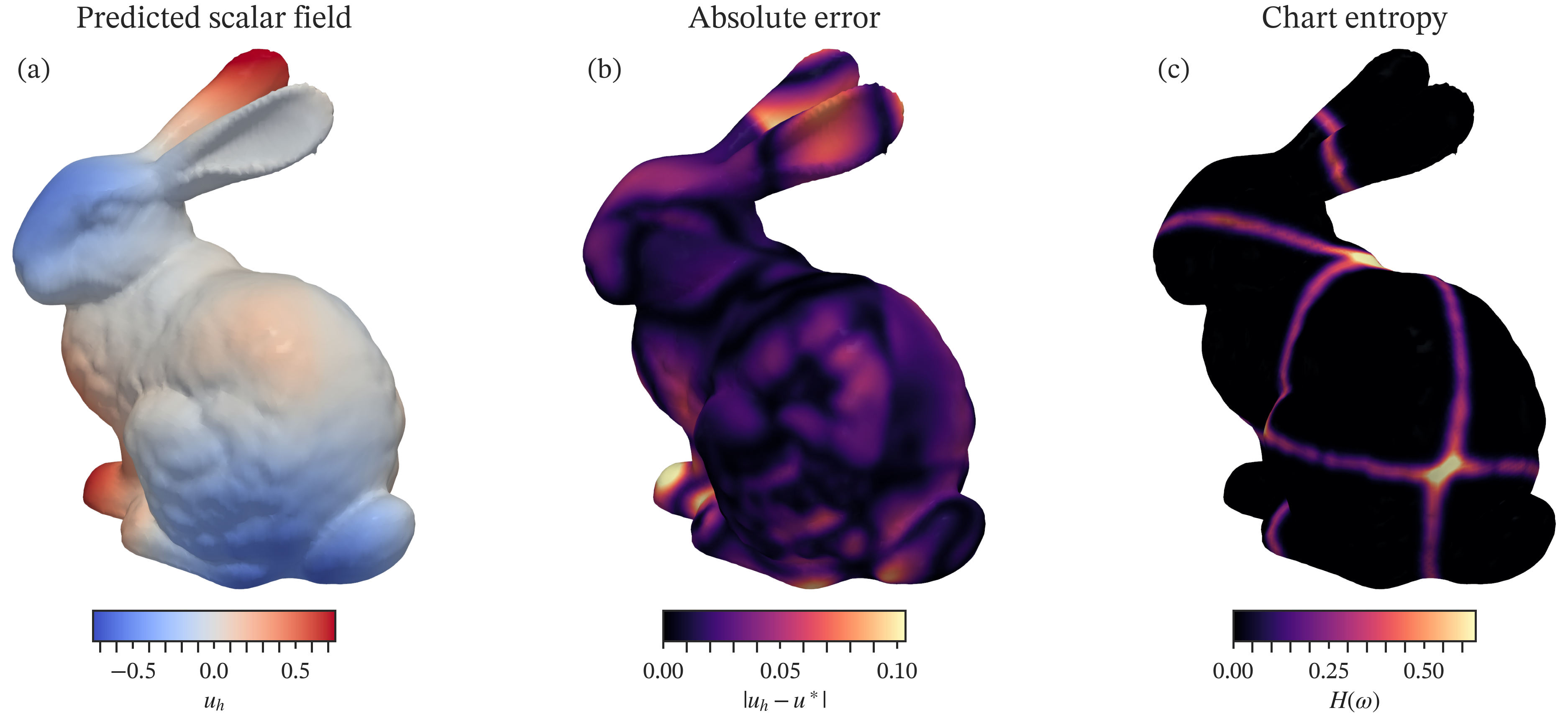}
\caption{Representative rabbit-surface visualization for the fixed-atlas SA-PINN solver. The surface is reconstructed from a dense boundary postprocessing pass of the rabbit Poisson solution family and shown from one fixed camera in all three panels. Panel (a) shows the recovered scalar field, panel (b) shows the absolute error, and panel (c) shows the chart entropy induced by the partition-of-unity weights. The first two panels show that the solved field remains smooth on the complex geometry, while the entropy panel shows that uncertainty is confined to narrow overlap traces rather than spreading over the whole surface.}
\label{fig:rabbit_poisson_example}
\end{figure}

\begin{figure}[htbp]
\centering
\includegraphics[width=0.98\textwidth]{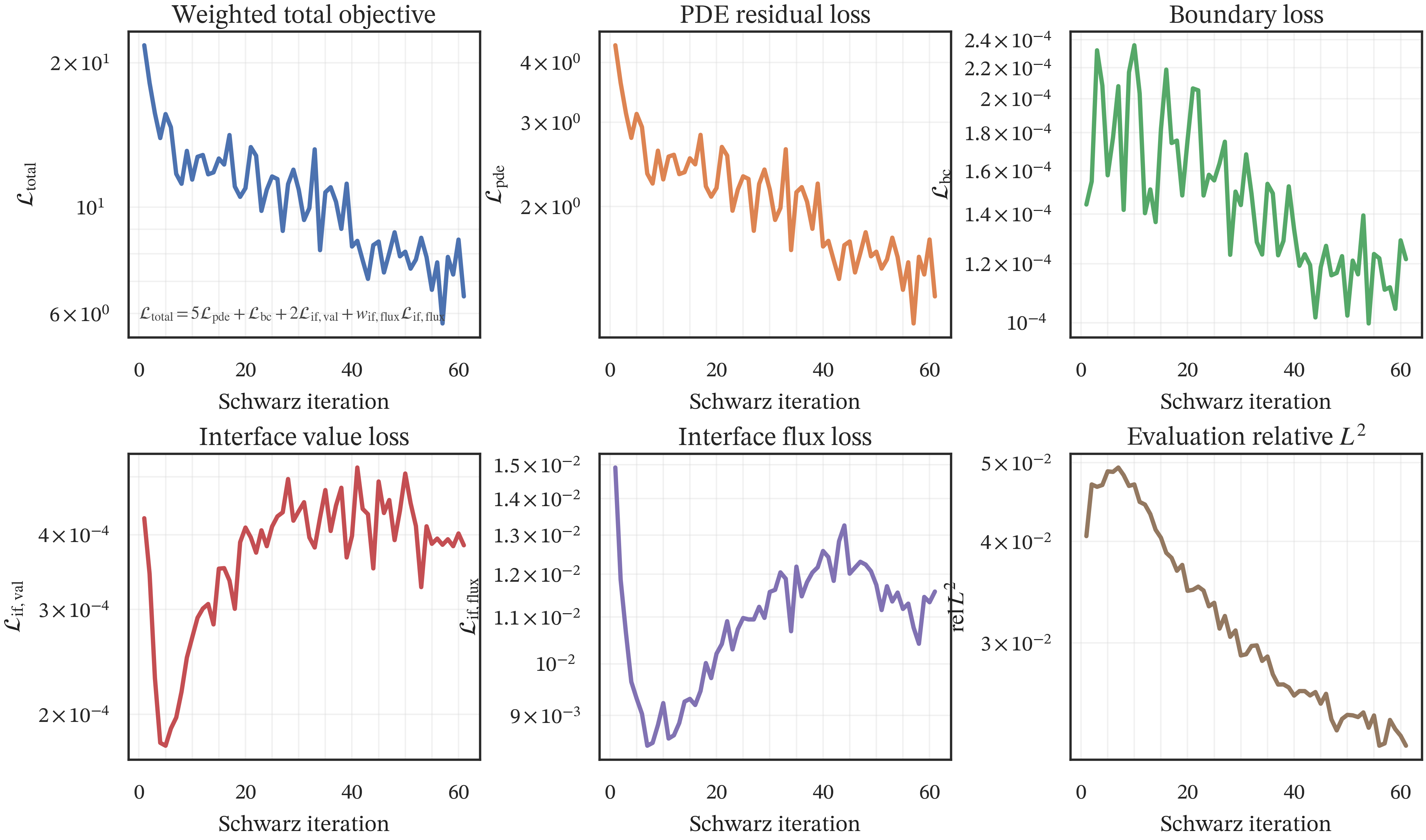}
\caption{Training diagnostics for the canonical rabbit Poisson compact PINN run. Each panel reports one scalar trace over Schwarz iterations: total objective, PDE residual, boundary loss, interface-value loss, interface-flux loss, and evaluation relative $L^2$ error.}
\label{fig:rabbit_poisson_training_losses}
\end{figure}

The compact PINN run attains relative $L^2$ error $2.21\times 10^{-2}$, absolute $L^2$ error $8.57\times 10^{-3}$, maximum pointwise error $6.78\times 10^{-2}$, final interface-flux mismatch $1.16\times 10^{-2}$, and runtime $3659$\,s. The training history records $61$ Schwarz iterations, and the selected checkpoint stores twelve compact chart networks with roughly $1.31\times 10^{5}$ trainable chart-local weights in total. The solver terminates after a fixed sweep budget ($K=100$); within that budget, the trust-region filter (Appendix~\ref{sec:appendix_stabilization}) rejects any sweep whose blended relative-$L^2$ error exceeds that of the previous accepted state by more than a prescribed tolerance, rolling back to the snapshot and reducing the local learning rate. The best checkpoint is selected a~posteriori as the sweep with the lowest blended relative-$L^2$ error. The diagnostics in Figures~\ref{fig:rabbit_poisson_example} and \ref{fig:rabbit_poisson_training_losses} show the same behavior spatially and temporally: the field remains smooth on the rabbit surface and the objective components contract over Schwarz iterations. A shorter dense-MLP run reaches best-checkpoint relative $L^2$ error $2.53\times 10^{-2}$ and maximum pointwise error $5.30\times 10^{-2}$ in $404$\,s, with a tighter final interface-flux mismatch of $2.52\times 10^{-3}$. The compact architecture therefore improves the best global relative $L^2$ metric and convergence smoothness. 

Given the high dimensionality of the Schwarz-PINN hyperparameter space, we used a coding assistant (Claude Sonnet 4.6) to explore tuning options for 24 hours. The results are summarized in Table~\ref{tab:rabbit_reinforcement_progression} in Appendix~\ref{sec:appendix_rabbit_details}. They show that the combination of 
direct-coordinate evaluation, manufactured-solution anchoring, plateau detection, stronger overlap coupling, and PCGrad may all improve the dense-MLP relative $L^2$ from $3.70\%$ to $2.53\%$; CompactChartNet then reaches $2.21\%$. Nevertheless, it should be noted that, in a realistic setting, the solution is often not known a priori and hence the results obtained with a manufactured-solution anchor are not representative. We believe further hyperparameter tuning may reduce the PINN errors; see, for instance, \citep{wang2023expert,bahmani2021studentteacher}. However, such an endeavor is outside the scope of this study. 

\begin{figure}[htbp]
\centering
\includegraphics[width=0.68\textwidth]{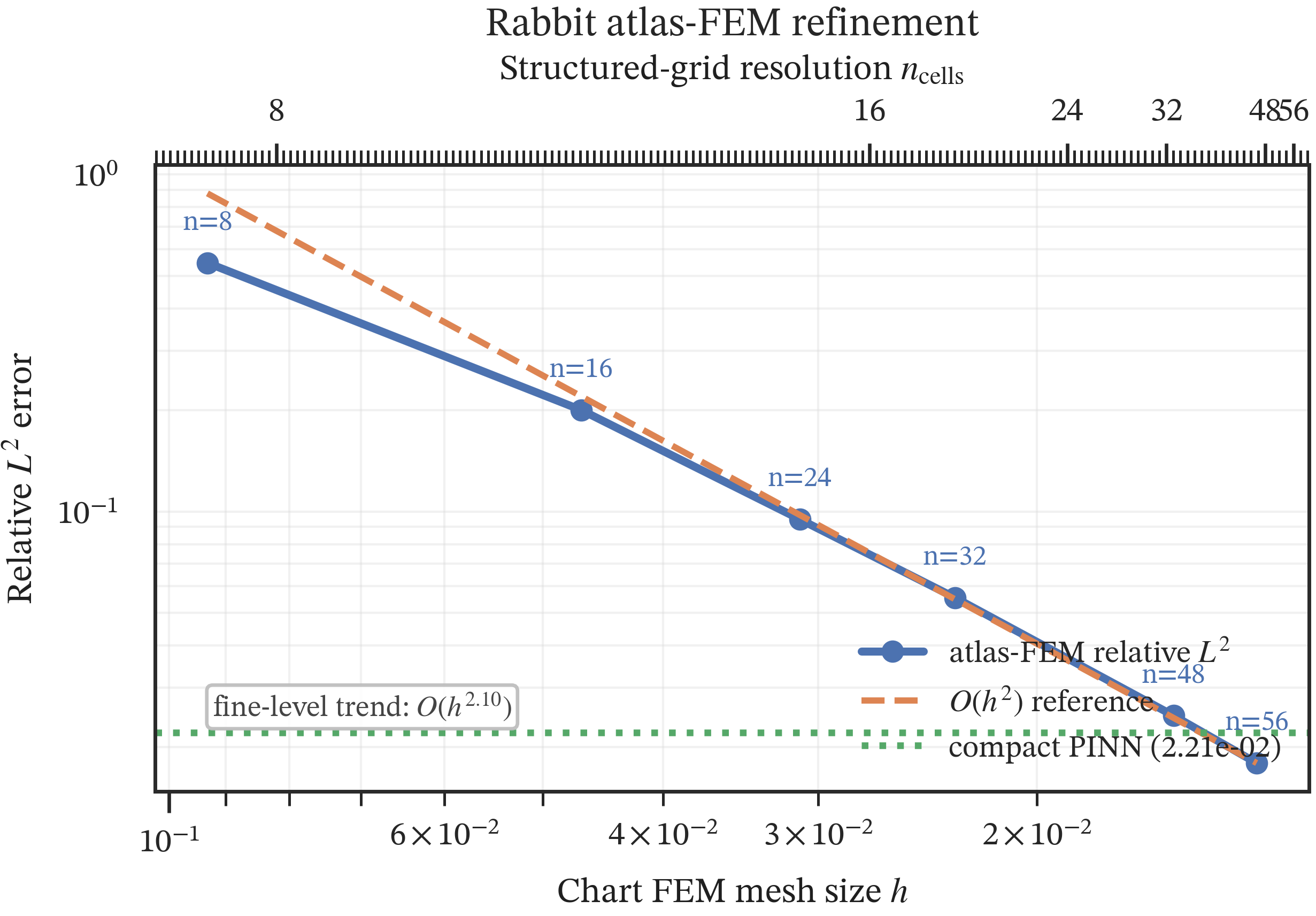}
\caption{Atlas-based FEM convergence on the rabbit benchmark. The dashed line gives an $O(h^2)$ reference slope; the dotted horizontal line marks the compact PINN accuracy $2.21\times10^{-2}$. The FEM trajectory crosses the PINN level near $n_{\mathrm{cells}}=48$.}
\label{fig:rabbit_poisson_fem_refinement}
\end{figure}

Table~\ref{tab:rabbit_fem_convergence} reports the FEM $h$-refinement study from $n_{\mathrm{cells}}=8$ to $56$. The relative $L^2$ error drops from $5.47\times 10^{-1}$ to $1.79\times 10^{-2}$, with rates consistent with the theoretical $O(h^2)$ behavior of mapped P1 elements (reaching $2.10$ at the finest level). The FEM curve (Figure~\ref{fig:rabbit_poisson_fem_refinement}) approaches the compact PINN reference at $n_{\mathrm{cells}}=48$ and crosses below it at $n_{\mathrm{cells}}=56$.

\begin{table}[htbp]
\centering
\scriptsize
\setlength{\tabcolsep}{3.0pt}
\caption{Atlas-based FEM $h$-refinement study for the rabbit Poisson benchmark. Rates are computed between successive refinement levels using the relative-$L^2$ error.}
\label{tab:rabbit_fem_convergence}
\begin{tabular}{>{\centering\arraybackslash}p{0.72cm}>{\centering\arraybackslash}p{1.0cm}>{\centering\arraybackslash}p{1.48cm}>{\centering\arraybackslash}p{1.28cm}>{\centering\arraybackslash}p{0.82cm}>{\centering\arraybackslash}p{1.18cm}>{\centering\arraybackslash}p{0.74cm}}
\toprule
$n_{\mathrm{cells}}$ & $h$ & DOFs & rel.\ $L^2$ & Rate & Max error & Time \\
\midrule
8  & $9.31\times10^{-2}$ & 5{,}832     & $5.47\times10^{-1}$ & ---  & $2.42\times10^{-1}$ & 7\,s \\
16 & $4.65\times10^{-2}$ & 39{,}304    & $2.00\times10^{-1}$ & 1.45 & $9.04\times10^{-2}$ & 45\,s \\
24 & $3.10\times10^{-2}$ & 125{,}000   & $9.49\times10^{-2}$ & 1.84 & $4.47\times10^{-2}$ & 5\,min \\
32 & $2.33\times10^{-2}$ & 287{,}496   & $5.55\times10^{-2}$ & 1.86 & $2.64\times10^{-2}$ & 21\,min \\
48 & $1.55\times10^{-2}$ & 941{,}192   & $2.48\times10^{-2}$ & 1.99 & $1.12\times10^{-2}$ & 49\,min \\
56 & $1.33\times10^{-2}$ & 1{,}481{,}544 & $\mathbf{1.79\times10^{-2}}$ & 2.10 & $\mathbf{8.72\times10^{-3}}$ & 140\,min \\
\bottomrule
\end{tabular}
\end{table}

\begin{table}[htbp]
\centering
\scriptsize
\setlength{\tabcolsep}{3.0pt}
\caption{Workflow-level rabbit comparison. The compact PINN uses a frozen 12-chart atlas; the FEM uses a frozen 8-chart atlas built by the same pipeline with a different chart-count parameter. Both atlases pass the same quality gates. The PINN row reports the selected best-rel-$L^2$ checkpoint; the FEM rows report final metrics after five Schwarz sweeps.}
\label{tab:rabbit_solver_comparison}
\begin{tabular}{>{\raggedright\arraybackslash}p{1.45cm}>{\centering\arraybackslash}p{0.75cm}>{\raggedright\arraybackslash}p{1.95cm}>{\centering\arraybackslash}p{1.12cm}>{\centering\arraybackslash}p{1.12cm}>{\centering\arraybackslash}p{0.72cm}>{\centering\arraybackslash}p{0.88cm}}
\toprule
Local solver & Charts & Size & rel.\ $L^2$ & Max error & Sweeps & Runtime \\
\midrule
PINN compact & 12 & $1.31\times10^{5}$ weights & $2.21\times10^{-2}$ & $6.78\times10^{-2}$ & 61 & 3659\,s \\
\shortstack[l]{FEM P1\\($n=48$)} & 8 & $9.41\times10^{5}$ DOFs & $2.48\times10^{-2}$ & $1.12\times10^{-2}$ & 5 & 2939\,s \\
\shortstack[l]{FEM P1\\($n=56$)} & 8 & $1.48\times10^{6}$ DOFs & $\mathbf{1.79\times10^{-2}}$ & $\mathbf{8.72\times10^{-3}}$ & 5 & 8395\,s \\
\bottomrule
\end{tabular}
\end{table}

Table~\ref{tab:rabbit_solver_comparison} clarifies the tradeoff along several axes. The PINN and FEM use atlases with different chart counts (12 and 8, respectively), reflecting the different granularity preferences of the two solver families: finer charts simplify per-chart neural approximation, while coarser charts maximize per-chart DOF counts for efficient Newton solves. Both atlases are produced by the same four-stage pipeline and satisfy the same quality gates; the chart count is a user parameter analogous to mesh resolution.

At $n_{\mathrm{cells}}=48$, the atlas-FEM error $2.48\times 10^{-2}$ is already comparable to the compact PINN's $2.21\times 10^{-2}$, and at $n_{\mathrm{cells}}=56$ it improves that metric by about $19\%$. In wall-clock time the two methods are broadly competitive at matched accuracy: the FEM reaches $2.48\times 10^{-2}$ in $2{,}939$\,s versus the PINN's $2.21\times 10^{-2}$ in $3{,}659$\,s. Surpassing the PINN with the FEM requires the $n=56$ level at $8{,}395$\,s, indicating that the atlas-plus-Schwarz infrastructure is the dominant cost driver in both cases, not the local PDE discretization itself.

The Schwarz behavior is qualitatively different. The FEM branch converges in five sweeps because each chart solves a sparse linear system by direct factorization, while inter-chart Schwarz boundary data require Newton inversion of the nonlinear decoder map $\varphi_j^{-1}$. These inversions are precomputed once per resolution level and cached, since they depend only on the atlas geometry and not on the evolving solution. The PINN branch instead evaluates its field directly in the rigid local coordinates, avoiding decoder inversion entirely, but at the expense of a longer gradient-based Schwarz history ($61$ sweeps) and ${\sim}200$ local optimizer steps per chart per sweep.

A revealing comparison metric is the total number of neural-network forward passes through the shared decoder and mask networks. The FEM solver queries these networks only during the one-time geometric precomputation---Newton-based map inversions at artificial boundary nodes and mask-logit evaluations for partition-of-unity weights---amounting to approximately $23$ million forward passes at $n=48$, performed once and cached. The SA-PINN evaluates its chart-local field networks and their autograd derivatives at every training step across all Schwarz sweeps: with $61$ sweeps, ${\sim}200$ local steps per chart, $12$ active charts, and ${\sim}2{,}000$ collocation points per step, the total reaches approximately $195$ million neural-network evaluations, roughly $8.5\times$ the FEM count. This disparity quantifies the fundamental cost difference: the PINN pays for approximation power through iterative training, while the FEM amortizes geometry queries into a one-time precomputation and pays instead through mesh refinement.

The FEM convergence study in Table~\ref{tab:rabbit_fem_convergence} serves as an independent validation of the atlas-based coordinate pullback. The observed $O(h^2)$ rates---approaching the theoretical expectation for P1 elements---confirm that the mapped diffusion tensor $\mathbf{A}_i = j_i \mathbf{J}_i^{-1}\mathbf{J}_i^{-T}$, the Jacobian-weighted load vector, and the Schwarz boundary exchange introduce no spurious accuracy degradation. This is a statement about the geometric infrastructure, not about either solver: the atlas is mathematically faithful as a coordinate substrate. The rabbit benchmark therefore supports a sharper conclusion: the atlas is independent of the local discretization, and the FEM convergence independently certifies the correctness of the mapped-operator formulation.

\begin{figure}[htbp]
\centering
\includegraphics[width=0.98\textwidth]{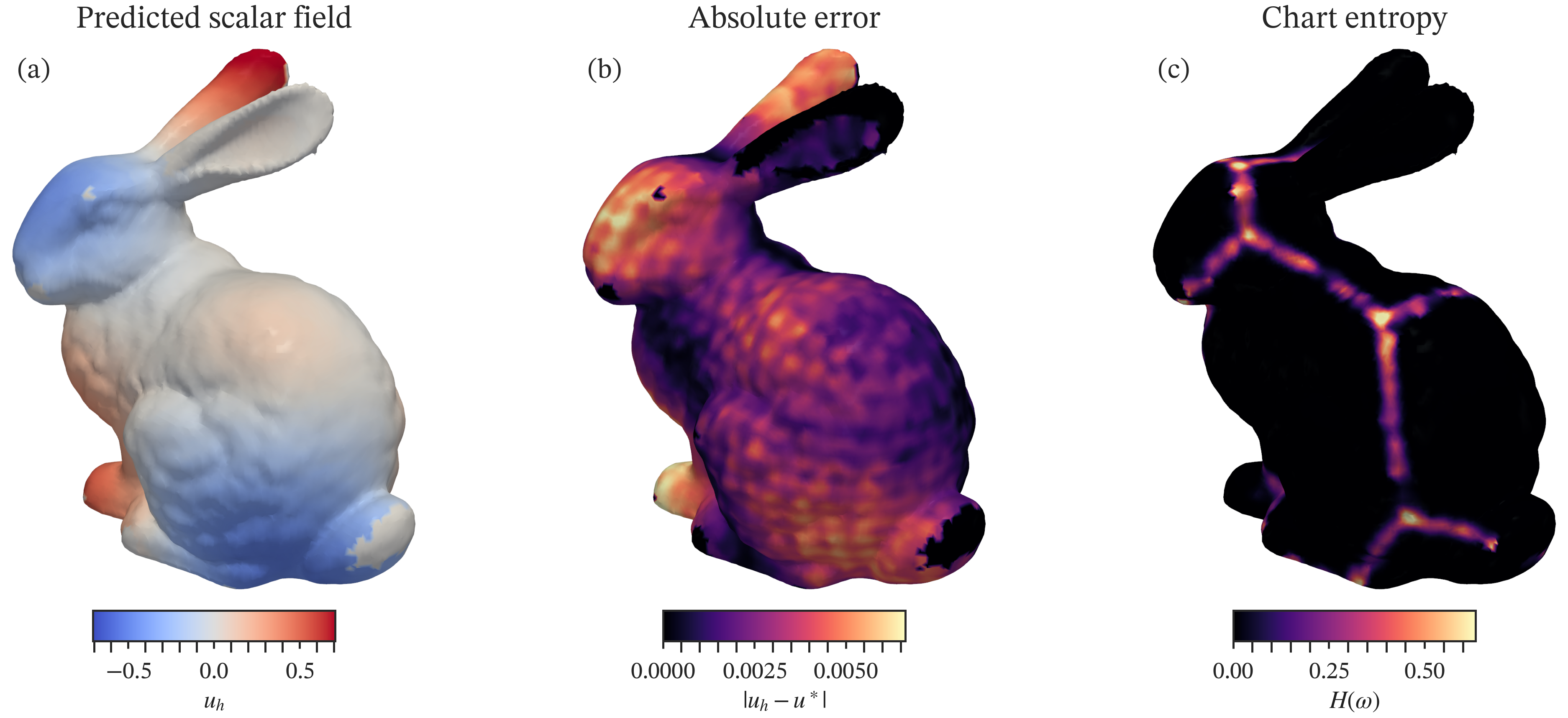}
\caption{Rabbit atlas-FEM surface rendering at the finest refinement level ($n_{\mathrm{cells}}=56$). The figure uses the same reconstructed rabbit surface and the same fixed camera as Figure~\ref{fig:rabbit_poisson_example}. Panel (a) shows the FEM scalar field, panel (b) shows the absolute error, and panel (c) shows chart entropy reconstructed from the frozen rabbit atlas masks on the shared surface postprocessor. The first two panels show that the atlas-FEM branch recovers the same large-scale field as the compact PINN while producing a visibly smaller surface error.}
\label{fig:rabbit_poisson_fem_surface}
\end{figure}

Figure~\ref{fig:rabbit_poisson_fem_surface} repeats the same surface-rendering protocol for the finest atlas-FEM baseline ($n_{\mathrm{cells}}=56$). Using the same rabbit shell and the same camera angle as Figure~\ref{fig:rabbit_poisson_example} makes the comparison direct: the recovered scalar field is visually consistent with the compact PINN at the scale of the rendered surface, while the FEM absolute-error map is markedly smaller and more uniform. This qualitative contrast matches Table~\ref{tab:rabbit_solver_comparison}: the FEM improves the relative $L^2$ error from $2.21\times10^{-2}$ to $1.79\times10^{-2}$ and reduces the maximum pointwise error from $6.78\times10^{-2}$ to $8.72\times10^{-3}$. The entropy panel is retained only to preserve the same atlas-view convention as the PINN rendering, so that the reader can compare the two surface solutions under identical visual framing.

% Example 3 source traceability (comment only; not visible in the compiled PDF).
% Paths below are relative to the PINN_coordinate_chart_3Dgeometry/ project root.
% Code:
%   - src/train_sdf_rabbit.py
%     Main global volumetric Bunny SDF trainer used for the canonical geometry-learning benchmark.
%   - experiments/run_poisson_rabbit_atlas_schwarz.py
%     Reused for the downstream 8-chart Bunny Poisson continuation study discussed in this example.
% Data / run artifacts:
%   - runs/atlas_schwarz_20260212_210412/downloads/bunny/reconstruction/bun_zipper.ply
%     Imported watertight Stanford Bunny surface mesh used as the raw geometry input.
%   - runs/bunny_sdf_v3/
%     Canonical archived global Bunny volumetric SDF run (checkpoint, history, slice plot, metadata).
%   - runs/atlas_bunny_repaired_8chart/
%     Repaired 8-chart Bunny atlas used for the downstream continuation study.
%   - runs/bunny_poisson_8chart*
%     Family of downstream Bunny Poisson continuation runs that vary interior supervised pretraining.
%   - paraview/bunny_poisson_8chart_intpre20k/
%     Best archived continuation export package used for the Example 3 surface visualization.
% Notes:
% Bunny example moved to Appendix~\ref{sec:appendix_bunny}.

% ======================================================================
% EXAMPLE 3: Forward elastoplastic BVP on the torus (Schwarz FEM)
% Code location: experiments/torus_elastoplastic/run_forward_bvp_schwarz.py
% Figure scripts: manuscript/scripts_figures/example_forward_bvp.py
% ======================================================================
\subsection{Example 3: forward elastoplastic boundary-value problem on the torus}
\label{sec:example3_bvp}

This example solves a forward boundary-value problem for finite-strain $J_2$ elastoplasticity with kinematic hardening on the full torus, using the eight-chart atlas and multiplicative Schwarz coupling. Unlike the inverse benchmarks that follow, the displacement field is \emph{unknown} and must be determined from equilibrium: at each load step the Newton--Raphson iteration solves $\mathbf{R}(\mathbf{u})=\mathbf{f}_{\mathrm{int}}(\mathbf{u})-\mathbf{f}_{\mathrm{ext}}=\mathbf{0}$ on every chart, with Schwarz interface conditions transmitting displacement continuity across chart boundaries. This is the first example in the manuscript where nonlinear material response, incremental history dependence, mapped PDE operators, and multi-chart Schwarz coupling are all active simultaneously.

\subsubsection{Problem definition}

The torus geometry ($R=1.0$, $r=0.35$) is covered by eight azimuthal charts with $\Delta\phi=\pi/4$ half-width, as in the Neo-Hookean identification benchmark. The material is governed by the compressible $J_2$ elastoplastic model with kinematic hardening detailed in Appendix~\ref{sec:appendix_constitutive}: elastic moduli $E=200$, $\nu=0.3$ ($\mu=76.92$, $K=166.67$), yield stress $\tau_y=0.5$, and kinematic hardening modulus $H_{\mathrm{kin}}=20$, with the smooth softplus return mapping \eqref{eq:smooth_return} at $\varepsilon_s=0.01$.

Cyclic opposite displacements are prescribed on two cross-sections of the torus to squeeze the central hole:
\begin{equation}
u_x\big|_{\phi\approx 0} = +\delta(t),
\qquad
u_x\big|_{\phi\approx \pi} = -\delta(t),
\label{eq:bvp_bc}
\end{equation}
where $\delta(t)$ follows a triangular cyclic profile: $0\to+\delta_{\max}\to-\delta_{\max}\to0$ with $\delta_{\max}=0.02\,r$. All other torus-surface nodes are traction-free, and Schwarz interface displacement continuity is enforced on chart boundaries. Each chart uses a P1 tetrahedral mesh on a structured hexahedral grid with Freudenthal decomposition, and Newton--Raphson equilibrium is solved with an autograd-computed consistent tangent through the smooth return mapping.

At each load step, 8~multiplicative Schwarz sweeps are performed. Within each sweep, charts are visited by color groups from the overlap adjacency graph: non-overlapping charts may be updated concurrently, while overlapping neighbours exchange frozen displacement traces at their shared interface nodes. The per-chart Newton solver typically converges in 1--4 iterations with residual tolerance $10^{-7}$.

\subsubsection{Mesh refinement study}

Three mesh resolutions are tested with an identical cyclic loading protocol (30~steps per half-cycle, 120~total steps, 1~cycle):

\begin{table}[htbp]
\centering
\scriptsize
\setlength{\tabcolsep}{3pt}
\caption{Mesh refinement study for the 8-chart Schwarz forward elastoplastic BVP. All runs use the same cyclic loading protocol ($\delta_{\max}=0.02\,r$, 120~steps).}
\label{tab:bvp_mesh}
\begin{tabular}{cccccccc}
\toprule
$n_{\mathrm{cells}}$ & Nodes & Elements & DOF & Final $\bar{\varepsilon}_p$ & Interface jump & Time (s) & s/step \\
\midrule
4  & 1{,}000  & 3{,}072  & 3{,}000  & $3.18\times10^{-2}$ & $2.21\times10^{-7}$ & 189  & 1.6 \\
6  & 2{,}744  & 10{,}368 & 8{,}232  & $5.41\times10^{-2}$ & $4.83\times10^{-8}$ & 502  & 4.2 \\
8  & 5{,}832  & 24{,}576 & 17{,}496 & $7.83\times10^{-2}$ & $1.97\times10^{-7}$ & 1{,}394 & 11.6 \\
\bottomrule
\end{tabular}
\end{table}

Table~\ref{tab:bvp_mesh} shows that the solver converges at all three resolutions with Newton requiring only 1--4 iterations per chart per sweep. The accumulated plastic strain $\bar{\varepsilon}_p$ increases with mesh refinement because finer meshes resolve sharper strain gradients near the loaded cross-sections where plastic flow concentrates. The Schwarz interface displacement jump remains $O(10^{-7})$ or better throughout, confirming tight inter-chart coupling under nonlinear, history-dependent loading.

\subsubsection{Load-step sensitivity}

To verify temporal convergence, the $n_{\mathrm{cells}}=6$ mesh is run with $\text{steps per half-cycle}\in\{10,20,40\}$ (Table~\ref{tab:bvp_timestep}).

\begin{table}[htbp]
\centering
\scriptsize
\setlength{\tabcolsep}{3pt}
\caption{Load-step sensitivity for the forward BVP at $n_{\mathrm{cells}}=6$. The final accumulated plastic strain varies by $0.16\%$ across a $4\times$ range of step sizes.}
\label{tab:bvp_timestep}
\begin{tabular}{ccccc}
\toprule
Steps/half & Total steps & Final $\bar{\varepsilon}_p$ & Interface jump & Time (s) \\
\midrule
10 & 40  & $5.407\times10^{-2}$ & $1.40\times10^{-7}$ & 171 \\
20 & 80  & $5.413\times10^{-2}$ & $7.20\times10^{-8}$ & 330 \\
40 & 160 & $5.416\times10^{-2}$ & $3.62\times10^{-8}$ & 657 \\
\bottomrule
\end{tabular}
\end{table}

The final accumulated plastic strain varies by only $0.16\%$ across the tested range, confirming that the smooth return mapping and incremental load-stepping produce temporally converged results. The interface jump halves with each step-size refinement, consistent with first-order splitting in the multiplicative Schwarz scheme.

\begin{figure}[htbp]
\centering
\includegraphics[width=0.98\textwidth]{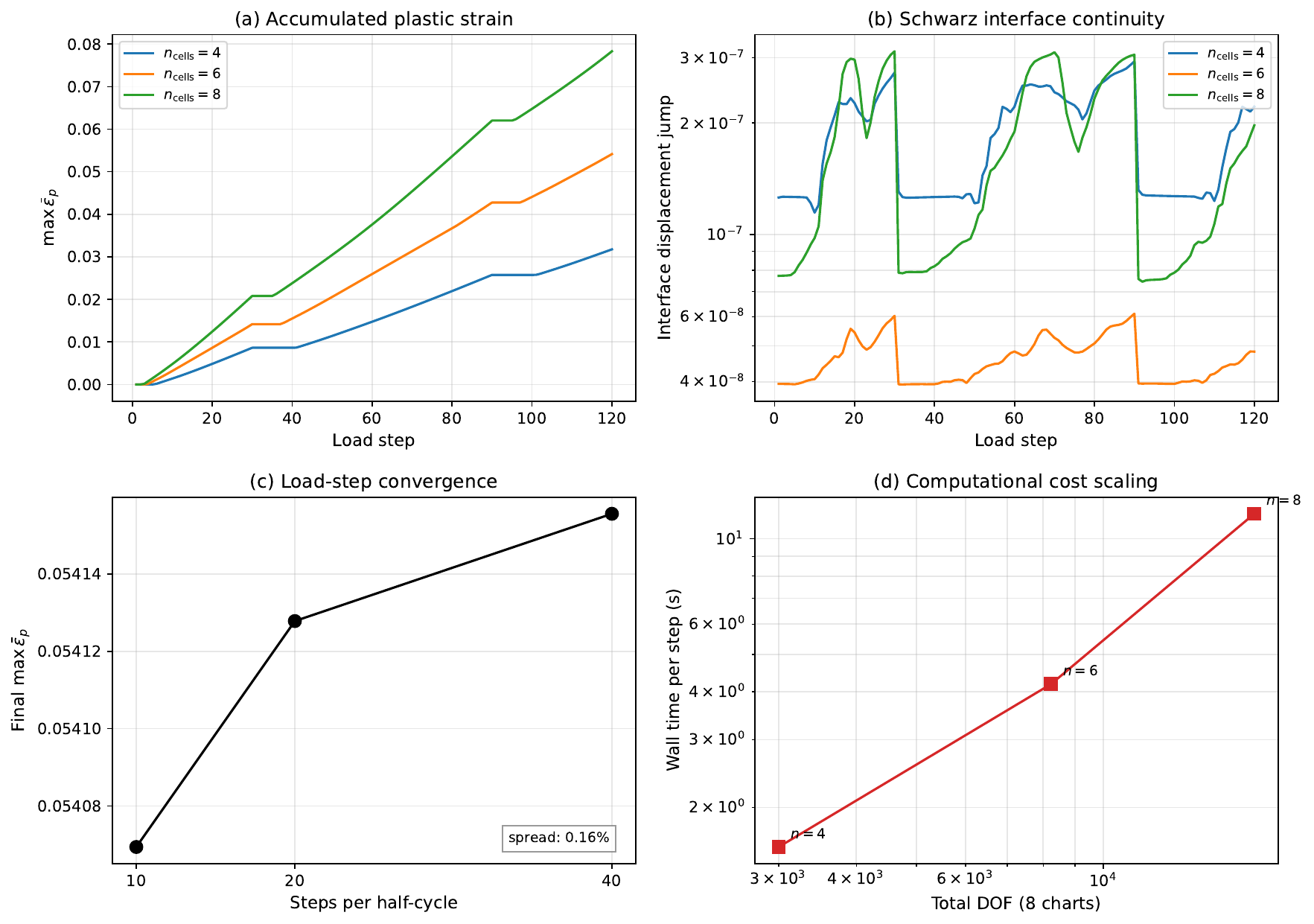}
\caption{Diagnostics for the 8-chart Schwarz forward elastoplastic BVP. (a)~Accumulated plastic strain $\bar{\varepsilon}_p$ vs.\ load step for three mesh resolutions; finer meshes resolve larger peak strains near the loaded sections. (b)~Schwarz interface displacement jump remains $O(10^{-7})$ throughout. (c)~Load-step convergence: the final plastic strain varies by $0.16\%$ across a $4\times$ range of step sizes. (d)~Computational cost scaling with total DOF.}
\label{fig:bvp_diagnostics}
\end{figure}

\begin{figure}[htbp]
\centering
\includegraphics[width=0.98\textwidth]{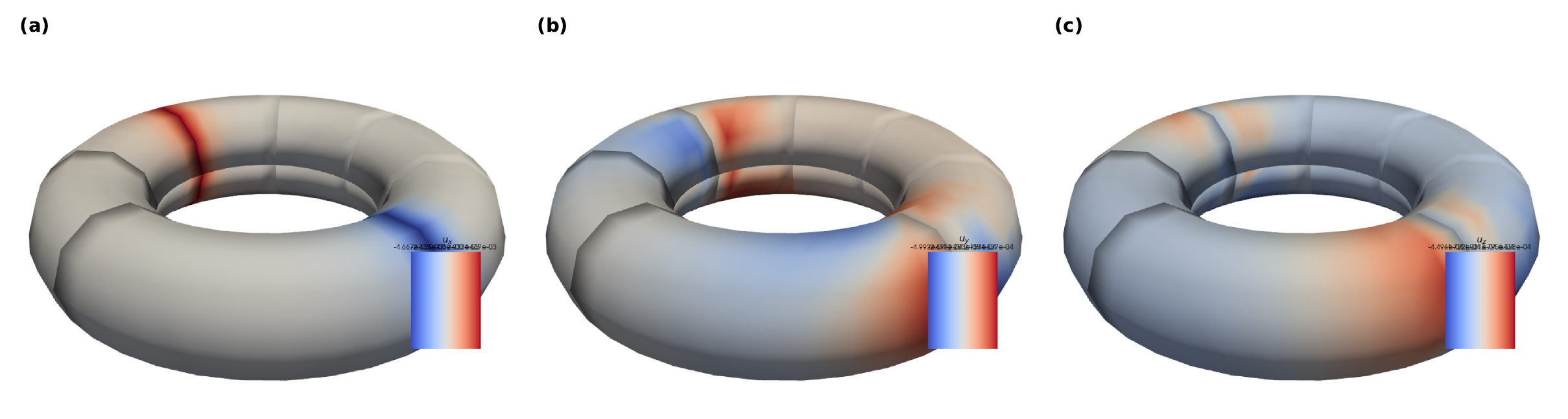}
\caption{Displacement field components on the torus at step~100 of the cyclic loading protocol ($n_{\mathrm{cells}}=10$, $\delta\approx-\tfrac{2}{3}\delta_{\max}$, recovery phase). (a)~$u_x$, (b)~$u_y$, (c)~$u_z$. The opposite $x$-displacement boundary conditions at $\phi\approx0$ and $\phi\approx\pi$ produce the dominant $u_x$ pattern, while $u_y$ and $u_z$ arise from the curved torus geometry coupling in-plane compression to out-of-plane deformation. All fields are the direct P1 finite-element nodal solution with no postprocessing interpolation.}
\label{fig:bvp_displacement_fields}
\end{figure}

Figure~\ref{fig:bvp_displacement_fields} shows the displacement field on a finer mesh ($n_{\mathrm{cells}}=10$, 1{,}331~nodes per chart, 6{,}000~elements per chart) at load step~100, which falls in the recovery phase of the cycle ($\delta\approx-\tfrac{2}{3}\delta_{\max}$). By this point the torus has completed the full loading--unloading--reverse-loading sequence and is partially recovering toward the undeformed configuration, so the displacement field reflects the combined effects of elastic recovery and accumulated plastic deformation with kinematic hardening. The finer mesh is used here for visualization quality; the quantitative convergence metrics are reported at $n_{\mathrm{cells}}\leq 8$ in Tables~\ref{tab:bvp_mesh}--\ref{tab:bvp_timestep}.

The forward BVP benchmark shows that the atlas-based Schwarz infrastructure extends from scalar diffusion (Example~2) and hyperelastic identification (Example~4) to nonlinear equilibrium with history-dependent plasticity. The Newton--Raphson solver converges robustly across all mesh sizes and loading steps, the interface displacement jump remains $O(10^{-7})$, and the temporal convergence study confirms that the smooth return mapping produces consistent plastic strain accumulation independent of step size.

% ======================================================================
% EXAMPLE 4: Torus inverse Neo-Hookean identification
% ======================================================================
% Notes:
%   - The torus geometry, prescribed kinematics, and synthetic traction observations are generated analytically inside the code.
%   - There is therefore no external raw-data directory; the run directories contain the generated observations,
%     learned parameters, metric logs, and VTU exports.
% Code location (global optimizer):  manuscript_experiments/example3_torus_inverse_original/  (dir name predates example renumbering)
% Code location (Schwarz dual):      manuscript_experiments/example4_torus_inverse_schwarz_dual/
% Figure scripts: manuscript/scripts_figures/example3_torus_inverse_original.py  (dir name predates example renumbering)
%                 manuscript/scripts_figures/example4_torus_inverse_schwarz_dual.py
\subsection{Example 4: torus inverse Neo-Hookean identification}
\label{sec:example4}

This example demonstrates constitutive parameter identification on a three-dimensional torus ($R=1.0$, $r=0.35$) under prescribed torsional kinematics. Unlike the rabbit and ellipsoid, the torus is \emph{not simply connected}: its fundamental group $\pi_1(\mathbb{T}^2)\cong\mathbb{Z}\times\mathbb{Z}$ prevents any single homeomorphism from a reference ball $B^3$ onto the toroidal solid. A multi-chart atlas is therefore topologically necessary, not merely algorithmically convenient. The domain is covered by eight azimuthal charts with 20\% overlap, each mapping a reference cube to a sector of the torus. The unknowns are the global shear modulus $\mu$ and bulk modulus $K$ of a compressible Neo-Hookean material. Three solver configurations are compared: (a)~a single-optimizer global solve, (b)~multiplicative Schwarz with per-chart parameters and traction observations, and (c)~multiplicative Schwarz with per-chart parameters and displacement observations.

\subsubsection{Problem formulation}

Let
\begin{equation}
\mathbf{u}_{\tau}^{\mathrm{pres}}:\Omega_{\mathrm{torus}}\rightarrow \mathbb{R}^3
\label{eq:torus_prescribed_u}
\end{equation}
be the prescribed torsional displacement field. In physical coordinates $x=(x_1,x_2,x_3)$, the displacement takes the form
\begin{equation}
\mathbf{u}_{\tau}^{\mathrm{pres}}(x) = \tau\,w(\phi)\,(1+\alpha\, x_3/r)\,(-x_2,\;x_1,\;0)^T,
\label{eq:torus_torsion_formula}
\end{equation}
where $\phi=\arctan(x_2/x_1)$ is the azimuthal angle, $\tau=0.085$ is the torsion amplitude, $\alpha=0.75$ is a through-thickness scaling factor, and $w(\phi)=\tfrac{1}{2}[1+\cos(\pi\,(\phi-\phi_c)/\phi_w)]$ is a smooth cosine window centred at $\phi_c=\pi/2$ with half-width $\phi_w=0.35\pi$ (${\approx}63^{\circ}$). The window localizes the torsion to one part of the torus; outside the window the displacement vanishes.
The constitutive law is the compressible Neo-Hookean first Piola--Kirchhoff stress
\begin{equation}
\mathbf{P}(\mathbf{F};\mu,K)
=
\mu\left(\mathbf{F}-\mathbf{F}^{-T}\right)
+
K\ln(\det\mathbf{F})\,\mathbf{F}^{-T}.
\label{eq:nh_stress}
\end{equation}
Synthetic boundary observations are generated on a loaded surface subset $\Gamma_{\mathrm{obs}}$ spanning a ${\sim}126^{\circ}$ azimuthal arc, with $N_{\mathrm{obs}}=6{,}000$ observation points. For traction-mode experiments the observed quantity is
\begin{equation}
\mathbf{t}^{\mathrm{obs}}(x)
=
\mathbf{P}\!\left(\mathbf{F}_{\tau}^{\mathrm{pres}}(x);\mu_{\mathrm{true}},K_{\mathrm{true}}\right)\mathbf{n}(x),
\qquad x\in \Gamma_{\mathrm{obs}},
\label{eq:torus_obs_traction}
\end{equation}
with $(\mu_{\mathrm{true}},K_{\mathrm{true}})=(1.8,25.0)$. Because the deformation gradient is prescribed rather than learned, the benchmark isolates the constitutive identification from the forward equilibrium solve.

\subsubsection{Solver configuration (a): global atlas optimizer}

In the global configuration, a single pair $(\mu,K)$ is optimized by minimizing the chart-weighted traction mismatch
\begin{equation}
\mathcal{L}^{(4)}_{\mathrm{trac}}(\mu,K)
=
\frac{1}{M}\sum_{i=1}^{M}
\frac{\sum_{k=1}^{N_{\mathrm{obs}}}\omega_i(x_k)\,
\left\|
\mathbf{t}(x_k;\mu,K)-\mathbf{t}^{\mathrm{obs}}(x_k)
\right\|_2^2}
{\sum_{k=1}^{N_{\mathrm{obs}}}\omega_i(x_k)+\varepsilon},
\label{eq:torus_original_loss}
\end{equation}
where $\omega_i$ are fixed azimuthal partition-of-unity weights from the eight-chart atlas. A determinant barrier $C_{\det}=\mathbb{E}[\mathrm{softplus}(\delta-\det\mathbf{F}_{\tau}^{\mathrm{pres}})^2]$ is included for kinematic admissibility monitoring but is constant with respect to $(\mu,K)$ since the kinematics are prescribed:
\begin{equation}
\mathcal{L}^{(4)}_{\mathrm{tot}}(\mu,K)
=
\mathcal{L}^{(4)}_{\mathrm{trac}}(\mu,K)
+
\lambda_{\det}\,C_{\det}.
\label{eq:torus_original_total_loss}
\end{equation}

Starting from $(\mu_0,K_0)=(1.746,26.0)$, the canonical run converges to $(\mu,K)=(1.800000,25.000006)$ after 300 epochs (201\,s), with final relative errors of $9.27\times 10^{-6}\%$ for $\mu$ and $2.29\times 10^{-5}\%$ for $K$, and a traction relative $L^2$ error of $2.35\times 10^{-7}$. The per-chart evaluation MSE values are uniformly below $1.3\times10^{-13}$, confirming that the chart-weighted aggregation introduces no observable inconsistency.

% Code location: manuscript_experiments/example4_torus_inverse_schwarz_dual/
% Figure scripts: manuscript/scripts_figures/example4_torus_inverse_schwarz_dual.py
\subsubsection{Solver configurations (b)--(c): multiplicative Schwarz with per-chart parameters}

The Schwarz variant replaces the single $(\mu,K)$ pair with per-chart unknowns $\{(\mu_i,K_i)\}_{i=1}^{8}$, updated through red--black alternating sweeps (even charts, then odd charts). Each chart~$i$ minimizes a local loss comprising three terms: (i)~a chart-weighted data-mismatch term against observations within chart~$i$'s support, (ii)~an interface consensus penalty
\begin{equation}
\mathcal{L}_{\mathrm{if},i}
=
\left(\frac{\mu_i - \bar{\mu}_{\mathrm{nb}(i)}}{\mu_{\mathrm{true}}}\right)^2
+
\left(\frac{K_i - \bar{K}_{\mathrm{nb}(i)}}{K_{\mathrm{true}}}\right)^2,
\label{eq:torus_schwarz_interface}
\end{equation}
where $\bar{\mu}_{\mathrm{nb}(i)}$ and $\bar{K}_{\mathrm{nb}(i)}$ are the frozen neighbor-chart parameter averages from the previous Schwarz sweep, and (iii)~a weak regularization toward prior estimates. Two inverse data modes are tested:

\begin{itemize}
\item \textbf{Traction mode (b)}: observations are the synthetic traction vectors from \eqref{eq:torus_obs_traction}. After 300 epochs (76.9\,s), the chart-averaged parameters converge to $\bar{\mu}=1.7938\pm 9.93\times10^{-4}$ and $\bar{K}=25.14\pm 2.80\times10^{-2}$, with mean relative errors of $0.35\%$ for $\mu$ and $0.54\%$ for $K$. The blended traction relative $L^2$ error is $5.80\times10^{-3}$.

\item \textbf{Displacement mode (c)}: observations are surrogate displacement values derived from the prescribed kinematics. After 405 epochs (38.6\,s), the chart-averaged parameters converge to $\bar{\mu}=1.8000\pm 1.47\times10^{-13}$ and $\bar{K}=25.25\pm 1.53\times10^{-2}$, with $\mu$ recovered to machine precision across all eight charts ($\mu$ relative error $< 10^{-11}\%$) and $K$ relative error of $1.0\%$. The displacement relative $L^2$ error is $1.74\times10^{-3}$.
\end{itemize}

\begin{table}[htbp]
\centering
\scriptsize
\setlength{\tabcolsep}{3pt}
\caption{Torus inverse identification: comparison of three solver configurations on the same eight-chart atlas. All runs use the same prescribed torsional kinematics and true parameters $(\mu_{\mathrm{true}},K_{\mathrm{true}})=(1.8,25.0)$. The Schwarz variants report mean $\pm$ standard deviation across eight per-chart parameter estimates.}
\label{tab:torus_comparison}
\begin{tabular}{lccccc}
\toprule
Configuration & $\mu$ error (\%) & $K$ error (\%) & Obs.\ rel-$L^2$ & Chart $\mu$ std & Time (s) \\
\midrule
(a) Global optimizer & $9.27\times10^{-6}$ & $2.29\times10^{-5}$ & $2.35\times10^{-7}$ & --- & 201 \\
(b) Schwarz traction & $3.45\times10^{-1}$ & $5.41\times10^{-1}$ & $5.80\times10^{-3}$ & $9.93\times10^{-4}$ & 76.9 \\
(c) Schwarz displacement & $4.08\times10^{-12}$ & $1.00\times10^{0\phantom{-}}$ & $1.74\times10^{-3}$ & $1.47\times10^{-13}$ & 38.6 \\
\bottomrule
\end{tabular}
\end{table}

Table~\ref{tab:torus_comparison} reveals two complementary findings. First, the global optimizer (a) reaches near-exact parameter recovery because all $6{,}000$ observation points contribute to a single well-conditioned traction-matching objective. Second, the Schwarz configurations (b) and (c) demonstrate that the eight-chart atlas can support \emph{distributed} parameter identification, with each chart maintaining its own material constants and achieving inter-chart consensus through the interface penalty \eqref{eq:torus_schwarz_interface}. In displacement mode~(c), the chart-to-chart standard deviation of $\mu$ is $1.47\times10^{-13}$---effectively machine precision---confirming that the Schwarz coupling drives all eight charts to identical parameter values despite independent optimization. The higher $K$ error in mode~(c) reflects a well-known challenge in displacement-based inverse identification: the surrogate displacement field has weaker sensitivity to volumetric stiffness than traction data, making bulk modulus recovery inherently harder from kinematic observations alone.

\begin{figure}[htbp]
\centering
\includegraphics[width=0.98\textwidth]{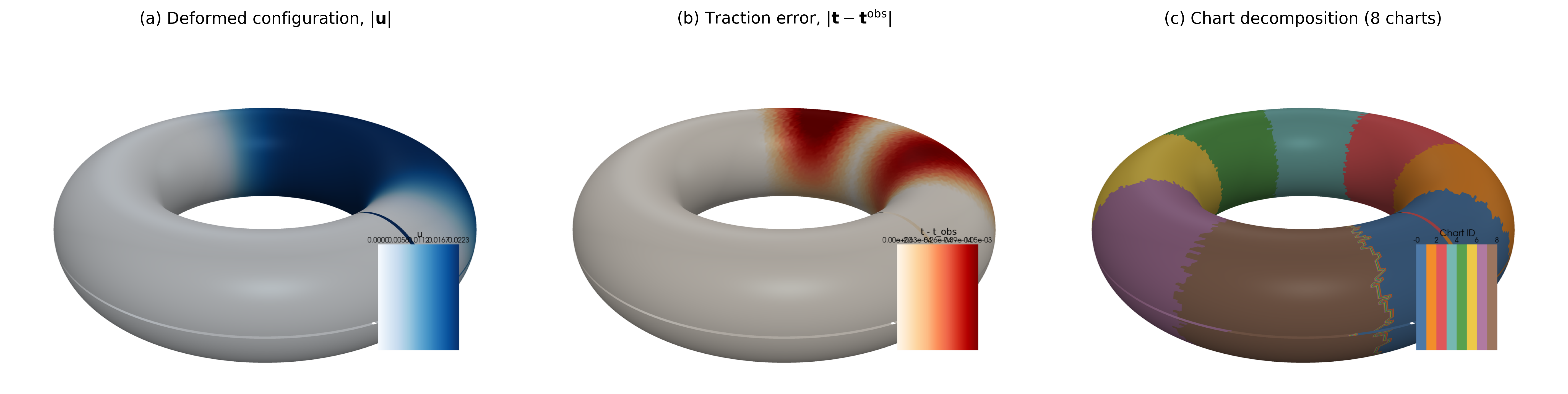}
\caption{Torus inverse benchmark (global optimizer), rendered from VTU exports on the three-dimensional surface. (a)~Prescribed deformed configuration colored by displacement magnitude~$|\mathbf{u}|$; the torsional load is localized over a ${\sim}126^{\circ}$ azimuthal arc. (b)~Traction-error magnitude $|\mathbf{t}-\mathbf{t}^{\mathrm{obs}}|$; the mismatch is concentrated in the loaded region and remains below $1.2\times10^{-3}$ everywhere. (c)~Eight-chart atlas decomposition of the torus, with each color denoting a distinct azimuthal chart; the multi-chart atlas is topologically necessary because the solid torus is not simply connected.}
\label{fig:torus_inverse_original_fields}
\end{figure}

\begin{figure}[htbp]
\centering
\includegraphics[width=0.62\textwidth]{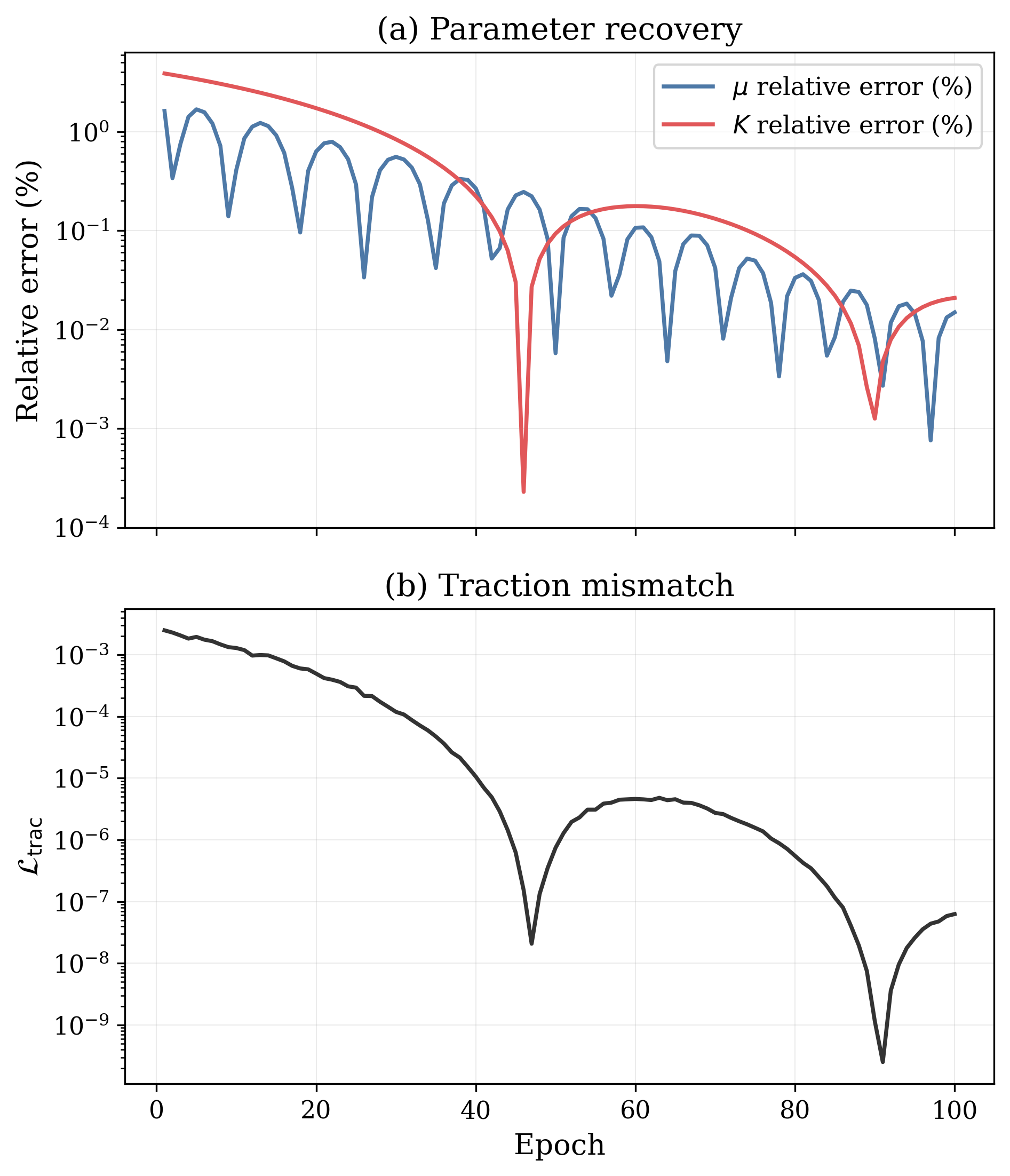}
\caption{Parameter convergence history for the global torus optimizer. (a)~Relative errors of~$\mu$ and~$K$ decrease monotonically over 100 epochs, reaching ${\sim}10^{-2}\%$ for both parameters. (b)~Traction mismatch~$\mathcal{L}_{\mathrm{trac}}$ decreases by five orders of magnitude from initialization to the $10^{-8}$ level. The oscillatory pattern in~(a) reflects the alternating descent directions in the two-parameter landscape.}
\label{fig:torus_inverse_original_convergence}
\end{figure}

\begin{figure}[htbp]
\centering
\includegraphics[width=0.98\textwidth]{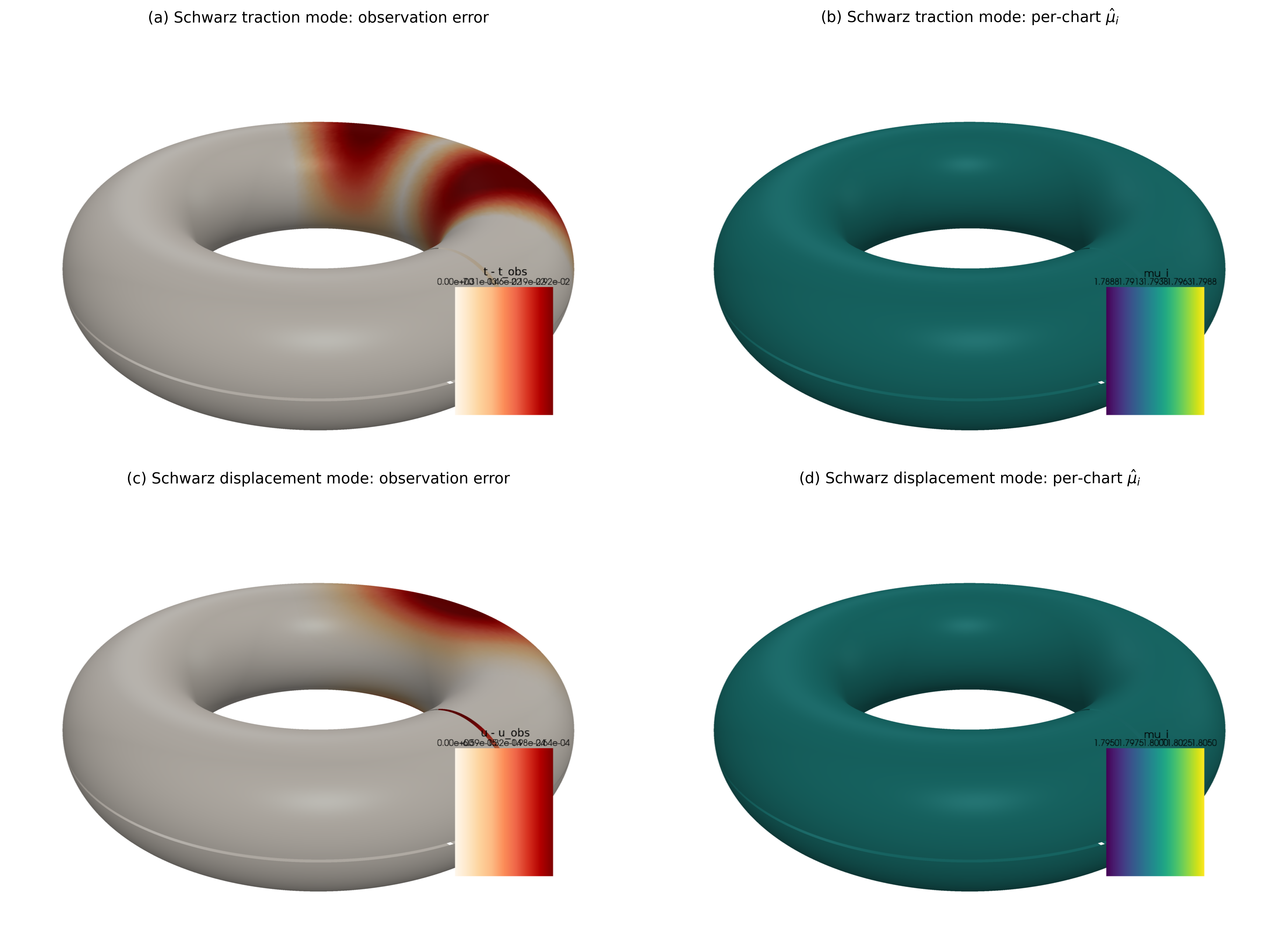}
\caption{Schwarz-based distributed parameter identification on the torus, rendered from VTU exports. Top row: traction-mode Schwarz. Bottom row: displacement-mode Schwarz. (a,\,c)~Observation-error magnitude on the torus surface; the traction-mode error is roughly $10\times$ larger than the displacement-mode error. (b,\,d)~Per-chart shear-modulus estimate~$\hat{\mu}_i$ mapped onto the torus surface; the color range is $[1.79,\,1.81]$ centered on $\mu_{\mathrm{true}}=1.8$. In displacement mode~(d), the surface is nearly uniform, consistent with the machine-precision inter-chart consensus ($\mu$ std $= 1.47\times10^{-13}$) reported in Table~\ref{tab:torus_comparison}.}
\label{fig:torus_schwarz_3d}
\end{figure}

The torus benchmark establishes three points relevant to the paper's thesis. First, the atlas-based constitutive operator is not limited to Poisson diffusion: the Neo-Hookean stress evaluation and its derivatives are correctly mediated through the chart decomposition (Figures~\ref{fig:torus_inverse_original_fields}a--b). Second, the multiplicative Schwarz framework extends naturally from field coupling (Example~2) to parameter-level consensus enforcement across charts, as evidenced by the near-uniform $\hat{\mu}_i$ field in Figure~\ref{fig:torus_schwarz_3d}d. In displacement mode, the chart-averaged $\mu$ converges to the true value within ${\sim}50$ epochs and the inter-chart standard deviation collapses to machine precision; in traction mode, $\mu$ oscillates more strongly before settling near $1.794$, reflecting the richer but noisier gradient signal from traction observations. Both modes drive $K$ monotonically downward from $26.0$ toward the true value $25.0$, though traction data reaches closer ($25.14$ vs.\ $25.25$) within the allotted epochs. Third, the per-chart standard deviation provides a built-in consistency diagnostic: large inter-chart parameter disagreement would signal atlas or observation pathologies before any external validation is needed.

\subsection{Example 5: torus elastoplastic parameter identification}
\label{sec:example5_ep}

This example extends the torus inverse benchmark from elastic to \emph{history-dependent elastoplastic} constitutive parameter estimation. While Example~4 recovers constant elastic moduli $(\mu,K)$ from a single prescribed deformation, Example~5 introduces a two-stage problem in which the yield stress $\tau_y$ and kinematic hardening modulus $H_{\mathrm{kin}}$ are recovered from incremental loading histories that include cyclic plastic deformation. The key methodological contribution is a \emph{smooth return mapping} that replaces the non-differentiable yield-surface check with a softplus regularisation, enabling end-to-end gradient-based optimization through the full elastoplastic load-stepping solver. The forward solver uses the same chart-local P1 finite-element infrastructure validated in Example~3 (Section~\ref{sec:example3_bvp}), operating in a single reference chart of the torus atlas; this demonstrates that the Piola-mapped operator framework extends naturally from forward equilibrium to gradient-based inverse identification.

\subsubsection{Constitutive model and solver setup}
The material is governed by finite-strain $J_2$ elastoplasticity with kinematic hardening in the multiplicative framework of \citet{simo_computational_1998}. The constitutive equations---multiplicative decomposition $\mathbf{F}=\mathbf{F}^e\mathbf{F}^p$ \eqref{eq:ep_multiplicative}, logarithmic-strain Kirchhoff stress \eqref{eq:kirchhoff_stress}, von Mises yield function with back-stress \eqref{eq:von_mises_yield}, and the smooth softplus return mapping \eqref{eq:smooth_return}---are detailed in Appendix~\ref{sec:appendix_constitutive}. The smooth return mapping replaces the non-differentiable yield-surface check with a $C^\infty$ softplus regularization, enabling end-to-end gradient-based optimization through the full elastoplastic load-stepping solver. At sharpness $\varepsilon_s=0.001\,\tau_y$ the smooth response is visually indistinguishable from the classical radial return (Figure~\ref{fig:ep_constitutive}a). A comparison with central finite differences confirms that the autograd gradient through the smooth return mapping is correct (both methods recover $\tau_y$ to within $0.25\%$) while providing a $2.9\times$ wall-clock speedup. With the classical hard-max return mapping, autograd gradients are zero in the plastic regime and the optimizer stalls.

The forward solver is a chart-local P1 FEM (Algorithm~\ref{alg:atlas_fem_schwarz}) on a structured hexahedral grid with Freudenthal six-tet decomposition ($4^3 = 64$ hexahedra, 384~elements, 125~nodes), operating on a single torus chart sector ($R=1.0$, $r=0.35$, azimuthal half-width $\Delta\phi=\pi/4$). In chart-local reference coordinates $\boldsymbol{\zeta}\in[-1,1]^3$, the boundary conditions are:
\begin{equation}
\mathbf{u}\big|_{\zeta_1=-1}=\mathbf{0} \quad\text{(clamped)},
\qquad
u_1\big|_{\zeta_1=+1}=\varepsilon_{\mathrm{app}}\cdot 2r \quad\text{(prescribed extension)},
\label{eq:ep_bc}
\end{equation}
where $r$ is the chart support radius and $\varepsilon_{\mathrm{app}}$ is the applied nominal strain, incremented over the loading history. All other boundary faces are traction-free. At each load step, Newton--Raphson iteration solves the chart-local residual \eqref{eq:fem_residual} with an autograd-computed consistent tangent through the smooth return mapping (Remark~\ref{rem:tangent_modularity}). Gradients with respect to material parameters $\theta=(\tau_y, H_{\mathrm{kin}})$ are obtained via the implicit function theorem.

\subsubsection{Two-stage inverse identification}

\begin{figure}[htbp]
\centering
\includegraphics[width=0.98\textwidth]{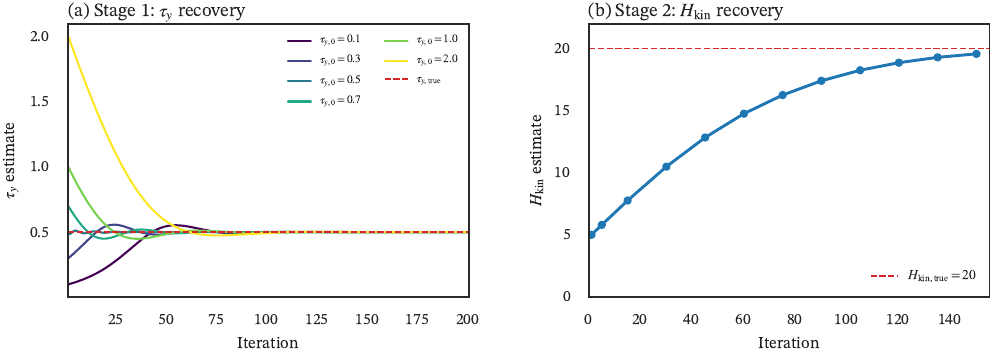}
\caption{Inverse parameter recovery convergence. (a)~Stage~1: $\tau_y$ trajectories from six initial guesses; all converge to $\tau_{y,\mathrm{true}}=0.5$ within 100 iterations. (b)~Stage~2: $H_{\mathrm{kin}}$ recovery from $H_{\mathrm{kin},0}=5$ using reaction-force observations; the estimate reaches $19.58$ ($2.11\%$ error) after 150 iterations.}
\label{fig:ep_convergence}
\end{figure}

\paragraph{Stage~1: yield stress.}
The elastic moduli are assumed known ($\mu=76.92$, $K=166.67$, corresponding to $E=200$, $\nu=0.3$) and kinematic hardening is inactive ($H_{\mathrm{kin}}=0$). A monotonic uniaxial tension test is applied: $\varepsilon_{\mathrm{app}}$ increases linearly from $0$ to $0.08$ in 5~load steps. The unknown yield stress $\tau_y$ is identified from the relative displacement mismatch at $n_s=20$ interior sensor nodes:
\begin{equation}
\mathcal{L}_{\mathrm{disp}} = \sum_{t=1}^{N_{\mathrm{step}}} \frac{\left\|\mathbf{u}^{(t)}_{\mathrm{pred}} - \mathbf{u}^{(t)}_{\mathrm{obs}}\right\|^2}{\left\|\mathbf{u}^{(t)}_{\mathrm{obs}}\right\|^2}.
\label{eq:disp_loss}
\end{equation}
Starting from $\tau_{y,0}=1.0$ (100\% above the true value $\tau_{y,\mathrm{true}}=0.5$), 100 Adam iterations recover $\tau_y=0.4988$, a \textbf{0.25\% relative error}. The optimizer converges from initial guesses spanning $\tau_{y,0}\in\{0.1,\ldots,2.0\}$ (Figure~\ref{fig:ep_convergence}a), indicating a wide basin of attraction.

\paragraph{Stage~2: kinematic hardening.}
With $\tau_y$ fixed at its Stage~1 estimate, the unknown is $H_{\mathrm{kin}}$. Two full symmetric load--unload--reverse cycles are applied: $\varepsilon_{\mathrm{app}}$ ramps from $0\to+\varepsilon_{\mathrm{peak}}$, reverses to $-\varepsilon_{\mathrm{peak}}$, and returns to $0$, with $\varepsilon_{\mathrm{peak}}=0.03$ and 15~steps per half-cycle (120~total load steps). The observations are \emph{reaction forces} at the loaded face $\xi_1=+1$ rather than interior displacements, because the back-stress $\boldsymbol{\beta}$ shifts the yield surface in stress space and interior displacements are insensitive to this shift under prescribed kinematics. The objective is
\begin{equation}
\mathcal{L}_{\mathrm{RF}} = \frac{1}{|\mathcal{S}|}\sum_{t\in\mathcal{S}} \left(R_{x,\mathrm{pred}}^{(t)} - R_{x,\mathrm{obs}}^{(t)}\right)^2,
\label{eq:rf_loss}
\end{equation}
where $R_x^{(t)}=\sum_{a\in\Gamma_{\mathrm{load}}} f_{\mathrm{int},x}^{(a)}$ is the total $x$-reaction force at load step~$t$. Starting from $H_{\mathrm{kin},0}=5.0$ (75\% below $H_{\mathrm{kin,true}}=20.0$), 150 Adam iterations recover $H_{\mathrm{kin}}=19.58$, a \textbf{2.11\% relative error} (Figure~\ref{fig:ep_convergence}b).

\begin{table}[htbp]
\centering
\scriptsize
\setlength{\tabcolsep}{2.5pt}
\caption{Two-stage elastoplastic inverse identification results. Stage~1 identifies $\tau_y$ from monotonic displacement data; Stage~2 identifies $H_{\mathrm{kin}}$ from cyclic reaction-force data. Material: $E=200$, $\nu=0.3$. Mesh: $4^3$ hexahedra (384~P1 tetrahedra, 125~nodes).}
\label{tab:ep_results}
\begin{tabular}{lcccccc}
\toprule
Stage & Unknown & Observation type & Loading & True & Identified & Error (\%) \\
\midrule
1 & $\tau_y$ & interior displacements & monotonic, 5 steps & 0.500 & 0.4988 & 0.25 \\
2 & $H_{\mathrm{kin}}$ & boundary reaction forces & cyclic, 120 steps & 20.0 & 19.58 & 2.11 \\
\bottomrule
\end{tabular}
\end{table}

\subsubsection{Sensitivity analysis}
\begin{figure}[htbp]
\centering
\includegraphics[width=0.98\textwidth]{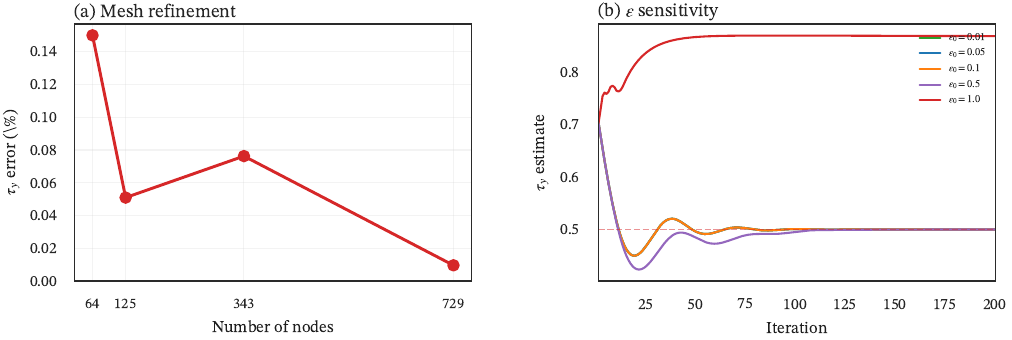}
\caption{Sensitivity studies. (a)~$\tau_y$ identification error versus mesh resolution; accuracy is consistent across all mesh sizes ($n_{\mathrm{cells}}\in\{3,4,6,8\}$). (b)~$\tau_y$ convergence trajectories for different initial smoothing parameters $\varepsilon_{s,\mathrm{start}}$; all values except $\varepsilon_{s,\mathrm{start}}=1.0$ converge to the correct yield stress.}
\label{fig:ep_sensitivity}
\end{figure}

The robustness of the inverse procedure is assessed through three studies (Figure~\ref{fig:ep_sensitivity}). \emph{Initial-guess sensitivity}: $\tau_y$ is recovered from initial guesses spanning an order of magnitude ($0.1$--$2.0$), all converging to within $0.1\%$ of the true value. \emph{Mesh refinement}: the inverse is solved on meshes with $n_{\mathrm{cells}}\in\{3,4,6,8\}$ (64--729~nodes); the $\tau_y$ error is consistent across all mesh sizes. \emph{Smoothing parameter sensitivity}: annealing schedules with $\varepsilon_{s,\mathrm{start}}\in\{0.01,0.05,0.1,0.5\}$ all converge to the correct $\tau_y$; only $\varepsilon_{s,\mathrm{start}}=1.0$ fails, producing a biased estimate because the overly smooth yield surface does not develop sufficient plastic deformation to generate a meaningful gradient signal. Additive Gaussian noise at levels up to $5\%$ yields $\tau_y$ errors below $5\%$.

\begin{figure}[htbp]
\centering
\includegraphics[width=0.98\textwidth]{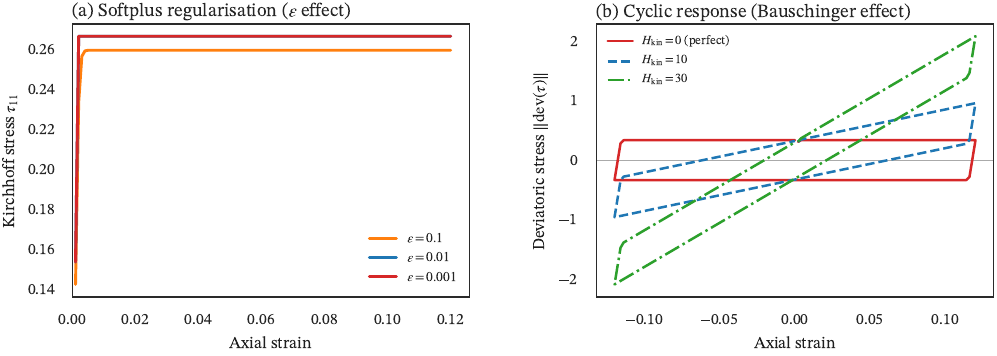}
\caption{Elastoplastic constitutive response. (a)~Uniaxial stress--strain curves for three softplus sharpness values $\varepsilon_s$; at $\varepsilon_s=0.001$, the smooth response is indistinguishable from the sharp classical yield point ($\tau_y=0.5$). (b)~Cyclic stress--strain loops for three kinematic hardening moduli: $H_{\mathrm{kin}}>0$ shifts the reverse yield point (Bauschinger effect), enabling $H_{\mathrm{kin}}$ identification from cyclic data.}
\label{fig:ep_constitutive}
\end{figure}

\begin{figure}[htbp]
\centering
\includegraphics[width=0.98\textwidth]{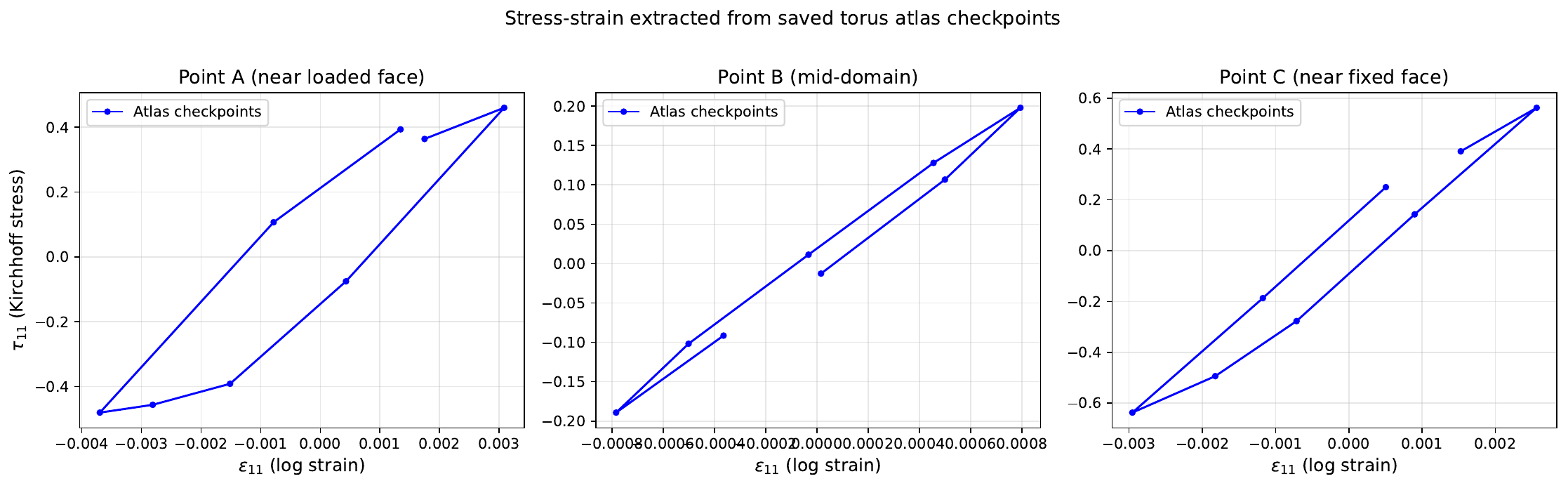}
\caption{Recovered versus true cyclic stress--strain response at three sample locations on a single torus chart sector. At all three locations the hysteresis loops generated by the true parameters (solid) and the identified parameters (dashed) are visually indistinguishable.}
\label{fig:ep_hysteresis_multipoint}
\end{figure}

The elastoplastic benchmark establishes two points. First, the chart-local P1 finite-element infrastructure extends without modification to nonlinear, history-dependent constitutive laws: the smooth return mapping \eqref{eq:smooth_return}, the autograd-computed consistent tangent, and the implicit-function-theorem sensitivity transfer all operate in the same reference-chart coordinates used for diffusion, confirming that the atlas is an operator-agnostic geometric substrate. Second, the two-stage decoupling exploits physically distinct sensitivity profiles: $\tau_y$ is identifiable from monotonic displacements because it controls the onset of permanent deformation, while $H_{\mathrm{kin}}$ is identifiable from cyclic reaction forces because it governs the stress-space shift of the yield surface during load reversal.

\begin{remark} (Single-chart usage of inverse problem) Example~5 operates on a single torus chart sector rather than the full eight-chart atlas used in Example~3. This scope restriction is deliberate: the purpose of the elastoplastic \emph{inverse} example is to verify that gradient-based parameter identification extends from hyperelastic to history-dependent constitutive laws, not to test multi-chart Schwarz coupling. Note that multi-chart Schwarz coupling under elastoplasticity for the \emph{forward} problem is already demonstrated in Example~3 (Section~\ref{sec:example3_bvp}), where displacement continuity at chart interfaces suffices. Extending the multi-chart Schwarz \emph{inverse} framework to elastoplasticity would additionally require transmitting internal-variable fields (back-stress, accumulated plastic strain) across chart interfaces for consistent parameter sensitivity---a nontrivial extension that is left to future work.
\end{remark}

\section{Discussion}
\label{sec:discussion}
A number of key observations on the performance of the proposed methods and the relations between the proposed method 
and other alternative workflows are both reported in this section. 

\subsection{Comparison with alternative workflows}
In conventional finite-element workflows a difficult three-dimensional body is first turned into a volumetric mesh and then coupled to mature local basis functions and linear or nonlinear solvers \citep{de_souza_neto_computational_2011}. The present atlas-based approach replaces the global volumetric meshing step with overlapping chart maps and level-set geometry, and the rabbit FEM baseline now shows concretely that classical FEM can operate \emph{within} this atlas framework with full theoretical convergence: $O(h^2)$ rates consistent with theoretical expectations (from $1.45$ to $2.10$ across refinement levels). This makes the atlas a general-purpose tool for geometric representations used in PDE solvers. 

In particular, three specific findings from the rabbit head-to-head merit discussion. First, at matched accuracy the two methods are broadly competitive in wall-clock time ($2{,}939$\,s for FEM at $n=48$ versus $3{,}659$\,s for the compact PINN), indicating that the atlas-plus-Schwarz infrastructure, not the local discretization, is the dominant cost driver. Second, the FEM requires roughly $8.5\times$ fewer neural-network evaluations than the PINN (${\sim}23$M cached versus ${\sim}195$M iterative), because the decoder networks are queried only during one-time geometric precomputation rather than at every training step. Third, the FEM provides a mesh-refinement convergence guarantee that the PINN does not, while the PINN avoids reference-domain mesh generation, stiffness-matrix assembly, and decoder inversion, and extends more naturally to nonlinear operators where matrix reassembly would be required at every Newton step.

The closest PINN alternatives in the literature, such as XPINNs and FBPINNs \citep{shukla2021parallel,moseley2021fbpinns}, also use decomposition to improve training, yet their partitions are primarily algorithmic. In contrast, the present atlas is geometry-first: the same chart structure is used to represent the domain, to generate collocation or quadrature, and to organize Schwarz communication. The rabbit benchmark and its FEM counterpart are the clearest evidence that this idea works in practice, because the same quality gates, the same Schwarz iteration logic, and the same partition-of-unity blending support both neural and classical local solvers on atlases with different chart counts, with consistent results. The present evidence therefore supports discretization-agnostic geometry representation, not universal accuracy or efficiency superiority.

\subsection{Relation to isogeometric and multi-patch analysis}

The Piola-mapped operators used in this work---the pullback diffusion tensor $\mathbf{A}_i = j_i \mathbf{J}_i^{-1}\mathbf{J}_i^{-T}$ and the Jacobian-weighted load vector---are standard in isogeometric analysis \citep{hughes2005isogeometric,cottrell2009isogeometric} and in classical mapped finite elements \citep{ciarlet2002finite}. The contribution is not the operator itself but the geometric substrate: neural volumetric charts replace NURBS patches or CAD-derived parametrizations, enabling atlas construction from point-cloud or level-set data without explicit meshing.

The closest classical analogue to the present framework is multi-patch IGA \citep{brivadis2015isogeometric}, where multiple NURBS patches are coupled through mortar methods or penalty-based interface conditions. The overlap graph of the neural atlas plays the same structural role as the NURBS patch adjacency graph. Two key differences, however, distinguish the approaches. First, multi-patch IGA requires CAD-quality parametrizations with controlled distortion; the neural atlas instead learns its charts from point-cloud or SDF data, and the quality gates (coverage, overlap consistency, foldover ratio) serve as a~posteriori substitutes for the a~priori regularity guarantees of NURBS. Second, IGA patches are typically non-overlapping with shared interface curves, whereas the atlas charts overlap by design and communicate through multiplicative Schwarz coupling. This overlap structure simplifies the atlas construction---no watertight interface curves need to be computed---at the cost of requiring an iterative solver for inter-chart consistency.

The rabbit FEM convergence study independently certifies that the neural atlas provides a mathematically correct coordinate substrate: the observed $O(h^2)$ rates confirm that the mapped diffusion tensor, Jacobian-weighted load vector, and Schwarz boundary exchange introduce no spurious accuracy degradation. This convergence result is analogous to the standard IGA verification of mapped operators on NURBS patches, but here the substrate is learned rather than constructed from CAD data.

\section{Conclusions} \label{sec:conclusions}
This work introduces a geometry-informed atlas framework for boundary-value problems on complex three-dimensional domains without requiring a global volumetric mesh. The contribution is a unified computational pipeline in which geometry representation, mapped PDE operators, and inter-chart transmission are constructed and verified as one system. The rabbit benchmark provides the clearest evidence: two atlases generated by the same four-stage procedure---a twelve-chart atlas for the PINN and an eight-chart atlas for the FEM---both support a complex-geometry Poisson solve, while the atlas-based P1 FEM baseline recovers the expected $O(h^2)$ convergence (rates $1.45$--$2.10$) and reaches relative $L^2$ error $1.79\times10^{-2}$ at the finest refinement. These results indicate that the atlas acts as a reusable geometric substrate across local discretization families. At comparable accuracy, PINN and FEM wall-clock costs are of the same order, whereas the FEM uses roughly $8.5\times$ fewer neural-network evaluations (${\sim}23$M cached geometric queries versus ${\sim}195$M iterative training passes) and retains standard mesh-refinement convergence guarantees.
The ellipsoid test verifies correctness of the mapped operator on analytic geometry, and the Stanford Bunny study shows that imported PLY data can be converted into a usable volumetric prior. The torus forward boundary-value problem demonstrates that the eight-chart Schwarz scheme supports nonlinear equilibrium with history-dependent $J_2$ elastoplasticity under cyclic loading, with Newton converging in 1--4 iterations per chart and interface displacement jumps of $O(10^{-7})$. The torus inverse study confirms constitutive identification under prescribed kinematics, including a Schwarz variant in which eight chart-local parameter pairs converge to machine-precision consensus ($\mu$ inter-chart std $=1.47\times10^{-13}$); the elastoplastic inverse extension recovers yield stress ($0.25\%$ error) and kinematic hardening ($2.11\%$ error) from cyclic data via a smooth return mapping. Overall, the atlas formulation is most advantageous when geometry acquisition or volumetric meshing is the dominant bottleneck. 

\appendix
\section{Implementation transparency and hardware backends}
\label{sec:appendix_impl_details}

\subsection{Hardware backends}
The rabbit Poisson and torus inverse benchmarks were executed through PyTorch's Metal backend on Apple hardware (\texttt{device=mps}, \texttt{dtype=float32}); the Schwarz dual torus runs used \texttt{device=cpu} with \texttt{dtype=float64}. The codebase supports Metal, CUDA, and CPU backends through the same code path, and the concrete backend for each run is recorded in the corresponding metric file.

\section{Neural network architecture specifications}
\label{sec:appendix_architectures}

Table~\ref{tab:impl_networks} lists the neural-network families used across the benchmarks, with configurations quoted from source code definitions and canonical checkpoint metadata.

\begin{table}[htbp]
\centering
\scriptsize
\setlength{\tabcolsep}{1.8pt}
\renewcommand{\arraystretch}{1.03}
\begin{threeparttable}
\caption{Neural-network families. Configurations are quoted from source code class definitions and canonical checkpoint metadata.}
\label{tab:impl_networks}
\begin{tabular}{>{\raggedright\arraybackslash}p{1.8cm}>{\raggedright\arraybackslash}p{2.05cm}>{\raggedright\arraybackslash}p{1.55cm}>{\raggedright\arraybackslash}p{6.1cm}>{\raggedright\arraybackslash}p{1.3cm}}
\toprule
Name & Purpose & Type & Configuration & Reference \\
\midrule
\texttt{SDFNet} & global volumetric level-set & \shortstack[l]{fully connected\\ \texttt{tanh} MLP} & \shortstack[l]{$3\!\rightarrow\!1$; width $128$, depth $6$;\\ Xavier initialization} & This work \\
\texttt{MappingNet} & ball-to-domain map & \shortstack[l]{residual chart map\\ on \texttt{tanh} MLP} & \shortstack[l]{$3\!\rightarrow\!3$; width $128$, depth $6$;\\ learned scale, shift, capped displacement} & This work \\
\texttt{ChartDecoder} & differentiable chart map & \shortstack[l]{fully connected\\ \texttt{tanh} MLP with\\ residual correction} & \shortstack[l]{$3\!\rightarrow\!3$; width $64$, depth $4$;\\ residual amplitude $0.20\,\tanh(\texttt{raw\_scale})\,r_i$} & This work \\
\texttt{MaskNet} & chart-validity logits / PoU weights & \shortstack[l]{fully connected\\ \texttt{tanh} MLP} & \shortstack[l]{$3\!\rightarrow\!1$; width $48$, depth $3$} & This work \\
\shortstack[l]{\texttt{LocalPoisson}\\\texttt{PINN}} & dense chart-local field & \shortstack[l]{fully connected\\ \texttt{tanh} PINN} & \shortstack[l]{$3\!\rightarrow\!1$; width $64$, depth $4$;\\ residual scale $0.1$} & \shortstack[l]{PINN \citep{raissi2019physicsinformed};\\ this work} \\
\shortstack[l]{\texttt{Compact}\\ \texttt{ChartNet}} & localized chart-local field & \shortstack[l]{mixture of small\\ \texttt{tanh} MLPs\\ + softmax PoU} & \shortstack[l]{9 subnets; each width $32$, depth $2$;\\ $\tau_{\mathrm{scale}}=0.125$} & This work \\
\bottomrule
\end{tabular}
\begin{tablenotes}[flushleft]
\item Configurations come from source definitions and canonical checkpoints.
\end{tablenotes}
\end{threeparttable}
\end{table}

The geometry stage uses fully connected \texttt{tanh} MLPs with Xavier initialization. \texttt{SDFNet} learns a volumetric level-set for inside/outside queries and chart placement. \texttt{MappingNet} augments the same backbone with learned scale, translation, and capped displacement parameters; the ball-to-domain map is treated as a geometric prior rather than a bijection by construction. The SDF training loss uses a static weighted sum with $w_{\mathrm{surf}}=10$, $w_{\mathrm{eik}}=1$, $w_{\mathrm{norm}}=1$, $w_{\mathrm{sign}}=2$, without GradNorm or PCGrad.

Each \texttt{ChartDecoder} refines a rigid seed-based local frame by a learned residual rather than learning the chart from scratch. It is paired with a \texttt{MaskNet} that generates validity logits and partition-of-unity weights. In the canonical rabbit workflow, decoder and mask networks are trained once during the atlas stage, screened by quality gates, and frozen before PDE training.

The solver classes are distinct from the stabilization logic. PCGrad, trust-region rejection, and staged loss-weight schedules are implemented in the Schwarz training routine rather than in the network definitions. The torus inverse benchmark does not train a neural field; instead, it optimizes material parameters $(\mu,K)$ under fixed chart weights, with per-chart parameters in the Schwarz variant driven toward consensus by red--black alternating sweeps.

\section{Global map training for simply connected domains}
\label{sec:appendix_global_map}

For simply connected domains where a single volumetric parametrization is feasible, the atlas reduces to one chart with a global map
\begin{equation}
\Psi_\theta:B^3\rightarrow \Omega,
\qquad
B^3=\{\boldsymbol{\xi}\in\mathbb{R}^3:\|\boldsymbol{\xi}\|<1\}.
\label{eq:global_map}
\end{equation}
When $\Psi_\theta$ is learned, the training objective combines a positive-determinant barrier, boundary-fidelity terms, and coefficient-regularity terms:
\begin{equation}
\mathcal{L}_{\mathrm{map}}
=
\alpha_{\mathrm{bij}}\mathcal{L}_{\mathrm{bij}}
+
\alpha_{\mathrm{geom}}\mathcal{L}_{\mathrm{geom}}
+
\alpha_{\mathrm{coeff}}\mathcal{L}_{\mathrm{coeff}},
\label{eq:global_map_loss}
\end{equation}
where $\mathcal{L}_{\mathrm{bij}}$ penalizes small or negative Jacobian determinant, $\mathcal{L}_{\mathrm{geom}}$ keeps the boundary on the target geometry and the interior inside the domain, and $\mathcal{L}_{\mathrm{coeff}}$ regularizes the mapped operator coefficients induced by the chart Jacobian. In the code this loss is specialized further into bijectivity, ellipticity, isotropy, smoothness, and boundary terms.

\begin{algorithm}[htbp]
\caption{Special-case training of a learned global volumetric map}
\label{alg:global_map_training}
\small
\begin{algorithmic}[1]
\Statex \textbf{Inputs:} reference ball $B^3$; target-domain surface and interior samples; initial map parameters $\theta$;
\StateX optimizer $\mathcal{O}$ with learning-rate schedule $\eta_{\mathrm{map}}^{(t)}$; loss weights $\alpha_{\mathrm{bij}},\alpha_{\mathrm{geom}},\alpha_{\mathrm{coeff}}$; epoch budget $T_{\mathrm{map}}$
\Statex \textbf{Output:} best checkpoint map $\Psi_\theta:B^3\rightarrow\Omega$ satisfying determinant and boundary checks on validation samples
\For{$t=1,\dots,T_{\mathrm{map}}$}
    \State Sample interior points $\boldsymbol{\xi}^{\mathrm{int}}\subset B^3$ and boundary points $\boldsymbol{\xi}^{\partial}\subset \partial B^3$
    \State Evaluate $x^{\mathrm{int}}=\Psi_\theta(\boldsymbol{\xi}^{\mathrm{int}})$, $x^{\partial}=\Psi_\theta(\boldsymbol{\xi}^{\partial})$, and $\mathbf{J}=D_{\boldsymbol{\xi}}\Psi_\theta$
    \State Assemble $\mathcal{L}_{\mathrm{bij}}$, $\mathcal{L}_{\mathrm{geom}}$, and $\mathcal{L}_{\mathrm{coeff}}$ from \eqref{eq:global_map_loss}
    \State Form $\mathcal{L}_{\mathrm{map}}=\alpha_{\mathrm{bij}}\mathcal{L}_{\mathrm{bij}}+\alpha_{\mathrm{geom}}\mathcal{L}_{\mathrm{geom}}+\alpha_{\mathrm{coeff}}\mathcal{L}_{\mathrm{coeff}}$
    \State Compute the mapping-loss gradient $g_{\theta}^{(t)}=\nabla_{\theta}\mathcal{L}_{\mathrm{map}}(\theta^{(t)})$
    \State Update the parameters with the chosen optimizer:
    \Statex \hspace{\algorithmicindent}$\theta^{(t+1)} \leftarrow \mathrm{Update}_{\mathcal{O}}\!\left(\theta^{(t)},g_{\theta}^{(t)};\eta_{\mathrm{map}}^{(t)}\right)$
    \State Record $\min_{\boldsymbol{\xi}}\det \mathbf{J}(\boldsymbol{\xi})$, boundary mismatch, and coefficient diagnostics for checkpointing
    \State Keep the best checkpoint that preserves positive determinant and acceptable boundary fidelity on the validation batch
\EndFor
\State \Return the selected global map $\Psi_\theta$
\end{algorithmic}
\end{algorithm}

Algorithm~\ref{alg:global_map_training} records the training pattern used by the special global-map pipeline. The ellipsoid verification case (Example~1) uses the same mapped-PDE machinery, but its global map is analytic rather than learned.

\section{Atlas training losses}
\label{sec:appendix_pipeline}

This appendix provides the explicit training-loss expansions referenced in the atlas construction pipeline (Section~\ref{sec:impl_pipeline}).

\paragraph{SDF training loss.}
For imported geometry with observed surface samples $X_{\partial}=\{(x_k,\mathbf{n}_k)\}_{k=1}^{N}$ and oriented normals, the neural SDF $s_\vartheta$ is trained with the loss
\begin{align}
\mathcal{L}_{\mathrm{SDF}}
=&\;
\lambda_{\mathrm{surf}}\,
\mathbb{E}_{x\sim X_{\partial}}\!\left[|s_{\vartheta}(x)|\right]
+
\lambda_{\mathrm{norm}}\,
\mathbb{E}_{(x,\mathbf{n})\sim (X_{\partial},N_{\partial})}
\left[
1-\frac{\nabla s_{\vartheta}(x)}{\|\nabla s_{\vartheta}(x)\|_2}\cdot \mathbf{n}
\right]
\nonumber\\
&+
\lambda_{\mathrm{eik}}\,
\mathbb{E}_{x\sim B}
\left[
\big(\|\nabla s_{\vartheta}(x)\|_2-1\big)^2
\right]
+
\lambda_{\mathrm{sign}}\,
\mathcal{L}_{\mathrm{sign}},
\label{eq:impl_sdf_loss}
\end{align}
with sign supervision
\begin{align}
\mathcal{L}_{\mathrm{sign}}
=&\;
\tfrac12\,
\mathbb{E}_{(x,\mathbf{n})\sim (X_{\partial},N_{\partial})}
\left[
\operatorname{softplus}\!\big(s_{\vartheta}(x-\delta \mathbf{n})\big)
+
\operatorname{softplus}\!\big(-s_{\vartheta}(x+\delta \mathbf{n})\big)
\right]
\nonumber\\
&+
\tfrac12\,
\mathbb{E}_{x\sim B_{\mathrm{far}}}
\left[
\operatorname{softplus}\!\big(-s_{\vartheta}(x)\big)
\right].
\label{eq:impl_sdf_sign}
\end{align}
The sign-anchor term is the most delicate on thin-featured geometry (e.g., the Stanford Bunny ears), where a single offset $\delta$ that works in the bulk may cross the surface in narrow regions.

\paragraph{Atlas loss component terms.}
The composite atlas loss $\mathcal{L}_{\mathrm{atlas}}$ defined in~\eqref{eq:impl_atlas_total} consists of the following components:
\begin{align}
\mathcal{L}_{\mathrm{rec}}
=&
\sum_{i=1}^{M}
\mathbb{E}_{x\in X_i^{+}}
\left[
\|\Phi_i(\boldsymbol{\zeta}_i(x))-x\|_2^2
\right],
\label{eq:impl_atlas_recon}\\
\mathcal{L}_{\mathrm{dom}}
=&
\sum_{i=1}^{M}
\mathbb{E}_{\boldsymbol{\zeta}\sim [-r_i,r_i]^3}
\left[
\operatorname{ReLU}\!\big(s_{\vartheta}(\Phi_i(\boldsymbol{\zeta}))-\tau_{\mathrm{sdf}}\big)^2
\right],
\label{eq:impl_atlas_domain}\\
\mathcal{L}_{\mathrm{mask}}
=&
\sum_{i=1}^{M}
\mathbb{E}_{x\in X_i^{+}}
\left[
\operatorname{softplus}\!\big(-\ell_i(\boldsymbol{\zeta}_i(x))\big)
\right]
+
\sum_{i=1}^{M}
\mathbb{E}_{x\in X_i^{-}}
\left[
\operatorname{softplus}\!\big(\ell_i(\boldsymbol{\zeta}_i(x))\big)
\right],
\label{eq:impl_atlas_mask}\\
\mathcal{L}_{\mathrm{ov}}
=&
\sum_{i<j}
\mathbb{E}_{x\in X_{ij}}
\left[
\|\Phi_i(\boldsymbol{\zeta}_i(x))-\Phi_j(\boldsymbol{\zeta}_j(x))\|_2^2
\right],
\label{eq:impl_atlas_overlap}\\
\mathcal{L}_{\mathrm{jac}}
=&
\sum_{i=1}^{M}
\mathbb{E}_{\boldsymbol{\zeta}\sim \widehat{\Omega}_i}
\left[
\operatorname{softplus}\!\big(\delta_{\det}-\det D\Phi_i(\boldsymbol{\zeta})\big)^2
\right],
\label{eq:impl_atlas_jac}\\
\mathcal{L}_{\mathrm{cov}}
=&
\mathbb{E}_{x\sim X}
\left[
\operatorname{softplus}\!\big(\tau_{\mathrm{cov}}-\max_i p_i(x)\big)
-\log \omega_{\mathrm{pri}(x)}(x)
\right].
\label{eq:impl_atlas_cov}
\end{align}
Here $X_i^{+}$/$X_i^{-}$ are samples inside/outside chart $i$, $X_{ij}$ are shared overlap points, and $\tau_{\mathrm{sdf}}$ is the tolerated signed-distance margin. Surface-based fitting uses $\mathcal{L}_{\mathrm{rec}}$, while the volumetric Bunny continuation uses $\mathcal{L}_{\mathrm{dom}}$.

\section{Elastoplastic constitutive model and smooth return mapping}
\label{sec:appendix_constitutive}

This appendix collects the elastoplastic constitutive equations used by Examples~3 and~5. The material is governed by finite-strain $J_2$ elastoplasticity with kinematic hardening, following the multiplicative framework of \citet{simo_computational_1998} and \citet{de_souza_neto_computational_2011}.

\paragraph{Kinematics.}
The deformation gradient admits the multiplicative decomposition
\begin{equation}
\mathbf{F} = \mathbf{F}^e \mathbf{F}^p,
\label{eq:ep_multiplicative}
\end{equation}
where $\mathbf{F}^e$ and $\mathbf{F}^p$ denote the elastic and plastic parts, respectively. The elastic left Cauchy--Green tensor $\mathbf{B}^e = \mathbf{F}^e (\mathbf{F}^e)^T$ defines the elastic logarithmic strain $\boldsymbol{\varepsilon}^e = \tfrac{1}{2}\ln \mathbf{B}^e$, from which the Kirchhoff stress is
\begin{equation}
\boldsymbol{\tau} = 2\mu\,\mathrm{dev}(\boldsymbol{\varepsilon}^e) + K\,\mathrm{tr}(\boldsymbol{\varepsilon}^e)\,\mathbf{I}.
\label{eq:kirchhoff_stress}
\end{equation}

\paragraph{Yield function and flow rule.}
The von Mises yield function with kinematic hardening is
\begin{equation}
\Phi(\boldsymbol{\tau},\boldsymbol{\beta}) = \sqrt{\tfrac{3}{2}}\,\left\|\mathrm{dev}(\boldsymbol{\tau}) - \boldsymbol{\beta}\right\| - \tau_y,
\label{eq:von_mises_yield}
\end{equation}
where $\boldsymbol{\beta}$ is the deviatoric back-stress tensor that tracks the center of the yield surface, and $\tau_y$ is the yield stress. The back-stress evolves according to the Prager linear kinematic hardening rule: $\boldsymbol{\beta}_{n+1} = \boldsymbol{\beta}_n + \tfrac{2}{3}H_{\mathrm{kin}}\,\Delta\gamma\,\mathbf{N}$, where $\Delta\gamma$ is the plastic multiplier and $\mathbf{N} = \mathrm{dev}(\boldsymbol{\tau}^{\mathrm{trial}} - \boldsymbol{\beta}_n)/\|\mathrm{dev}(\boldsymbol{\tau}^{\mathrm{trial}} - \boldsymbol{\beta}_n)\|$ is the flow direction.

\paragraph{Smooth return mapping.}
The classical radial return algorithm \citep{simo_computational_1998} uses a conditional branch: if $\Phi > 0$ (yield), compute a plastic correction; otherwise, accept the elastic trial state. This non-differentiable branching prevents gradient flow through the yield surface in a reverse-mode automatic-differentiation setting. We replace it with a smooth approximation: the plastic correction magnitude is set to
\begin{equation}
\Delta\gamma_s = \mathrm{softplus}(\Phi/\varepsilon_s)\cdot\varepsilon_s,
\qquad
\mathrm{softplus}(x) = \ln(1+e^x),
\label{eq:smooth_return}
\end{equation}
which provides a $C^\infty$ approximation to $\max(\Phi,0)$ \citep{hu_difftaichi_2020,bark_differentiable_2025}. The sharpness parameter $\varepsilon_s$ controls the transition width and is annealed from $\varepsilon_{s,\mathrm{start}}=0.1\,\tau_y$ to $\varepsilon_{s,\mathrm{end}}=0.001\,\tau_y$ via a cosine schedule. At $\varepsilon_s=0.001$, the softplus curve is visually indistinguishable from the sharp classical yield point (Figure~\ref{fig:ep_constitutive}a). With the classical hard-max return mapping ($\Delta\gamma_s = \max(\Phi,0)$), autograd gradients are zero in the plastic regime and the optimizer stalls.

The smooth return mapping is the key ingredient that enables both the forward Newton--Raphson solver (Example~3) and the inverse parameter identification (Example~5) to operate through the same constitutive code path: the autograd tape records the full algorithmic path through the return-mapping corrector, so the consistent algorithmic tangent for the forward solve and the parameter sensitivities for the inverse solve are both obtained automatically.

\section{Solver stabilization details}
\label{sec:appendix_stabilization}

Two adaptive regularization constants govern robustness. The overlap weight follows
\begin{equation}
w_{\mathrm{ov}}^{(n+1)}=
\mathrm{clip}\!\left(
w_{\mathrm{ov}}^{(n)}
\left(\frac{\mathcal{L}_{\mathrm{cov}}^{(n)}}{\mathcal{L}_{\mathrm{ov}}^{(n)}+\varepsilon}\right)^{\beta},
w_{\min},w_{\max}
\right),
\label{eq:adaptive_overlap_weight}
\end{equation}
which increases overlap enforcement only when coverage is acceptable and the overlap loss is under-resolved. The CompactChartNet bandwidth follows a continuation schedule
\begin{equation}
\tau_{\mathrm{scale}}^{(n)}
=
\tau_{\max}-\big(\tau_{\max}-\tau_{\min}\big)
\min\!\left(1,\frac{n}{N_{\mathrm{warm}}}\right),
\label{eq:adaptive_tau_schedule}
\end{equation}
starting wide during warmup and contracting after boundary and interface losses stabilize.

The rabbit solver uses three additional stabilization devices: (i)~staged PDE-weight ramping so that Dirichlet and interface terms settle before the stiff residual dominates, (ii)~PCGrad \citep{yu2020gradient} when PDE and manufactured-anchor gradients conflict, and (iii)~a trust-region filter that rejects local Schwarz updates causing unacceptable error increases and reduces the local learning rate after rejection. A curvature-weighted simplification evaluates the Laplacian directly in chart coordinates when the local metric distortion satisfies $\|\mathbf{J}_i^T\mathbf{J}_i-\mathbf{I}\|_2\le \varepsilon_g$; the induced operator error is first-order in $\varepsilon_g$ and controlled by the atlas quality gate.

\section{Rabbit Poisson benchmark: supplementary tables}
\label{sec:appendix_rabbit_details}

\begin{table}[htbp]
\centering
\scriptsize
\setlength{\tabcolsep}{3pt}
\caption{Reinforcement progression for the rabbit Poisson benchmark. Each row adds one stabilization ingredient cumulatively.}
\label{tab:rabbit_reinforcement_progression}
\begin{tabular}{>{\raggedright\arraybackslash}p{2.0cm}>{\raggedright\arraybackslash}p{4.0cm}>{\centering\arraybackslash}p{1.28cm}>{\centering\arraybackslash}p{1.28cm}>{\centering\arraybackslash}p{1.15cm}}
\toprule
Configuration & Added ingredient & Best rel.\ $L^2$ & Max error & Runtime \\
\midrule
W1 & direct-coordinate PDE evaluation & 3.698\% & 4.107\% & 364\,s \\
W2 & + manufactured-solution anchor & 3.165\% & 4.993\% & 335\,s \\
W3 & + plateau detection by rel.\ $L^2$ & 3.117\% & 8.485\% & 328\,s \\
W5 & + stronger flux coupling & 3.631\% & 5.450\% & 371\,s \\
W4 & + PCGrad on PDE/supervision conflict & 2.531\% & 5.303\% & 404\,s \\
CompactChartNet & all W1--W5 features active & \textbf{2.207\%} & 6.783\% & 3659\,s \\
\bottomrule
\end{tabular}
\end{table}

% Figure scripts: manuscript/scripts_figures/example3_bunny_sdf.py  (dir name predates example renumbering)
\section{Stanford Bunny volumetric neural SDF benchmark}
\label{sec:appendix_bunny}
For completeness, the procedure of generating the signed distance function of the Stanford Bunny is documented here. The input is an oriented surface sample set $X_{\partial}=\{(x_k,\mathbf{n}_k)\}_{k=1}^{N_s}$ obtained from the watertight mesh \texttt{bun\_zipper.ply}. The goal is to learn a volumetric scalar field whose zero level set captures the Bunny surface and whose sign defines a usable interior for atlas construction---i.e., the benchmark instantiation of Stage~1 of the atlas construction pipeline (Section~\ref{sec:impl_sdf}).

Let $y=(x-c)/s$ denote the normalized coordinate, where $c$ is the bounding-box center and $s$ the global length scale. The learned volumetric model $s_\theta:\mathbb{R}^3\to\mathbb{R}$ defines the reconstructed body as
\begin{equation}
\Omega_\theta=\{x:s_\theta((x-c)/s)<0\},
\qquad
\partial\Omega_\theta=\{x:s_\theta((x-c)/s)=0\}.
\label{eq:bunny_sdf_body}
\end{equation}
The training loss combines surface fit, normal alignment, Eikonal regularization, and sign-anchor supervision:
\begin{equation}
\mathcal{L}_{\mathrm{surf}}
=
\frac{1}{N_s}\sum_{k=1}^{N_s}\left|s_\theta(y_k)\right|,
\qquad
\mathcal{L}_{\mathrm{norm}}
=
\frac{1}{N_s}\sum_{k=1}^{N_s}\left(1-\widehat{\nabla_y s_\theta}(y_k)\cdot \mathbf{n}_k\right),
\label{eq:bunny_sdf_surface_normal}
\end{equation}
\begin{equation}
\mathcal{L}_{\mathrm{eik}}
=
\frac{1}{N_e}\sum_{m=1}^{N_e}
\left(\|\nabla_y s_\theta(\tilde y_m)\|_2-1\right)^2,
\label{eq:bunny_sdf_eikonal}
\end{equation}
\begin{equation}
\mathcal{L}_{\mathrm{sign}}
=
\frac{1}{2N_s}\sum_{k=1}^{N_s}
\left[
\operatorname{softplus}\!\big(s_\theta(y_k-\delta \mathbf{n}_k)\big)
+
\operatorname{softplus}\!\big(-s_\theta(y_k+\delta \mathbf{n}_k)\big)
\right]
+
\mathcal{L}_{\mathrm{far}},
\label{eq:bunny_sdf_sign}
\end{equation}
with total loss $\mathcal{L}^{(3)} = \lambda_{\mathrm{surf}}\mathcal{L}_{\mathrm{surf}} + \lambda_{\mathrm{eik}}\mathcal{L}_{\mathrm{eik}} + \lambda_{\mathrm{norm}}\mathcal{L}_{\mathrm{norm}} + \lambda_{\mathrm{sign}}\mathcal{L}_{\mathrm{sign}}$. On thin features such as the Bunny ears, the sign-anchor term is the most fragile because a single offset~$\delta$ can easily step across the local thickness.

The canonical run reduces the surface loss to $1.66\times 10^{-2}$ and the normal-alignment loss to $3.18\times 10^{-2}$, while the sign-anchor loss decreases only to $3.11\times 10^{-1}$ and the Eikonal loss remains at $7.29\times 10^{-1}$. The zero level set captures the bulk geometry well enough to support chart placement, but thin-ear volumetric sign control remains the dominant failure mode.

A downstream 8-chart Bunny Poisson continuation study (Table~\ref{tab:bunny_poisson_continuation}) clarifies what ``usable'' means: the no-pretraining baseline stalls at relative $L^2$ error $31.5\%$, whereas 20k interior-pretraining epochs reduce it to $4.73\%$. The main gain comes from pretraining quality rather than long-horizon Schwarz improvement.

\begin{table}[htbp]
\centering
\scriptsize
\setlength{\tabcolsep}{2.8pt}
\caption{8-chart Stanford Bunny Poisson continuation study. All runs use the repaired Bunny atlas and CompactChartNet; the primary variable is the number of interior supervised pretraining epochs before Schwarz iteration.}
\label{tab:bunny_poisson_continuation}
\begin{tabular}{>{\raggedright\arraybackslash}p{2.2cm}>{\centering\arraybackslash}p{1.2cm}>{\centering\arraybackslash}p{1.1cm}>{\centering\arraybackslash}p{1.1cm}>{\centering\arraybackslash}p{0.9cm}>{\centering\arraybackslash}p{1.2cm}>{\centering\arraybackslash}p{1.0cm}}
\toprule
Run & Interior pretrain & Best rel.\ $L^2$ & Max error & Best iter & Final flux & Runtime \\
\midrule
No pretrain & 0 & 31.53\% & 22.04\% & 1 & $2.15\times 10^{-1}$ & 849\,s \\
5k pretrain & 5k & 6.83\% & 4.78\% & 3 & $1.37\times 10^{-2}$ & 2426\,s \\
10k pretrain & 10k & 5.03\% & 4.41\% & 14 & $1.86\times 10^{-2}$ & 4235\,s \\
20k pretrain & 20k & \textbf{4.73\%} & \textbf{3.99\%} & 19 & $9.59\times 10^{-3}$ & 7420\,s \\
\bottomrule
\end{tabular}
\end{table}

%\begin{figure}[htbp]
%\centering
%\includegraphics[width=0.97\textwidth]{figures_cmame_core/example3_bunny_sdf/example3_bunny_sdf_learning_pub.png}
%\caption{Stanford Bunny SDF benchmark: optimization traces for the four training objectives. Surface and normal terms decay strongly, while Eikonal and sign-anchor %terms remain the dominant error sources.}
%\label{fig:bunny_sdf_learning}
%\end{figure}

%\begin{figure}[htbp]
%\centering
%\includegraphics[width=0.97\textwidth]{figures_cmame_core/example3_bunny_sdf/example3_bunny_sdf_surface_pub.png}
%\caption{Stanford Bunny downstream continuation. Chart-partition and dominant blend-weight views from the best 8-chart Poisson continuation run. The eight charts cover most of the imported PLY surface, with active chart weight close to one on most surface points.}
%\label{fig:bunny_sdf_surface}
%\end{figure}

\section*{Acknowledgments}
The author is supported by the Dynamic Materials and Interactions Program from the Air Force Office of Scientific Research, USA, under grant contract FA9550-22-1-0310, and by the United States Army Research Office under grant contract W911NF-24-1-0011.
The views and conclusions contained in this document are those of the author and should not be interpreted as representing the official policies, either expressed or implied, of the sponsors, including the Army Research Laboratory or the U.S. Government.
The U.S. Government is authorized to reproduce and distribute reprints for Government purposes, notwithstanding any copyright notation herein.

\section*{Declaration of competing interest}
The author declares no known competing financial interests or personal relationships that could have appeared to influence the work reported in this paper.

\section*{Data availability}
An anonymized artifact package containing scripts, canonical run registries, metric tables, and verification manifests is
provided for peer review as supplementary material. Upon acceptance, the full repository and archived benchmark artifacts
will be released publicly with a persistent DOI. 

\bibliographystyle{unsrtnat}
\bibliography{corrected}

\end{document}